\documentclass[a4paper,11pt]{article}

\usepackage[utf8]{inputenc}
\usepackage[T1,T2A]{fontenc}
\usepackage{amsmath,amsfonts,amssymb}
\usepackage{mathtools} % For dcases and multlined
\usepackage{bbold} % For \mathbb{1} and \mathbb{0}
\usepackage{bm} % For bold math symbols
\usepackage{multirow}
\usepackage{diagbox}
\usepackage{hhline}
\usepackage{cite}
\usepackage{graphicx}
\usepackage{wrapfig}

\usepackage{xcolor}
\definecolor{darkGreen}{RGB}{2,156,43}
\definecolor{darkBlue}{RGB}{44,105,200}

\usepackage{hyperref}
\hypersetup{colorlinks=true,linkcolor=darkBlue,citecolor=darkGreen}
\usepackage{geometry}
\numberwithin{equation}{section}

\newcommand{\lb}{\left(}
\newcommand{\rb}{\right)}
\newcommand{\lsb}{\left[}
\newcommand{\rsb}{\right]}
\newcommand{\veps}{\varepsilon}
\newcommand{\vk}{\varkappa}
\newcommand{\pz}{\partial_z}
\newcommand{\qrq}{\quad\Rightarrow\quad}
\newcommand{\vphfrac}{\vphantom{\frac{1}{2}}}

\newcommand{\hx}{\hat{x}}
\newcommand{\hy}{\hat{y}}
\newcommand{\hz}{\hat{z}}
\newcommand{\hg}{\hat{g}}

\newcommand{\ukk}{U_\text{KK}}
\newcommand{\mkk}{M_\text{KK}}

\allowdisplaybreaks

\begin{document}

\begin{titlepage}

\begin{center}
  \textbf{\LARGE Sakai-Sugimoto Model in an Off-Shell:\\Chiral Lagrangian to All Orders}
\end{center}

\vspace{1cm}

\begin{center}
  \textbf{Michael Lublinsky, Timofey Solomko}
\end{center}

\begin{center}
  \vspace*{0.15cm}
  \textit{Physics Department, Ben-Gurion University of the Negev, Beer Sheva 84105, Israel}
\end{center}

\vspace{2.5cm}

\begin{center}
  \textbf{Abstract}
\end{center}

\noindent The Sakai-Sugimoto holographic model is famous for implementing the approximate
chiral symmetry of QCD and reproducing the Chiral Lagrangian in a top-down approach. In
this manuscript, we revisit the model in a formalism that is somewhat different from the
original work by Sakai and Sugimoto:~We start by identifying boundary degrees of freedom
and splitting the bulk equations of motion into dynamical ones and constraints. The
former are then solved to all orders in derivatives of the boundary fields. The
constraints are left unsolved, leaving the dynamical degrees of freedom off-shell. This
approach enables us to systematically derive the effective action of the boundary theory.
The derived effective action is very rich in physics: it contains an \(U(N_f)\) multiplet
of massless pseudoscalars interacting (via trilinear and higher terms) with towers of
massive (axial-)vector mesons. In contrast to the previous studies, our effective action
is non-local. The original Chiral Lagrangian is recovered as its local expansion in small
\(\pi\)-meson momenta (derivative expansion). We particularly zoom in on the values of
four derivative terms couplings, the low energy constants, and compare those with the
ones reported in the literature.
\end{titlepage}

\tableofcontents

\section{Introduction}

The AdS/CFT-correspondence (or holography) has been a very fruitful research field since
its discovery within string theory~\cite{Maldacena:1997re,Witten:1998zw,Witten:1998qj,
Gubser:1998bc,Aharony:1999ti,Casalderrey-Solana:2011dxg}. The correspondence relates the
physics of a \emph{boundary} (gauge) theory with the physics of the spacetime of higher
dimensions (the \emph{bulk}; usually, a theory of gravity). One of the directions of
holographic research is the top-down holography where effective theories of different
physical phenomena on the boundary are derived from string theory (brane) constructions
in the bulk.

Multiple top-down models have been developed intended to capture various physical aspects
of quantum chromodynamics (QCD)~\cite{Babington:2003vm,Kruczenski:2003uq,Kruczenski:2004me}.
Perhaps one of the most famous top-down models is the (Witten-)Sakai-Sugimoto (SS) model~\cite{Witten:1998zw,
Sakai:2004cn,Sakai:2005yt}. This model consists of \(N_f\) pairs of \(D8\) branes put
into the spacetime background generated by \(N_c\gg1\) \(D4\) branes. Its action is given
by the Dirac-Born-Infeld (DBI) and Chern-Simons (CS) terms for the bulk gauge field
\(A_M\in\mathfrak{u}(N_f)\).\footnote{\(U(N_f)\) denotes the Lie group of gauge
transformations, while \(\mathfrak{u}(N_f)\) refers to the corresponding Lie algebra.}
We expand further on the model details in the following sections.

Arguably, the main result of the SS model is the holographic derivation of the Chiral
Lagrangian, the effective low energy theory of QCD which implements the approximate
chiral symmetry and is an expansion in powers of \(\pi\)-meson momentum~\cite{Gasser:1984gg,
Ecker:1994gg,Pich:2018ltt},
\begin{multline}\label{eq:chi_lagr}
  \mathcal{L}=-\frac{f_\pi^{2}}{4}\text{Tr}\lb\partial_\mu\Sigma^\dagger\partial^\mu\Sigma\rb+
  L_1\lb\text{Tr}\lb\partial_\mu\Sigma^\dagger\partial^\mu\Sigma\rb\rb^2+
  L_2\lb\text{Tr}\lb\partial_\mu\Sigma^\dagger\partial_\nu\Sigma\rb\rb^2\\-
  \frac{1}{2}\partial_\mu\eta'\partial^\mu\eta'-
  \frac{1}{2}\sum_n\text{Tr}\lb\partial_\mu V_{\nu,n}\partial^\mu V^\nu_n\rb-
  \frac{1}{2}\sum_nm^2_n\text{Tr}\lb V_{\mu,n}V^\mu_n\rb+\dots,
\end{multline}
where \(\Sigma\), \(\eta'\), and \(V_{\mu,n}\) are the low energy degrees of freedom.
\(\Sigma\) is the \(SU(N_f)\) matrix which contains the multiplet of massless
pseudoscalar \(\pi\)-mesons (Nambu-Goldstone bosons of the chiral symmetry),
\begin{equation}\label{eq:sigma_exp}
  \Sigma=\exp\lb\frac{2i}{f_\pi}\pi^aT^a\rb.
\end{equation}
\(\eta'\) is the massless pseudoscalar meson, which acquires its mass through \(U(1)_A\)
anomaly. The supergravity-based mechanism behind the mass term of the \(\eta'\)-meson was
discussed in the original work~\cite{Sakai:2004cn}. This mass-generation effect is
subleading in the large \(N_c\) limit, which is the limit usually assumed in holography
in general and in the SS model in particular. As such the massless \(\eta'\)-meson
alongside \(\Sigma\) combine into a larger \(U(N_f)\) multiplet. \(V_{\mu,n}\) are
massive (axial-)vector mesons (or rather meson multiplets since \(V_{\mu,n}\in\mathfrak{u}(N_f)\)),
with the lightest being \(\rho\), \(a_1\), etc. There is an infinite tower of such mesons
in the effective theory. In principle, the Lagrangian~\eqref{eq:chi_lagr} also involves
interactions between the vector and pseudoscalar sectors. Below we will explicitly derive
terms of the \(V\partial V\Sigma^{-1}\partial\Sigma\)-type.

Vector mesons are usually introduced into the Chiral Lagrangian via an approach based on
a Hidden Local Symmetry (HLS)~\cite{Bando:1984ej,Bando:1987br,Meissner:1987ge,Harada:2003jx,
Ma:2016npf}. Within the HLS, the action is assumed to be invariant under the chiral
symmetry and an additional, ``hidden'' local symmetry, groups. The vector mesons are
identified as gauge fields of the HLS, which, however, emerge without any kinetic terms.

The results for the SS model are frequently compared with those of the HLS approach,\footnote{It
is also worth mentioning the work~\cite{Son:2003et}, where it was demonstrated that the
infinite number of hidden local symmetries naturally gives rise to deconstruction of a
fifth dimension. This hints at further connection between holography and HLS.}
considering that the \(\Sigma\) matrix in the SS model is introduced in a way that
resembles the HLS (the details of this construction are reviewed in Appendix~\ref{app:bcond}).
An advantage of the SS model is that it does give rise to kinetic terms for the vector
mesons. Yet, the vector mesons of the SS have somewhat different transformation
properties under the hidden symmetry and different interaction terms with \(\pi\)-mesons,
as compared to HLS approach. Beyond HLS, the (axial-)vector mesons also appear in various
versions of the chiral perturbation theory, such as the Resonance Chiral Theory~\cite{Ecker:1988te,
Ecker:1989yg}, although there the vectors are commonly introduced as rank two
antisymmetric tensors~\cite{Gasser:1983yg}. In the present work, the gauge symmetry that
could be identified as the HLS is fixed (this point is discussed in the bulk of the paper).
Hence, no comparison with the HLS will be pursued any further.

The Chiral Lagrangian~\eqref{eq:chi_lagr} features several couplings: \(f_\pi\) is the
pion decay constant, \(L_1\) and \(L_2\) are referred to as low energy constants (LECs).
The form of the Chiral Lagrangian~\eqref{eq:chi_lagr} with two double-trace operators
denoted by the accompanying LECs \(L_{1,2}\) is the one customarily used for \(N_f=2\).
For arbitrary \(N_f\) instead of \(L_{1,2}\) operators the Chiral Lagrangian is naturally
expressed via two single-trace operators~\cite{Pich:2018ltt}:
\begin{equation}\label{eq:L3}
  L_3\,\text{Tr}\lb\partial_\mu\Sigma^{-1}\partial^\mu\Sigma\partial_\nu\Sigma^{-1}\partial^\nu\Sigma\rb,\qquad
  \tilde{L}_3\,\text{Tr}\lb\partial_\mu\Sigma^{-1}\partial_\nu\Sigma\partial^\mu\Sigma^{-1}\partial^\nu\Sigma\rb.
\end{equation}
For \(N_f\in\lbrace2,3\rbrace\) there exist identities that relate the single-trace and
double-trace operators. These points will be discussed in more detail later in this work.\footnote{In
fact, for a generic Chiral EFT with some underlying microscopic QFT, one would expect all
four independent 4-derivative operators, \(L_1\), \(L_2\), \(L_3\), and \(\tilde{L}_3\),
to be present in the Lagrangian. Yet, as we demonstrate below, it is a special feature of
the SS model that only \(L_3\) and \(\tilde{L}_3\) are non-vanishing.}

It is important to stress that the LECs are not defined uniquely as long as the pion
Lagrangian contains interactions with the vector mesons. In other words, it is possible
to change their values by redefining the vector meson fields, as was demonstrated in~\cite{Sakai:2004cn,
Sakai:2005yt}. We will refer to the couplings \(L_1\) and \(L_2\) which explicitly appear
in the Lagrangian with the vector meson interactions, as \emph{bare} LECs. Alternatively,
if only the pion interactions are of interest, one could integrate the vector mesons out
exactly or perturbatively. Upon the integration out, the bare couplings \(L_{1,2}\) get
dressed. We denote thus obtained \emph{effective} couplings as \(\mathbb{L}_{1,2}\).
Importantly, \(\mathbb{L}_{1,2}\) are ``physical'' as they can be uniquely related to
\(\pi\pi\) elastic scattering amplitudes (4\(\pi\) correlators).

Since the discovery of the SS model, an abundance of literature has followed devoted to
applications and improvements of the model. While we obviously cannot cover all the
literature here, examples include study of finite temperature behavior and deconfinement~\cite{Aharony:2006da},
analysis of baryon properties~\cite{Kim:2007zm,Hashimoto:2008zw}, efforts to include
quark masses and chiral condensate~\cite{Evans:2007jr,Bergman:2007pm,Aharony:2008an,
Hashimoto:2008sr,Dhar:2008um,McNees:2008km}, interpretation of the model results for
form-factor relations~\cite{Colangelo:2012ipa}, investigation of string theory aspects
such as backreaction of the probe D-branes~\cite{Burrington:2007qd,Bigazzi:2014qsa} and
open string tachyon condensation~\cite{Dhar:2007bz}, computation of bulk viscosity in the
SS model~\cite{Eling:2011ms}. The model remains highly relevant to this day; the most
recent works include study of glueballs and radiative decays of vector mesons~\cite{Hechenberger:2023ljn,
Hechenberger:2024piy}, improving description of baryonic matter~\cite{Kovensky:2023mye},
discussion of subtleties related to the infrared boundary terms with respect to the
bulk~\cite{Bartolini:2023eam}, analysis of electron-nucleon scatterings~\cite{Bigazzi:2023odl}.

Determining the values of \(L_1\) and \(L_2\) in the Chiral Lagrangian~\eqref{eq:chi_lagr}
from the SS model remains a subject of ongoing research, which is also one of the
motivations behind the present work.

The history begins with the original work~\cite{Sakai:2004cn} where it was shown that the
pion effective action (for \(N_f>2\)) takes the form of the Skyrme model. For \(N_f=3\)
the LECs \(L_1\), \(L_2\), and \(L_3\) were explicitly computed in the follow-up paper~\cite{Sakai:2005yt}.
As briefly mentioned in~\cite{Sakai:2005yt} and further elaborated in~\cite{Hoyos:2022ptd},
for \(N_f=2\), the \(L_3\) term can be reduced to the \(L_1\) term. Thus, for \(N_f=2\)
there are only two independent LECs, \(L_1\) and \(L_2\), which obey the relation (with
the explicit values to be quoted in Section~\ref{sect:lecs})
\begin{equation}
  L^{(l)}_2=-L^{(l)}_1.
\end{equation}
Within the framework of the SS model this result was derived from the leading order
expansion of the DBI action (marked by the superscript \((l)\)). In~\cite{Sakai:2004cn,
Sakai:2005yt}, the authors also brought the boundary action to the form prescribed by the
HLS. This was achieved by redefining the vector meson fields such that the new vectors
could be identified with the gauge fields of the HLS. As a result, the bare LECs got
modified and were found to vanish (as was mentioned above, the bare couplings are indeed
not unique). Below we consider only the bare LECs before any vector meson redefinition.

In the most recent work on the subject,~\cite{Hoyos:2022ptd}, the effective (not bare!)
LECs were determined through a holographic computation of the \(\pi\)-meson elastic
scattering amplitude (for \(N_f=2\)). In contrast to the original work~\cite{Sakai:2004cn},
the authors of~\cite{Hoyos:2022ptd} went beyond the lowest order in the expansion of the
DBI action, keeping higher order terms in the bulk field strength. The main result of~\cite{Hoyos:2022ptd}
is that the effective LECs are given by a sum of two contributions
\begin{equation}
  \mathbb{L}_{1,2}=L^{(l)}_{1,2}+L^{(nl)}_{1,2},
\end{equation}
where \(L^{(nl)}_{1,2}\) are new contributions induced by the non-linear effects in the
DBI. With the explicit values to be quoted below, \(L^{(nl)}_{1,2}\) obey the relation
\begin{equation}
  L^{(nl)}_2=2L^{(nl)}_1.
\end{equation}
The two contributions, \(L^{(l)}_{1,2}\) and \(L^{({nl})}_{1,2}\), have different
parametric dependence~\cite{Hoyos:2022ptd} (here, \(\lambda_\text{YM}\) is the 't Hooft
constant, one of the parameters of the SS model),
\begin{equation}
  L^{(l)}_{1,2}=\mathcal{O}\lb N_c\lambda_\text{YM}\rb,\qquad
  L^{(nl)}_{1,2}=\mathcal{O}\lb\frac{N_c}{\lambda_\text{YM}}\rb.
\end{equation}

The goal of the present work is to revisit the SS model in the off-shell formalism~\cite{Bu:2014sia,
Bu:2014ena,Crossley:2015tka}. This formalism, also known as partially on-shell, appeared
within the fluid/gravity-correspondence (a program aimed at studying hydrodynamics
through holographic methods). This formalism is the most adequate for deriving boundary
effective actions for low-energy degrees of freedom, keeping the latter off-shell.\footnote{The
original works~\cite{Sakai:2004cn,Sakai:2005yt} also employ off-shellness at certain
stages of the derivation. Yet, the advantage of our formalism is that it from the very
start, in a systematic way, identifies and separates the dynamical degrees of freedom of
the boundary effective theory.}

Holography relates a partition function of a higher dimensional theory of gravity to that
of a lower dimensional gauge theory~\cite{Witten:1998qj}. Traditionally, the action in
the bulk is ``put on-shell'' (denoted as \(S_0\)), i.e., computed on solutions of
equations of motion (EOMs) for the bulk fields. These fields are sourced at the boundary
by operators of the boundary theory. As a result, one obtains a generating functional for
the boundary theory which can be used to compute correlation functions~\cite{Witten:1998qj},
\begin{equation}
  Z_\text{boundary}=Z_\text{bulk}=e^{iS_0}.
\end{equation}
In the off-shell formalism, the EOMs are first divided into dynamical and constraints.
The dynamical equations are then solved and solutions are substituted back into the
action, while the constraints remain unsolved. This procedure leaves certain bulk degrees
of freedom unfixed. Below we explain how gauge fixing converts the unfixed bulk degrees
of freedom into dynamical fields of the boundary theory.

The main result of this work is a derivation of the effective action corresponding to the
SS model. Similarly to~\cite{Hoyos:2022ptd}, we keep the leading non-linear effects
induced by the DBI. Complete expression for the action is very lengthy and is presented
in Appendix~\ref{app:eff}, see~\eqref{eq:Seff}. The pions enter through derivative
interactions as expected for Nambu-Goldstone bosons. At quadratic order in the fields (\(\Sigma\),
\(\eta'\), and \(V_{\mu,n}\)), the action is local and consistent with~\eqref{eq:chi_lagr}.
Beyond the quadratic order, the effective action includes non-linear non-local
interaction terms, up to fourth combined order in \(\Sigma^{-1}\partial_\mu\Sigma\) and
\(V_{\mu,n}\). The non-locality originates from integrating out non-hadronic degrees of
freedom (such as quarks and gluons, that are not present in the boundary theory as the
dynamical degrees of freedom). The effective action can be alternatively represented as
an all order local derivative expansion, while the Chiral Lagrangian~\eqref{eq:chi_lagr}
corresponds to a small momenta truncation. Since all orders in the gradient expansion are
included, the effective action is relativistically causal and potentially UV complete,\footnote{Analogous
studies in the context of the fluid/gravity-correspondence refer to the all order
resummation as UV completion. We, however, avoid making such a strong statement.} in
sharp contrast to truncated theories. A wealth of hadronic physics is encoded in various
interaction terms of the effective action (we grouped them into 21 different categories!).
Novel results obtained in this work particularly include the following points.
\begin{itemize}
  \item The effective action for the boundary theory is more general compared to all
  previous results:
  \begin{itemize}
    \item[\(\circ\)] non-linear contributions from higher order bulk EOMs are computed,
    \item[\(\circ\)] the action is derived keeping all orders in the derivatives,
    \item[\(\circ\)] the value of \(N_f\) is kept arbitrary in contrast to the previous
    results limited to \(N_f=2,3\),
    \item[\(\circ\)] the abelian contributions (in the pseudoscalar sector corresponding
    to \(\eta'\)-meson) are fully included.
  \end{itemize}
  \item The LECs and the pion elastic scattering amplitude are derived for arbitrary \(N_f\).
  We highlight the difference between bare and effective LECs and clarify the role of the
  vector mesons. For \(N_f=2,3\), our results for LECs are in agreement with those of~\cite{Hoyos:2022ptd}.
  \item While we have not worked out any phenomenological implications of the obtained
  boundary effective action (this is left as a future project), we wish to flash a few
  experimentally testable results which are encoded in the effective action. These are
  meson spectra, interactions between pseudoscalar and vector sectors (such as
  \(\omega\to\rho\pi\) and \(\rho(1450)\to \lb a_1(1260)+\pi\rb_{P\text{-wave}}\)),
  \(\pi\pi\)-scattering amplitudes.
\end{itemize}

\subsection{Walkthrough}

Below we describe the strategy and major steps for deriving the effective action.

\noindent \(\mathbf{1.}\) \textbf{The model.}

The spacetime geometry of the model is provided by a stack of \(D4\) branes. The matter
is introduced via insertion of pairs of \(D8\) branes into this background. The resulting
bulk theory is a theory of a non-abelian \(\mathfrak{u}(N_f)\) vector gauge field \(A_M\)
in a 5D curved spacetime. The action of the model consists of the DBI action for the
field \(A_M\) and a CS term. The DBI action is expanded in the limit \(\alpha'=l_s^2\to0\)
which corresponds to expansion in powers of the field strength. The leading term in the
expansion is quadratic in \(A_M\). The present work will be limited to expansion up to
the fourth order. This expansion does not apply to the CS term, which is treated as is.
The model setup is reviewed in detail in the next Section~\ref{sect:mod_rec}.

\noindent \(\mathbf{2.}\) \textbf{Gauge fixing and boundary conditions.}

The model has 5D \(U(N_f)\) gauge symmetry which at the boundary is realized as
\(U(N_f)_L\times U(N_f)_R\) chiral symmetry (left and right correspond to two different
boundaries with respect to the holographic coordinate, \(z\to+\infty\) and \(z\to-\infty\)).
The \(U(N_f)\) gauge freedom is partially fixed by imposing the ``axial'' gauge,
\begin{equation}\label{eq:ax_g}
  A_z=0.
\end{equation}
The Wilson line of the \(z\)-component of the bulk gauge field (before the imposition of
the axial gauge) stretched between the boundaries is identified as the \(\Sigma\) matrix
and \(\eta'\)-meson. This identification is based on the line's transformation properties
under the \(U(N_f)_L\times U(N_f)_R\) group at the boundary. The \(\Sigma\) matrix and
the \(\eta'\)-meson emerge as boundary conditions for the bulk gauge fields after a
residual gauge transformation, which leaves the axial gauge intact, is performed,
\begin{equation}\label{eq:intro_bc}
  A_\mu(x,z)\underset{z\to+\infty}{\to}B_\mu(x),\qquad
  A_\mu(x,z)\underset{z\to-\infty}{\to}0.
\end{equation}
where
\begin{equation}\label{eq:lo_sols_amp}
  \begin{gathered}
    b_\mu(x)\equiv-\sqrt{\frac{2}{N_f}}\frac{1}{f_\pi}\partial_\mu\eta',\qquad
    B^a_\mu(x)\equiv-2i\,\text{Tr}\lb\Sigma^{-1}\partial_\mu\Sigma T^a\rb,\\[0.2cm]
    B_\mu(x)\equiv b_\mu(x)\,\mathbb{1}_{N_f}+B^a_\mu(x)\,T^a=
    -\sqrt{\frac{2}{N_f}}\frac{1}{f_\pi}\partial_\mu\eta'\mathbb{1}_{N_f}-
    i\Sigma^{-1}\partial_\mu\Sigma.
  \end{gathered}
\end{equation}
Here, \(\eta'\) is identified as the \(\eta'\)-meson and \(\Sigma\) as the matrix of the
\(\pi\)-meson multiplet~\eqref{eq:sigma_exp}. We will occasionally refer to the set of
the boundary fields (\(b_\mu\), \(B^a_\mu\), and \(B_\mu\)) as \(\mathit{B}\)\emph{-fields}.
From the expansion of \(\Sigma\) in~\eqref{eq:sigma_exp} to first order in the pion field,
\begin{equation}
  \Sigma=1+\frac{2i}{f_\pi}\pi^aT^a+\mathcal{O}\lb(\pi^a)^2\rb,
\end{equation}
follow the expansions of the \(B\)-fields,
\begin{equation}\label{eq:B_exp}
  B^a_\mu(x)\approx\frac{2}{f_\pi}\partial_\mu\pi^a-
  \frac{8i}{f^2_\pi}\pi^b\partial_\mu\pi^c\,\text{Tr}\lb T^aT^bT^c\rb,\quad
  B_\mu\approx-\sqrt{\frac{2}{N_f}}\frac{1}{f_\pi}\partial_\mu\eta'\mathbb{1}_{N_f}+
  \frac{2}{f_\pi}\partial_\mu\pi^aT^a.
\end{equation}
Details of the gauge fixing and its relation to the boundary conditions, which are the
key parts of the off-shell formalism, are presented in Appendix~\ref{app:bcond}.
In the bulk theory, fixing the axial gauge amounts to changing the integration variable
from \(A_z\) to \(\Sigma\) and \(\eta'\) in the bulk path integral,
\begin{equation}
  Z_\text{bulk}=\int DA_\mu\,DA_z\,e^{iS}\sim\int DA_\mu\,D\Sigma\,D\eta'\,e^{iS\lsb A_z=0\rsb}.
\end{equation}
Importantly, \(\Sigma\) and \(\eta'\) remain dynamical variables in the action.

\noindent \(\mathbf{3.}\) \textbf{The off-shell action.}

After the axial gauge is fixed, there remain four components \(A_\mu\) of the bulk gauge
field. These are found by solving four dynamical EOMs. The remaining fifth equation is a
constraint. It governs the on-shell dynamics of \(\Sigma\) and \(\eta'\). The dynamical
equations for \(A_\mu\) could be split into longitudinal and transverse. The latter are
identified with the dynamical equations for the heavy vector mesons \(V_{\mu,n}\) --- in
this formalism the mesons always appear on-shell. Thus, solving these dynamical equations
integrates the vector mesons out. Instead, we identify the vector mesons and leave the
relevant equations unsolved. This procedure integrates these mesons back into the
effective action (this is described in more detail in Section~\ref{sect:sols}).

Solutions of the dynamical EOMs are substituted into the bulk action. The resulting
action is known as an ``off-shell'' action, since there are dynamical degrees of freedom
left. The bulk partition function takes the form (only \(V_{\mu,n}\) remain as dynamical
variables out of the bulk field \(A_\mu\)),
\begin{equation}
  Z_\text{bulk}=\int DV_{\mu,n}\,D\Sigma\,D\eta'\,e^{iS_\text{off-shell}\lsb\Sigma,\eta',V_{\mu,n}\rsb}.
\end{equation}
Finally, integrating over the bulk coordinate in the off-shell action,
\begin{equation}
  S_\text{off-shell}=\int d^4x\,dz\,\mathcal{L}_\text{off-shell}\lsb z\rsb=
  \int d^4x\,\mathcal{L}_\text{eff}=S_\text{eff},
\end{equation}
produces the off-shell effective action for the low energy degrees of freedom of the
boundary theory and the vector mesons. The partition function of the boundary theory
eventually reads
\begin{equation}\label{eq:4d_pt}
  Z_\text{boundary}=\int DV_{\mu,n}\,D\Sigma\,D\eta'\,e^{iS_\text{eff}\lsb\Sigma,\eta',V_{\mu,n}\rsb}.
\end{equation}

\noindent \(\mathbf{4.}\) \textbf{EOMs and the perturbative expansion.}

The EOMs are non-linear. They will be solved perturbatively in a weak field approximation
by expanding the gauge field in an auxiliary parameter \(\veps\to1\), which counts the
order of non-linearity (this expansion is different from the \(\alpha'\to0\) expansion of
the DBI action),
\begin{equation}\label{eq:f_pert_intro}
  A_M=\veps A^{(1)}_M+\veps^2 A^{(2)}_M+\veps^3 A^{(3)}_M+\mathcal{O}(\veps^4).
\end{equation}
This weak field expansion is applied to both the DBI and CS actions. To obtain quartic
terms in the effective action, it is necessary to have the expansion~\eqref{eq:f_pert_intro}
kept up to \(\veps^3\) order. As a result, the quintic contributions of the CS term are
not considered here. Furthermore, the bulk gauge field is split into abelian and
non-abelian parts,
\begin{equation}
  A_M=a_M\mathbb{1}_{N_f}+A^a_MT^a.
\end{equation}
At each order \(\veps^i\), there is a system of equations for \(a^\nu\) and \(A^{\nu,a}\),
\begin{equation}
  K^\nu\lsb f^{(i)}\rsb=j^{\nu,(i)},\qquad
  \mathcal{K}^\nu\lsb f^{a,(i)}\rsb=J^{\nu,a,(i)},
\end{equation}
where \(K^\nu\) and \(\mathcal{K}^\nu\) are differential operators acting on the abelian
\(f\) and the linear part \(f^a\) of the non-abelian field strength tensors. \(j^{\nu,(i)}\)
and \(J^{\nu,a,(i)}\) are sources built from solutions of the EOMs at \(i-1\) and lower
orders. The boundary conditions~\eqref{eq:intro_bc} are completely saturated by the
lowest order solution \(A^{(1)}_\mu\).

\noindent \(\mathbf{5.}\) \textbf{The effective action.}

The expression for the effective action in~\eqref{eq:4d_pt} is the main result of this
paper, which can be found in Appendix~\ref{app:eff}, see~\eqref{eq:Seff}. Its degrees of
freedom are the \(B\)-field~\eqref{eq:intro_bc}, containing the pions and \(\eta'\)-meson,
and a multiplet of heavy vector mesons \(V_{\mu,n}\). As expected the pions and \(\eta'\)
are pseudoscalars while both positive and negative spatial parities emerge in the vector
meson sector.

As an example, one of the terms in the effective action (written in momentum space),
\begin{equation}\label{eq:seff_ex}
  S_\text{eff}\sim i\veps^{z\mu\nu\rho\sigma}\int d^4q\,d^4p\,
  \lb\frac{\delta^\vk_\nu-D^\vk_\nu(q)}{q^2-\lambda_k}-\frac{D^\vk_\nu(q)}{\lambda_k}\rb
  p_\mu\,\text{STr}\lsb B_\vk(-q)\,B_\rho(q-p)\,V_{\sigma,n}(p)\rsb,
\end{equation}
is an interaction term between \(\pi\) (or \(\eta'\)) mesons with the vector mesons \(V_{\mu,n}\).
Its origin is the CS term of the bulk action. Here, \(\veps^{z\mu\nu\rho\sigma}\) is the
Levi-Civita symbol, \(\text{STr}\) is a symmetrized trace (described in the next section),
\(D^\vk_\nu(q)\) is the longitudinal projector~\eqref{eq:proj_def}, and \(\lambda_k\) are
the eigenvalues of the spectral problem of the transverse linear (first order in \(\veps\))
equation. This expression illustrates an important property of the effective action,
namely its non-locality, contained in the momentum-dependent interaction vertex. In order
to extract local contributions from the non-local terms, a \emph{local expansion} is
introduced. This is an expansion of the vertex in powers of \(q^2\) in the limit of \(q\to0\).
In coordinate space this corresponds to gradient expansion. The local expansion can be
truncated at any order. Its leading term in~\eqref{eq:seff_ex} becomes
\begin{equation}\label{eq:loc_exp}
  \frac{\delta^\vk_\nu-D^\vk_\nu(q)}{q^2-\lambda_k}-
  \frac{D^\vk_\nu(q)}{\lambda_k}\underset{q\to0}{\approx}
  -\frac{\delta^\vk_\nu}{\lambda_k}.
\end{equation}
The local approximation is an important step in making a connection between the obtained
effective action \(S_\text{eff}\) and the Chiral Lagrangian~\eqref{eq:chi_lagr}.

In Section~\ref{sect:mod_rec} the main elements of the SS model are recapitulated.
Section~\ref{sect:eoms} describes the non-linear EOMs and perturbative procedure for
solving them. Linear and higher order solutions are constructed in Section~\ref{sect:sols}.
The next three sections are devoted to the discussion of the resulting effective action.
In Section~\ref{sect:eff} we make general comments and study the free part of the action.
The LECs are derived and discussed in Section~\ref{sect:lecs}. The vector meson
interactions appearing in the effective action are studied in Section~\ref{sect:vmes_int}.
Finally, we conclude in Section~\ref{sect:conc}.

The main body of the text is accompanied by several appendices that provide technical
details and list explicit yet cumbersome expressions for the results. Appendix~\ref{app:eff}
contains the main result of our work --- the effective action. In Appendix~\ref{app:cs5form}
some details related to the CS action are discussed. Low energy expansion of the DBI
action is derived in Appendix~\ref{app:dbi_exp}. In Appendix~\ref{app:bcond} we focus on
the key part of the off-shell formalism, the gauge fixing and resulting boundary
conditions for the bulk gauge field. Appendix~\ref{app:rhs} contains expressions for the
sources of the non-linear EOMs. In Appendix~\ref{app:spec_prob} the spectral problem in
the bulk is discussed. Numerical results related to the eigensystem of the spectral
problem can be found in Appendix~\ref{app:num}. Appendix~\ref{app:parity} presents a
discussion of spatial parity of the degrees of freedom. The Green function used in higher
order solutions is described in Appendix~\ref{app:long_gf}. Appendix~\ref{app:ho_sols}
lists explicit expressions for the second (\(\veps^2\)) and third (\(\veps^3\)) order
solutions. In Appendix~\ref{app:op_reord} a mathematical subtlety that occurs in higher
order solutions is discussed. Certain details of the derivation of LECs are contained in
Appendix~\ref{app:lec_terms}. Appendix~\ref{app:ss_redef} illustrates the effect of the
vector meson redefinitions on the values of LECs. Symmetrized traces and conventions for
group generators are listed in Appendix~\ref{app:traces}. Appendix~\ref{app:fourier_proj}
provides notations for the Fourier transforms and the properties of the longitudinal
projector.

\section{Model Recap}\label{sect:mod_rec}

Let us start from a recap of the essential ingredients of the SS model: geometry of the
spacetime and ``matter'' content.

\clearpage

\noindent \(\bullet\) \textbf{The geometry.}

The background consists of \(N_c\) \(D4\) branes compactified on a circle in a 10D
spacetime. The metric for this background is obtained as a type IIA supergravity
solution\footnote{The coverage of the related string theory topics can be found in
various textbooks, for instance, in~\cite{Becker:2006dvp}.}
\begin{equation}\label{eq:spacetime}
  \begin{gathered}
    ds^2_{10}=\lb\frac{U}{R}\rb^{3/2}\lb\eta_{\mu\nu}dx^\mu dx^\nu+f(U)d\tau^2\rb+
    \lb\frac{R}{U}\rb^{3/2}\lb\frac{dU^2}{f(U)}+U^2d\Omega_4^2\rb,\\
    f(U)=1-\frac{\ukk^3}{U^3},\qquad
    \eta_{\mu\nu}=\text{diag}\left\lbrace+1,-1,-1,-1\right\rbrace.
  \end{gathered}
\end{equation}
Originally, this brane setup was proposed by Witten~\cite{Witten:1998zw} as a background
that provides a holographic dual of a 4D Yang-Mills theory at low energies which exhibits
confinement~\cite{Witten:1998zw,Sakai:2004cn}. In~\eqref{eq:spacetime} the fundamental
parameters of the theory are the number of colors \(N_c\), the string length \(l_s\), the
string coupling \(g_s\), and the horizon position \(\ukk\). The coordinate \(U\), which
corresponds to the energy scale, is bounded from below by the horizon, \(U\ge\ukk\),
which means that \(\ukk\) is related to the minimal energy scale for states in the dual
boundary theory and to the mass gap in the confining phase of that theory~\cite{Hoyos:2022ptd}.
The constant background curvature radius \(R\) is expressed through these parameters,
\(R^3=\pi g_sN_cl_s^3\)~\cite{Kruczenski:2003uq}. The compactified coordinate \(\tau\)
has period \(\delta\tau\):
\begin{equation}
  \tau\sim\tau+\delta\tau,\quad
  \delta\tau\equiv\frac{2\pi}{\mkk},\quad
  \mkk\equiv\frac{3}{2}\frac{\ukk^{1/2}}{R^{3/2}}.
\end{equation}
This compactification avoids the apparent singularity in the metric. The constant \(\mkk\)
is an important scale in this model. On the one hand, in the bulk it is the
compactification scale (\(\mkk^{-1}\) is the radius of the compactified coordinate). For
energies lower than \(\mkk\) the theory on \(D4\) branes is effectively four-dimensional~\cite{Kruczenski:2003uq}.
From the boundary perspective, \(\mkk\) is related to the mass of the lowest vector mode
(see~\eqref{eq:spectra}) which is the energy scale of the effective theory on the
boundary. Together with the 't Hooft coupling \(\lambda_\text{YM}\) it fixes the pion
decay constant \(f_\pi\).

Two other important ingredients, which emerge from the supergravity solution, are a
dilaton field \(\phi\) and a 3-form Ramond-Ramond (RR) field \(C_3\),
\begin{equation}\label{eq:dilaton_F4}
  e^{-\phi(U)}=\frac{1}{g_s}\lb\frac{R}{U}\rb^{3/4},\qquad
  F_4=dC_3=\frac{2\pi N_c}{V_4}\epsilon_4.
\end{equation}
Here, \(\epsilon_4\) is a volume form of a unit \(S_4\) sphere with the volume
\begin{equation}\label{eq:s4_vol}
  V_4=\frac{8\pi^2}{3}.
\end{equation}
It is customary to change the coordinates of the spacetime~\eqref{eq:spacetime} by first
switching \(\lb\tau, U\rb\to(r,\theta)\to(y,z)\),
\begin{equation}
  U^3=\ukk^3+\ukk r^2,\qquad
  \theta\equiv\mkk\tau,\qquad
  y\equiv r\cos\theta,\qquad
  z\equiv r\sin\theta,
\end{equation}
and then introducing dimensionless coordinates \((\hx^\mu,\hy,\hz)\) via rescaling
\begin{equation}\label{eq:resc_vars}
  \hz\equiv\frac{z}{\ukk},\qquad
  \hy\equiv\frac{y}{\ukk},\qquad
  \hx^\mu\equiv x^\mu\mkk,\qquad
  \hat{\partial}_\mu\equiv\frac{\partial}{\partial\hx^\mu}.
\end{equation}
In these new coordinates the 4D boundary theory exists at \(z\to\pm\infty\) and the
metric \(\hg_{\alpha\beta}\) of the 10D spacetime takes the form,
\begin{equation}
  \hg_{\mu\nu}=L^2\lb1+\hy^2+\hz^2\rb^{1/2}\eta_{\mu\nu},
\end{equation}
\begin{equation}
  \hg_{yy}=L^2\lb\frac{\hz^2f(\hy,\hz)\lb1+\hy^2+\hz^2\rb^{1/2}}{\lb\hy^2+\hz^2\rb^2}+
  \frac{\hy^2}{\lb1+\hy^2+\hz^2\rb^{11/6}f(\hy,\hz)}\rb,
\end{equation}
\begin{equation}
  \hg_{yz}=\hg_{zy}=L^2\lb\frac{\hy\hz}{\lb1+\hy^2+\hz^2\rb^{11/6}f(\hy,\hz)}-
  \frac{\hy\hz f(\hy,\hz)\lb1+\hy^2+\hz^2\rb^{1/2}}{\lb\hy^2+\hz^2\rb^2}\rb,
\end{equation}
\begin{equation}
  \hg_{zz}=L^2\lb\frac{\hy^2f(\hy,\hz)\lb1+\hy^2+\hz^2\rb^{1/2}}{\lb\hy^2+\hz^2\rb^2}+
  \frac{\hz^2}{\lb1+\hy^2+\hz^2\rb^{11/6}f(\hy,\hz)}\rb,
\end{equation}
\begin{equation}
  \hg_\mathcal{AB}=\frac{9}{4}L^2\lb1+\hy^2+\hz^2\rb^{1/6}\delta_\mathcal{AB},
\end{equation}
where the indices \(\mathcal{A}\) and \(\mathcal{B}\) label the four coordinates on the
\(4\)-sphere \(S_4\) (which are collectively referred to as \(\Omega_4\)) and
where we defined
\begin{equation}
  L^2\equiv\frac{4}{9}R^{3/2}\ukk^{1/2}.
\end{equation}

\noindent \(\bullet\) \textbf{The matter.}

The matter of the model is described by embedding \(N_f\) pairs of \(D8/\bar{D}8\) branes
into the 10D spacetime described above. These branes are taken in the probe limit,
meaning there is no backreaction from the \(D8\) branes on the 10D metric (see, for
example,~\cite{Burrington:2007qd} for discussions on the backreaction problem). Each
\(D8\) or \(\bar{D}8\) brane spans nine worldvolume coordinates: \(\hx^\mu\), \(\hz\),
and \(\Omega_4\). There is only one non-trivial (i.e., not an identity) embedding
function \(\hy=\hy\lb\hx^\mu,\hz,\Omega_4\rb\). We consider branes that are placed at
antipodal positions with respect to the \(\tau\) (and \(\theta\)) coordinate. With
respect to the coordinate \(\hy\) this placement fixes \(\hy=0\) for both \(D8\) and
\(\bar{D}8\) branes in each pair~\cite{Sakai:2004cn},
\begin{equation}
  \tau=\left\lbrace\frac{\pi}{2\mkk},\frac{\pi}{2\mkk}+\frac{\delta\tau}{2}\right\rbrace\qrq
  \theta=\left\lbrace\frac{\pi}{2},\frac{\pi}{2}+\pi\right\rbrace\qrq
  \hy=0.
\end{equation}
The induced metric on a \(D8\) brane in the rescaled coordinates~\eqref{eq:resc_vars} can
be calculated as
\begin{equation}
  G_{MN}=\hg_{\alpha\beta}\partial_M X^\alpha\partial_N X^\beta,
\end{equation}
where \(M,N\) label the nine coordinates on the \(D8\) branes, \(\hg_{\alpha\beta}\) is
the metric of the 10D spacetime, and \(X^\alpha(x_M)\) are the embedding functions.
Taking into account the embedding described above, the diagonal induced metric is
\begin{equation}\label{eq:metric}
  G_{\mu\nu}=L^2u(\hz)^{3/2}\eta_{\mu\nu},\qquad
  G_{zz}=\frac{L^2}{u(\hz)^{5/2}},\qquad
  G_\mathcal{AB}=\frac{9}{4}L^2u(\hz)^{1/2}\delta_\mathcal{AB},
\end{equation}
where
\begin{equation}
  u(\hz)\equiv\lb1+\hz^2\rb^{1/3}.
\end{equation}
To simplify notations in what follows, instead of \(\hx^\mu\) and \(\hz\) we will write
\(x^\mu\) (or even just \(x\)) and \(z\). Finally, note the difference in the notations
with~\cite{Hoyos:2022ptd} where the metric is expressed in terms of \(\hat{G}_{MN}\)
which is related to \(G_{MN}\) as
\begin{equation}
  G_{MN}=L^2\hat{G}_{MN},\qquad
  d\hat{s}^2_{D8}=\hat{G}_{MN}d\hx^Md\hx^N.
\end{equation}
Open strings ending on the \(D8\) branes generate a non-abelian gauge field \(A_M\in\mathfrak{u}(N_f)\),
which depends on the coordinates along the branes. For most of this work we keep \(N_f\)
unspecified. In the string frame the action for this field is given by a non-abelian
generalization of the DBI action\footnote{There is no derivation of the non-abelian DBI
action from the string theory and various proposals for such an action exist~\cite{Tseytlin:1997csa,
Koerber:2002zb,Hagiwara:1981my,Argyres:1989qr,Park:1999gd}. The one used in the SS model
involves the symmetrized trace (discussed below)~\cite{Tseytlin:1997csa,Koerber:2002zb}.}
(the dilaton factor~\eqref{eq:dilaton_F4} is rewritten in the bulk coordinate~\(z\)),
\begin{equation}\label{eq:dbi}
  S_\text{DBI}=-T_8\int\limits_\text{D8} d^9x\,e^{-\phi}\,\text{STr}\sqrt{-\det\lb G_{MN}\mathbb{1}_{N_f}+2\pi\alpha'F_{MN}\rb},\quad
  e^{-\phi(z)}=\frac{1}{g_s}\frac{R^{3/4}}{\ukk^{3/4}u(z)^{3/4}}.
\end{equation}
Here, \(T_8\) is the tension of the \(D8\) branes and \(F_{MN}\) is the field strength of
the non-abelian bulk gauge field \(A_M\),
\begin{equation}
  F_{MN}\equiv\partial_MA_N-\partial_NA_M+i\lsb A_M,A_N\rsb.
\end{equation}
The gauge field can be decomposed into abelian \(\mathfrak{u}(1)\) and non-abelian
\(\mathfrak{su}(N_f)\) parts~\cite{Haber:2019sgz},
\begin{equation}\label{eq:bf_decomp}
  A_M=a_M\mathbb{1}_{N_f}+A^a_MT^a,
\end{equation}
where \(\mathbb{1}_{N_f}\) is the \(N_f\times N_f\) identity matrix,\footnote{A somewhat
more common choice for the normalization of the abelian generator is \(\mathbb{1}_{N_f}/\sqrt{2N_f}\).
We, however, chose to use the normalization that is more convenient for the intermediate
calculations when \(N_f\) is not fixed, although a compensating adjustment must be made
in the boundary conditions, see Appendix~\ref{app:bcond}.} \(T^a\) are the generators of
the \(\mathfrak{su}(N_f)\) algebra,\footnote{Various useful properties of the generators
\(T^a\) are listed in Appendix~\ref{app:traces}. The generators normalization is defined
by~\eqref{eq:1_2gen_tr}.} and \(a,b,\dots\) are the generator indices. The field strength
of the gauge field can also be split into abelian and non-abelian parts,
\begin{equation}
  \begin{gathered}
    F_{MN}=f_{MN}\mathbb{1}_{N_f}+F^a_{MN}T^a,\\
    f_{MN}\equiv\partial_Ma_N-\partial_Na_M,\qquad
    F^a_{MN}\equiv\partial_MA^a_N-\partial_NA^a_M-f_{abc}A^b_MA^c_N,
  \end{gathered}
\end{equation}
where \(f_{abc}\) are the structure constants of the \(\mathfrak{su}(N_f)\) algebra,
\begin{equation}\label{eq:str_const}
  \lsb T^a,T^b\rsb=if_{abc}T^c.
\end{equation}
\(\text{STr}\) in~\eqref{eq:dbi} denotes a symmetrized trace with respect to the
generator indices and is defined for the product of \(n\) matrices \(M_i\) via\footnote{\(\text{STr}\)
is just a compact notation and should not be considered a function of matrices. That said
this notation still exhibits the linearity property over its ``arguments''.}
\begin{equation}
  \text{STr}\lb M_1\dots M_n\rb\equiv\frac{1}{n!}\text{Tr}\lb M_1\dots M_n+\text{ all permutations}\rb.
\end{equation}
In the present work, just as in the original SS model~\cite{Sakai:2004cn}, only \(SO(5)\)
singlet states are considered which in practice means that the bulk gauge fields do not
depend on the coordinates on the \(S_4\) sphere. Additionally the field components along
the sphere are zero. This makes it possible to perform integration over the sphere
coordinates in the DBI action:
\begin{equation}
  \int\limits_\text{D8}d^9x=V_4\int\limits_{\text{D8}\setminus S_4} d^5x,\qquad
  d^5x=d^4x\,dz.
\end{equation}
From now on, the big Latin indices \(M,N,\dots\) will collectively label \(x^\mu\) and
\(z\) coordinates only, which are the coordinates on \(\text{D8}\setminus S_4\).

\noindent \(\bullet\) \textbf{The CS term.}

Another part of the action is the CS term. It is one of the sources of non-linearity in
the EOMs and the off-shell action. The CS term appears as an interaction term between the
gauge field \(A_M\) and the RR field \(C_3\)~\cite{Polchinski:1998rr},\footnote{Note that
CS term requires some adjustments if one wants to include baryons in the model~\cite{Lau:2016dxk}.
Since baryons are outside the scope of the present work, we use the orginal formulation
of the CS term.}
\begin{equation}\label{eq:cs_term}
  S_\text{CS}=\frac{1}{48\pi^3}\int\limits_\text{D8}F_4\wedge\omega_5,
\end{equation}
where \(F_4\) is the 4-form field strength of the RR field~\eqref{eq:dilaton_F4} and
\(\omega_5\) is the CS \(5\)-form,
\begin{equation}
  \omega_5(A)=\text{STr}\lb AF^2-\frac{i}{2}A^3F-\frac{1}{10}A^5\rb.
\end{equation}
The \(F_4\) form in~\eqref{eq:cs_term} can be integrated away via~\eqref{eq:dilaton_F4},~\eqref{eq:s4_vol},
and exploiting the fact that the bulk gauge fields components along the sphere are zero,
\begin{equation}
  S_\text{CS}=\frac{N_c}{24\pi^2}\int\limits_{\text{D8}\setminus S_4}\omega_5,
\end{equation}
where the remaining integration is performed over \(x^\mu\) and \(z\). The expression for
the CS \(5\)-form, that is used in practice, is obtained using~\eqref{eq:bf_decomp},
\begin{equation}\label{eq:omega5_Wexp}
  \omega_5=N_fa(da)^2+\frac{1}{2}\lb adA^adA^a+2A^ada\,dA^a\rb+
  \frac{1}{4}d_{abc}A^adA^bdA^c.
\end{equation}
Derivation of~\eqref{eq:omega5_Wexp} and other details pertinent to the CS \(5\)-form
can be found in Appendix~\ref{app:cs5form}. This expression for the CS \(5\)-form is more
general than the one found in~\cite{Hoyos:2022ptd} for \(N_f=2\) which does not have
neither the first nor the last terms. First, we keep the purely abelian contribution (the
first term in~\eqref{eq:omega5_Wexp}). Second, while the \(d\)-tensor vanishes for \(N_f=2\),
our derivation is carried for arbitrary \(N_f\). Hence the last term in~\eqref{eq:omega5_Wexp}
is retained.

\section{Non-linear EOMs}\label{sect:eoms}

Before discussing EOMs arising from the action of the SS model, let us make a couple of
comments about the DBI action. The DBI action will be expanded in the low energy limit,
which is equivalent to \(\alpha'=l_s^2\to0\), up to the fourth power in the field
strength of the bulk gauge field. The details of the expansion are provided in Appendix~\ref{app:dbi_exp},
the result is
\begin{equation}\label{eq:dbi_exp}
  S_\text{DBI}=-\tilde{T}_8\int\limits_{\text{D8}\setminus S_4}d^4x\,dz\,u(z)^2
  \lsb N_f+S^{(2)}+S^{(4)}+\mathcal{O}\lb\alpha'^6\rb\rsb,
\end{equation}
with the quadratic and quartic terms\footnote{The apparent differences in some of the
coefficients with the expressions available in the literature, e.g.~\cite{Hoyos:2022ptd},
are due to our different choice for the normalization of the abelian generator in the
decomposition~\eqref{eq:bf_decomp}.}
\begin{equation}
  S^{(2)}=N_f(\pi\alpha')^2f^{MN}f_{MN}+\frac{(\pi\alpha')^2}{2}F^{MN,a}F_{MN}^a,
\end{equation}
\begin{multline}
  S^{(4)}=(\pi\alpha')^4\lsb
  N_f\lb\frac{1}{2}f^{KL}f_{KL}f^{MN}f_{MN}-2f_{LM}f_{NK}f^{KL}f^{MN}\rb\right.\\\left.+
  \frac{1}{2}f_{KL}f^{KL}F^{MN,a}F^a_{MN}+f_{KL}f^{MN}F^{KL,a}F^a_{MN}\right.\\\left.-
  4f_{LM}f^{MN}F^{KL,a}F^a_{NK}-2f_{LM}f_{NK}F^{KL,a}F^{MN,a}\right.\\\left.+
  \lb\frac{1}{2}f_{KL}F^{KL,a}F^{MN,b}F^c_{MN}-2f^{KL}F^a_{LM}F^{MN,b}F^c_{NK}\rb d_{abc}\right.\\\left.+
  \lb\frac{1}{2}F^{KL,a}F^b_{KL}F^{MN,c}F^d_{MN}-2F^{KL,a}F^b_{LM}F^{MN,c}F^d_{NK}\rb\text{STr}\lb T^aT^bT^cT^d\rb\rsb.
\end{multline}
In~\cite{Hoyos:2022ptd} only purely non-abelian quartic terms (the last line in the
expression above) were considered, while we utilize the complete expression which
additionally includes both the abelian and mixed terms. In the expanded DBI action~\eqref{eq:dbi_exp},
\begin{equation}
  \tilde{T}_8\equiv\frac{1}{g_s}\lb\frac{9}{4}\rb^2\frac{R^{3/4}}{\ukk^{3/4}}L^9T_8V_4.
\end{equation}
The relation between the parameters of the bulk (string length \(l_s\), string coupling
\(g_s\), horizon \(\ukk\) and previously introduced combinations of them such as \(T_8\),
\(R\), \(\mkk\), \(L\)) and boundary ('t Hooft coupling \(\lambda_\text{YM}\) and number
of colors \(N_c\)) theory~\cite{Kruczenski:2003uq,Sakai:2004cn,Hoyos:2022ptd} reads:
\begin{equation}\label{eq:4d_params}
  \begin{gathered}
    R^3=\frac{1}{2}\frac{\lambda_\text{YM}l_s^2}{\mkk},\qquad
    \ukk=\frac{2}{9}\lambda_\text{YM}\mkk l_s^2,\qquad
    g_sN_c=\frac{1}{2\pi}\frac{\lambda_\text{YM}}{\mkk l_s},\\[0.2cm]
    T_8=\frac{1}{(2\pi)^8l_s^9},\qquad
    \alpha'=l_s^2,\qquad
    \frac{\pi\alpha'}{L^2}=\frac{27\pi}{4\lambda_\text{YM}}.
  \end{gathered}
\end{equation}
Using~\eqref{eq:4d_params} and the \(S_4\) volume~\eqref{eq:s4_vol}, in terms of the
boundary theory parameters,
\begin{equation}
  \tilde{T}_8=\frac{N_c\lambda_\text{YM}^3}{3^9\pi^5}.
\end{equation}
Finally, let us introduce a dimensionless constant,
\begin{equation}\label{eq:kappa}
  \kappa\equiv\frac{2\tilde{T}_8(\pi\alpha')^2}{L^4}=\frac{\lambda_\text{YM}N_c}{6^3\pi^3},
\end{equation}
which will be very convenient in future discussions.

Varying the DBI and CS actions, the EOMs for both abelian and non-abelian components of
the bulk gauge field are obtained:
\begin{equation}\label{eq:eoms}
  \begin{dcases}
    \partial_M\lb u(z)^2G^{MR}G^{\nu\sigma}e_{R\sigma}\rb+
    \veps^{MPRQ\nu}\delta\omega_{MPRQ,\text{ab}}=0,\\
    \partial_M\lb u(z)^2G^{MR}G^{\nu\sigma}E^a_{R\sigma}\rb-
    f_{abe}u(z)^2G^{MR}G^{\nu\sigma}A^b_ME^e_{R\sigma}+
    \veps^{MPRQ\nu}\delta\omega^a_{MPRQ,\text{non-ab}}=0,
  \end{dcases}
\end{equation}
where the following notations have been introduced for compactness,
\begin{equation}\label{eq:E_ab}
  e_{R\sigma}\equiv\tilde{T}_8(\pi\alpha')^2\lsb4N_ff_{R\sigma}+(\pi\alpha')^2\lb4N_f\mathcal{T}_{R\sigma,1}+2\mathcal{T}^{aa}_{R\sigma,3}+d_{abc}\mathcal{T}^{abc}_{R\sigma,4}\rb\rsb,
\end{equation}
\begin{equation}\label{eq:E_n_ab}
  E^a_{R\sigma}\equiv\tilde{T}_8(\pi\alpha')^2\lsb2F^a_{R\sigma}+(\pi\alpha')^2\lb2\mathcal{T}^a_{R\sigma,2}+d_{abc}\mathcal{T}^{bc}_{R\sigma,3}+4\,\text{STr}\lb T^aT^bT^cT^d\rb\mathcal{T}^{bcd}_{R\sigma,4}\rb\rsb,
\end{equation}
\begin{equation}\label{eq:T1}
  \mathcal{T}_{R\sigma,1}\equiv f_{R\sigma}f_{KL}f^{KL}-4f_{KR}f_{L\sigma}f^{KL},
\end{equation}
\begin{multline}
  \mathcal{T}^a_{R\sigma,2}\equiv f_{KL}f^{KL}F^a_{R\sigma}+2f_{KL}f_{R\sigma}F^{KL,a}\\-
  4f_{L\sigma}f^{KL}F^a_{KR}-4f_{KR}f^{KL}F^a_{L\sigma}-4f_{KR}f_{L\sigma}F^{KL,a},
\end{multline}
\begin{multline}
  \mathcal{T}^{ab}_{R\sigma,3}\equiv f_{R\sigma}F^{KL,a}F^b_{KL}+2f_{KL}F^{KL,a}F^b_{R\sigma}\\-
  4f_{KR}F^{KL,a}F^b_{L\sigma}-4f_{L\sigma}F^{KL,a}F^b_{KR}-4f^{KL}F^a_{KR}F^b_{L\sigma},
\end{multline}
\begin{equation}\label{eq:T4}
  \mathcal{T}^{abc}_{R\sigma,4}\equiv F^{KL,a}F^b_{KL}F^c_{R\sigma}-4F^{KL,a}F^b_{KR}F^c_{L\sigma},
\end{equation}
\begin{equation}\label{eq:delta_omega_ab}
  \delta\omega_{MPRQ,\text{ab}}\equiv\frac{N_c}{8\pi^2}\lb
  N_f\partial_Ma_P\partial_Ra_Q+\frac{1}{2}\partial_MA^a_P\partial_RA^a_Q\rb,
\end{equation}
\begin{equation}\label{eq:delta_omega_n_ab}
  \delta\omega^a_{MPRQ,\text{non-ab}}\equiv\frac{N_c}{8\pi^2}\lb
  \partial_Ma_P\partial_RA^a_Q+\frac{1}{4}d_{abc}\partial_MA^b_P\partial_RA^c_Q\rb.
\end{equation}
In~\eqref{eq:eoms} there are only four pairs of EOMs, while one expects to have five for
a 5D field. There is an additional pair of equations,
\begin{equation}
  \begin{dcases}
    \eta^{\mu\rho}\partial_\mu e_{\rho z}+
    \frac{L^4}{u(z)^3}\veps^{\mu\lambda\rho\sigma z}\delta\omega_{\mu\lambda\rho\sigma,\text{ab}}=0,\\
    \eta^{\mu\rho}\partial_\mu E^a_{\rho z}-
    f_{abe}\eta^{\mu\rho}A^b_\mu E^e_{\rho z}+
    \frac{L^4}{u(z)^3}\veps^{\mu\lambda\rho\sigma z}\delta\omega^a_{\mu\lambda\rho\sigma,\text{non-ab}}=0,
  \end{dcases}
\end{equation}
which does not have a second derivative in \(z\). This pair of equations is a
\emph{constraint}. Within the off-shell formalism the constraints are left unsolved.

The EOMs must be supplemented by boundary conditions. The precise form of the boundary
conditions is linked to the chosen gauge for the bulk field (see Appendix~\ref{app:bcond}
for a comprehensive discussion of this topic). In the axial gauge~\eqref{eq:ax_g} the
boundary conditions are the following:
\begin{equation}
  \begin{split}
    a_z&=0,\qquad
    a_\mu\underset{z\to-\infty}{\to}0,\qquad
    a_\mu\underset{z\to+\infty}{\to}b_\mu(x),\\[0.2cm]
    A^a_z&=0,\qquad
    A^a_\mu\underset{z\to-\infty}{\to}0,\qquad
    A^a_\mu\underset{z\to+\infty}{\to}B^a_\mu(x),
  \end{split}
\end{equation}
where the \(B\)-fields are given in~\eqref{eq:lo_sols_amp}. The EOMs are highly
non-linear. The solution method will rely on weak field approximation and employ the
perturbative expansion introduced in~\eqref{eq:f_pert_intro}. As discussed in the
Introduction, the EOMs have to be solved up to the third order in the fields only. In
order to carry out the perturbative expansion, an auxiliary parameter \(\veps\to1\) is
introduced,
\begin{equation}\label{eq:field_pertr}
  \begin{split}
    a_M&=\veps a^{(1)}_M+\veps^2 a^{(2)}_M+\veps^3 a^{(3)}_M+\mathcal{O}(\veps^4),\\[0.2cm]
    A^a_M&=\veps A^{a,(1)}_M+\veps^2 A^{a,(2)}_M+\veps^3 A^{a,(3)}_M+\mathcal{O}(\veps^4).
  \end{split}
\end{equation}
The boundary conditions read
\begin{equation}\label{eq:bcond_pert}
  \begin{aligned}
    a^{(i)}_z&=0,\quad
    a^{(i)}_\mu\underset{z\to-\infty}{\to}0,\quad
    &&a^{(1)}_\mu\underset{z\to+\infty}{\to}b_\mu(x),\quad
    &&&a^{(j)}_\mu\underset{z\to+\infty}{\to}0,\\[0.2cm]
    A^{a,(i)}_z&=0,\quad
    A^{a,(i)}_\mu\underset{z\to-\infty}{\to}0,\quad
    &&A^{a,(1)}_\mu\underset{z\to+\infty}{\to}B^a_\mu(x),\quad
    &&&A^{a,(j)}_\mu\underset{z\to+\infty}{\to}0,
  \end{aligned}
\end{equation}
for \(i\in\lbrace1,2,3\rbrace\) and \(j\in\lbrace2,3\rbrace\). Similarly,
\begin{equation}
  \begin{split}
    f_{MN}&=\veps f^{(1)}_{MN}+\veps^2f^{(2)}_{MN}+\veps^3f^{(3)}_{MN}+\mathcal{O}(\veps^4),\\[0.2cm]
    F^a_{MN}&=\veps F^{a,(1)}_{MN}+\veps^2F^{a,(2)}_{MN}+\veps^3F^{a,(3)}_{MN}+\mathcal{O}(\veps^4).
  \end{split}
\end{equation}
The linear part of the non-abelian field strength tensor,
\begin{equation}
  f^{a,(i)}_{MN}\equiv\partial_MA^{a,(i)}_N-\partial_NA^{a,(i)}_M.
\end{equation}
The non-abelian field strength can be written as
\begin{equation}\label{eq:n_ab_fs_pert}
  \begin{gathered}
    F^{a,(1)}_{MN}=f^{a,(1)}_{MN}=\partial_MA^{a,(1)}_N-\partial_NA^{a,(1)}_M,\qquad
    F^{a,(2)}_{MN}=f^{a,(2)}_{MN}-f_{abc}A^{b,(1)}_MA^{c,(1)}_N,\\[0.2cm]
    F^{a,(3)}_{MN}=f^{a,(3)}_{MN}-f_{abc}\lb A^{b,(1)}_MA^{c,(2)}_N+A^{b,(2)}_MA^{c,(1)}_N\rb.
  \end{gathered}
\end{equation}
The non-linear terms of the order \(i\) depend on the terms of the order \(j<i\) only.
The EOMs at each order can be split into differential operators and source terms,
\begin{equation}\label{eq:pert_eq}
  K^\nu\lsb f^{(i)}\rsb=j^{\nu,(i)},\qquad
  \mathcal{K}^\nu\lsb f^{a,(i)}\rsb=J^{\nu,a,(i)},
\end{equation}
where the abelian and non-abelian differential operators,
\begin{equation}\label{eq:dif_ops}
  K^\nu\lsb f^{(i)}\rsb\equiv\partial_M\lb u(z)^2G^{MR}G^{\nu\sigma}f^{(i)}_{R\sigma}\rb,\qquad
  \mathcal{K}^\nu\lsb f^{a,(i)}\rsb\equiv\partial_M\lb u(z)^2G^{MR}G^{\nu\sigma}f^{a,(i)}_{R\sigma}\rb.
\end{equation}
Explicit expressions for the source terms (right-hand side of the equations~\eqref{eq:pert_eq})
are given in Appendix~\ref{app:rhs}. The sources are decomposed similarly to~\eqref{eq:bf_decomp},
\begin{equation}
  J^{\nu,(i)}=j^{\nu,(i)}\,\mathbb{1}_{N_f}+J^{\nu,a,(i)}T^a.
\end{equation}

\section{Solutions of EOMs}\label{sect:sols}

\subsection{Linear Homogeneous Equations (First Order in \texorpdfstring{\(\bm{\veps}\)}{epsilon})}

\noindent \(\bullet\) \textbf{Differential operator.}

Let us start by making a couple of comments about the differential operator~\eqref{eq:dif_ops}.
First, the abelian and non-abelian operator~\eqref{eq:dif_ops} are identical, up to
the generator index, so we will mostly focus on the abelian one,
\begin{equation}
  K^\nu\lsb f^{(i)}\rsb=\partial_M\lb u(z)^2G^{MR}G^{\nu\sigma}f^{(i)}_{R\sigma}\rb.
\end{equation}
Using the metric~\eqref{eq:metric}, the axial gauge \(a^{(i)}_z=0\), and performing
the 4D Fourier transform (see Appendix~\ref{app:fourier_proj} for Fourier-related
notations used in this work), \(K^\nu\) can be rewritten as
\begin{equation}
  K^\nu\lsb f^{(i)}\rsb=\frac{1}{L^4u(z)}\lsb-q^2\eta^{\nu\sigma}a^{(i)}_\sigma+
  \eta^{\mu\rho}\eta^{\nu\sigma}q_\mu q_\sigma a^{(i)}_\rho+
  u(z)\,\pz\lb u(z)^3\eta^{\nu\sigma}\pz a^{(i)}_\sigma\rb\rsb,
\end{equation}
where~\eqref{eq:momsq} is used for momentum squared. The Helmholtz decomposition,
\begin{equation}
  a^{(i)}_\mu=\frac{q_\mu}{q^2}a^{(i)}_l+a^{(i)}_{\mu,t},\qquad
  \eta^{\mu\nu}q_\mu a^{(i)}_{\nu,t}=0,
\end{equation}
splits the original operator
\begin{equation}
  K^\nu\lsb f^{(i)}\rsb=\eta^{\nu\sigma}\frac{q_\sigma}{q^2}K_l\lsb a^{(i)}_l\rsb+K_t\lsb a^{\nu,(i)}_t\rsb
\end{equation}
into longitudinal and transverse differential operators,
\begin{equation}
  \begin{split}
    K_l\lsb a^{(i)}_l\rsb&=\frac{1}{L^4}\pz\lb u(z)^3\pz a^{(i)}_l\rb,\\[0.2cm]
    K_t\lsb a^{\nu,(i)}_t\rsb&=\frac{1}{L^4u(z)}\lsb-q^2a^{\nu,(i)}_t+
    u(z)\pz\lb u(z)^3\pz a^{\nu,(i)}_t\rb\rsb.
  \end{split}
\end{equation}
Similarly, the non-abelian operator is written as
\begin{equation}
  \mathcal{K}^\nu\lsb f^{a,(i)}\rsb=\eta^{\nu\sigma}\frac{q_\sigma}{q^2}K_l\lsb A^{a,(i)}_l\rsb+K_t\lsb A^{\nu,a,(i)}_t\rsb.
\end{equation}
The longitudinal equation can be isolated by multiplying both sides of~\eqref{eq:pert_eq}
with \(q_\nu\). Thus, a system of two decoupled equations, for the longitudinal and
transverse parts, is obtained,
\begin{equation}
  \begin{dcases}
    K_l\lsb a^{(i)}_l\rsb=j^{(i)}_l,\qquad&j^{(i)}_l\equiv q_\nu j^{\nu,(i)},\\
    K_t\lsb a^{\nu,(i)}_t\rsb=j^{\nu,(i)}_t,\qquad&j^{\nu,(i)}_t\equiv\lb\delta^\nu_\mu-D^\nu_\mu(q)\rb j^{\mu,(i)}.
  \end{dcases}
\end{equation}
Here, \(D^\nu_\mu\) is the \emph{longitudinal projector}~\eqref{eq:proj_def}
(\(\delta^\nu_\mu-D^\nu_\mu\) is the transverse projector), properties of which are
listed in Appendix~\ref{app:fourier_proj}.

\noindent \(\bullet\) \textbf{The abelian field.}

At first order in \(\veps\) (\(i=1\)), all sources vanish and the equations are
homogeneous,
\begin{subequations}
  \begin{align}
    \label{eq:lo_ab_long_eq}
    K_l\lsb a^{(1)}_l\rsb&=\pz\lb u(z)^3\pz a^{(1)}_l(q,z)\rb=0,\\[0.2cm]
    \label{eq:lo_ab_trans_eq}
    K_t\lsb a^{\nu,(1)}_t\rsb&=-\frac{q^2}{u(z)}a^{\nu,(1)}_t+\pz\lb u(z)^3\pz a^{\nu,(1)}_t(q,z)\rb=0.
  \end{align}
\end{subequations}
These equations are supplemented by the boundary conditions, which can be obtained
from~\eqref{eq:bcond_pert},
\begin{equation}\label{eq:lo_ab_bcond}
  a^{(1)}_l\underset{z\to+\infty}{\to}\eta^{\mu\nu}q_\mu b_\nu,\qquad
  a^{(1)}_l\underset{z\to-\infty}{\to}0,\qquad
  a^{(1)}_{\mu,t}\underset{z\to\pm\infty}{\to}0.
\end{equation}
The transverse equation has vanishing boundary conditions, which follows from~\eqref{eq:d_ab_prop}.

\noindent \(\bullet\bullet\) \textbf{The longitudinal part.}

The solution to the longitudinal equation can be easily obtained by integrating it twice,
\begin{equation}\label{eq:long_sol}
  a^{(1)}_l(q,z)=C_1(q)\,w(z)+C_2(q),
\end{equation}
where \(C_1\) and \(C_2\) are momentum-dependent integration constants. The function
\(w(z)\),
\begin{equation}\label{eq:w_def}
  w(z)\equiv\frac{1}{2}\lb1+\frac{2}{\pi}\arctan z\rb,
\end{equation}
has the following properties,
\begin{equation}\label{eq:w_prop}
  \pi\pz w(z)=u(z)^{-3},\qquad
  \lim_{z\to+\infty}w(z)=1,\qquad
  \lim_{z\to-\infty}w(z)=0.
\end{equation}
Hence, the integration constants are \(C_1=\eta^{\mu\nu}q_\mu b_\nu\) and \(C_2=0\). The
longitudinal solution takes the form
\begin{equation}
  a^{(1)}_l(q,z)=\eta^{\mu\nu}q_\mu b_\nu(q)\,w(z).
\end{equation}

\noindent \(\bullet\bullet\) \textbf{The transverse part.}

Assume that the differential operator of the transverse equation~\eqref{eq:lo_ab_trans_eq}
has a discrete eigensystem \(\lambda_n\) and \(\psi_n\),
\begin{equation}\label{eq:spec_prob}
  \pz\lb u(z)^3\pz\psi_n(z)\rb=\frac{\lambda_n}{u(z)}\psi_n(z),\qquad
  \psi_n(z)\underset{z\to\pm\infty}{\to}0.
\end{equation}
We are not aware of any analytical solutions for this spectral problem in the literature.
The main complicating factor is that the spectral problem is defined on an infinite
interval and thus cannot be treated as a regular Sturm-Liouville problem. Consequently,
some results from the Sturm-Liouville theory are not immediately apparent, although there
are important properties about the eigensystem that can be demonstrated (Appendix~\ref{app:spec_prob}).
This spectral problem can be treated numerically which was originally done with the
shooting method in~\cite{Sakai:2004cn}.

The solution to the abelian transverse equation takes the form
\begin{equation}
  a^{(1)}_{\mu,t}(q,z)\propto\tilde{v}_{\mu,n}(q)\,\psi_n(z),
\end{equation}
where the arbitrary coefficients \(\tilde{v}_{\mu,n}\) can be identified as the vector
mesons of the boundary theory. Substituting this ansatz into~\eqref{eq:lo_ab_trans_eq}
gives
\begin{equation}\label{eq:qsq_lambdan}
  \lb q^2-\lambda_n\rb\tilde{v}_{\mu,n}(q)=0,
\end{equation}
which implies \(q^2=\lambda_n\). The equation~\eqref{eq:qsq_lambdan} can be understood as
the on-shell condition for the vector mesons. Yet, we want to keep the mesons as
dynamical degrees of freedom, so instead of solving the transverse equation we decompose
\(a^{(1)}_{\mu,t}\) in the basis of \(\psi_n\),
\begin{equation}
  a^{(1)}_{\mu,t}(q,z)=\sum_{n=1}^\infty v_{\mu,n}(q)\,\psi_n(z).
\end{equation}
Here, the coefficients \(v_{\mu,n}\) designated as the ``off-shell'' vector meson fields
are introduced to distinguish them from the ``on-shell'' mesons \(\tilde{v}_{\mu,n}\).
The quadratic terms in the boundary effective action for the new coefficients \(v_{\mu,n}\)
generate the boundary EOMs that recover the on-shell condition~\eqref{eq:qsq_lambdan}.

\noindent \(\bullet\) \textbf{The non-abelian field.}

The process of solving the linear non-abelian EOMs,
\begin{equation}
  K_l\lsb A^{a,(1)}_l\rsb=0,\qquad K_t\lsb A^{\nu,a,(i)}_t\rsb=0,
\end{equation}
is nearly identical to that of the abelian case. The only difference is in the boundary
conditions,
\begin{equation}\label{eq:n_ab_bc}
  \begin{gathered}
    A^{a,(1)}_l\underset{z\to+\infty}{\to}\eta^{\mu\nu}q_\mu B^a_\nu,\qquad
    A^{a,(1)}_l\underset{z\to-\infty}{\to}0,\\[0.2cm]
    A^{a,(1)}_{\mu,t}\underset{z\to+\infty}{\to}\lb\delta^\nu_\mu-D^\nu_\mu\rb B^a_\nu,\qquad
    A^{a,(1)}_{\mu,t}\underset{z\to-\infty}{\to}0.
  \end{gathered}
\end{equation}
Notice the appearance of a non-trivial boundary condition for the transverse part. This
occurs due to the non-abelian structure of the matrix \(\Sigma\). The first term in the
expansion~\eqref{eq:B_exp} of \(B^a_\mu\) is proportional to \(q_\mu\) (in momentum space)
and evidently longitudinal. The next term is not proportional to the momentum (since it
is a convolution in the momentum space), and it contributes to the transverse part.

Proceeding to solution, the longitudinal part takes the form,
\begin{equation}
  A^{a,(1)}_l(q,z)=\eta^{\mu\nu}q_\nu B^a_\mu(q)\,w(z).
\end{equation}
In the transverse part, in order to satisfy the non-zero boundary condition~\eqref{eq:n_ab_bc},
the following ansatz is employed
\begin{equation}
  A^{a,(1)}_{\mu,t}(q,z)=\lb\delta^\sigma_\mu-D^\sigma_\mu\rb B^a_\sigma(q)\,w(z)+
  \tilde{A}^{a,(1)}_{\mu,t},
\end{equation}
where \(\tilde{A}^{a,(1)}_{\mu,t}\) is a solution to the equation
\begin{equation}\label{eq:Atilde_eq}
  -\frac{q^2}{u(z)}\eta^{\mu\nu}\lb\delta^\sigma_\mu-D^\sigma_\mu\rb B^a_\sigma(q)\,w(z)-
  \frac{q^2}{u(z)}\eta^{\mu\nu}\tilde{A}^{a,(1)}_{\mu,t}(q,z)+
  \pz\lb u(z)^3\eta^{\mu\nu}\pz\tilde{A}^{a,(1)}_{\mu,t}(q,z)\rb=0
\end{equation}
with \emph{zero} boundary conditions. The solution to~\eqref{eq:Atilde_eq} can be written
using the Green function method. In addition, there is a contribution from the
homogeneous transverse equation (with no source),
\begin{equation}
  \tilde{A}^{a,(1)}_{\mu,t}=-\frac{q^2}{L^4}\lb\delta^\sigma_\mu-D^\sigma_\mu\rb B^a_\sigma(q)
  \int\limits_{-\infty}^{+\infty}dz'\,G_t(z,z',q)\,\frac{w(z')}{u(z')}+
  \sum_{n=1}^{\infty}\tilde{V}^a_{\mu,n}(q)\,\psi_n(z),
\end{equation}
where \(\lambda_n\) and \(\psi_n\) are the eigensystem of the transverse equation~\eqref{eq:spec_prob},
\(G_t(z,z',q)\) is the transverse Green function~\eqref{eq:trans_gf}, that is constructed
in the next subsection, and the coefficients \(\tilde{V}^a_{\mu,n}\) are the non-abelian
counterparts of the abelian coefficients \(\tilde{v}_{\mu,n}\). Similarly to the abelian
case, the coefficients \(\tilde{V}^a_{\mu,n}\) can be identified as the vector mesons
with imposed on-shell condition,
\begin{equation}
  \lb q^2-\lambda_n\rb\tilde{V}^a_{\mu,n}(q)=0.
\end{equation}
In order to have the dynamical degrees of freedom for vector mesons we once again leave~\eqref{eq:Atilde_eq}
unsolved and introduce as a solution
\begin{equation}
  A^{a,(1)}_{\mu,t}(q,z)=\lb\delta^\sigma_\mu-D^\sigma_\mu\rb B^a_\sigma(q)\,w(z)+
  \sum_{n=1}^\infty V^a_{\mu,n}(q)\,\psi_n(z).
\end{equation}
By making this choice, we end up not solving the transverse equations in both abelian and
non-abelian cases. Later, the coefficients \(v_{\mu,n}\) and \(V^a_{\mu,n}\) will be
identified as the degrees of freedom, corresponding to the parts of the vector meson
multiplets. This replacement of the proper transverse solution with the tower of vector
mesons can be viewed as \emph{integrating in} the vector mesons. Conversely, if the
equations were completely solved, there would be no dynamical degrees of freedom for the
vector mesons in the resulting effective theory.

\noindent \(\bullet\) \textbf{Summary.}

The full solution of the EOMs in the momentum space takes the form
\begin{equation}\label{eq:lo_sols_mom}
  \begin{split}
    a^{(1)}_\mu(q,z)&=b_\mu(q)\,w(z)+\sum_{n=1}^\infty v_{\mu,n}(q)\,\psi_n(z),\\[0.2cm]
    A^{a,(1)}_\mu(q,z)&=B^a_\mu(q)\,w(z)+\sum_{n=1}^\infty V^a_{\mu,n}(q)\,\psi_n(z).
  \end{split}
\end{equation}
The sums in~\eqref{eq:lo_sols_mom} are commonly referred to as the Kaluza-Klein tower of
massive vector mesons~\cite{Sakai:2004cn,Hoyos:2022ptd}. The abelian and non-abelian
coefficients of the transverse parts of the solutions, \(v_{\mu,n}(q)\) and \(V^a_{\mu,n}(q)\)
are identified as parts of the multiplet of massive (axial-)vector mesons,
\begin{equation}\label{eq:vtilde_decomp}
  V_{\mu,n}(q)\equiv v_{\mu,n}(q)\,\mathbb{1}_{N_f}+V^a_{\mu,n}(q)\,T^a.
\end{equation}
In Section~\ref{sect:eff} it will be demonstrated that the fields \(V_{\mu,n}\) have both
kinetic and mass terms with the mass proportional to \(\sqrt{-\lambda_n}\), specifically,
\(m_n=\sqrt{-\lambda_n}\mkk\). This supports their identification as physical mesons.
Discussion of spatial parity of the multiplets can be found in Appendix~\ref{app:parity}.
Finally, since the vector mesons have been introduced as part of the transverse solution,
it is natural that they are also transverse:
\begin{equation}\label{eq:trans_cond}
  \eta^{\mu\nu}q_\nu v_{\mu,n}=
  \eta^{\mu\nu}q_\nu V^a_{\mu,n}=
  \eta^{\mu\nu}q_\nu V_{\mu,n}=0.
\end{equation}

\subsection{Green Functions and Higher Order Equations}\label{sect:ho}

In order to solve the higher order equations (\eqref{eq:pert_eq} with \(i>1\)), the Green
functions of the longitudinal~\eqref{eq:lo_ab_long_eq} and transverse~\eqref{eq:lo_ab_trans_eq}
linear equations will be constructed first. Since the abelian and non-abelian
differential operators, both the longitudinal and transverse (and the boundary conditions
for higher order equations), are the same, the abelian and non-abelian Green functions
are identical too. We focus on the abelian case only. The longitudinal and transverse
Green functions\footnote{Since there is no explicit momentum dependence in the
longitudinal differential operator, the longitudinal Green function is also independent
of the momentum.} are defined as follows
\begin{equation}\label{eq:gf_intro}
  K_l\lsb G_l(z,z')\rsb=\delta(z-z'),\qquad
  K_t\lsb G_t(z,z',q)\rsb=\delta(z-z'),
\end{equation}
with the homogeneous boundary conditions
\begin{equation}\label{eq:gf_bcond}
  G_l(z,z')\underset{z\to\pm\infty}{\to}0,\qquad
  G_t(z,z',q)\underset{z\to\pm\infty}{\to}0.
\end{equation}
The transverse Green function can be constructed from the eigensystem of the linear (\(i=1\))
equation via spectral representation,
\begin{equation}\label{eq:trans_gf}
  G_t(z,z',q)=-\kappa L^4\sum_{n=1}^\infty\frac{\psi_n(z)\,\psi_n(z')}{q^2-\lambda_n},
\end{equation}
Indeed, application of the differential operator \(K_t\) to the Green function~\eqref{eq:trans_gf}
recovers the completeness relation~\eqref{eq:comp_rel}. Appearance of \(q^2\) in the
denominator of the transverse Green function is one of the sources of the mentioned
non-locality in the resulting effective action.

Derivation of the longitudinal Green function is a bit more involved (see Appendix~\ref{app:long_gf}
for the details). The result is
\begin{equation}\label{eq:long_gf}
  G_l(z,z')=\pi L^4\lsb
  \theta(z'-z)\lb w(z')-1\rb w(z)+\theta(z-z')w(z')\lb w(z)-1\rb\rsb.
\end{equation}
In the zero momentum limit,
\begin{equation}\label{eq:tl_gf}
  G_l(z,z')=G_t(z,z',0)=\kappa L^4\sum_n\frac{\psi_n(z)\,\psi_n(z')}{\lambda_n}.
\end{equation}
The full abelian higher order solution takes the form,\footnote{A subtle mathematical
complication that may occur in the transverse solutions is discussed in Appendix~\ref{app:op_reord}.}
\begin{equation}\label{eq:ho_ab_sol_gen}
  a^{(i)}_\mu(q,z)=D^\lambda_\mu(q)\int\limits_{-\infty}^{+\infty}dz'\,G_l(z,z')\,j^{(i)}_\lambda(q,z')+
  \lb\delta^\lambda_\mu-D^\lambda_\mu(q)\rb\int\limits_{-\infty}^{+\infty}dz'\,G_t(z,z',q)\,j^{(i)}_\lambda(q,z').
\end{equation}
The sources \(j^{(i)}_\lambda(q,z')\) and their non-abelian counterparts, listed in Appendix~\ref{app:rhs},
can be factorized into \(z'\)- and \(q\)-dependent factors, making it possible to perform
the integration in~\eqref{eq:ho_ab_sol_gen}. Full set of solutions of the higher order
equations for \(i\in\lbrace2,3\rbrace\) can be found in Appendix~\ref{app:ho_sols},
presented in the momentum space.

Let us comment on the difference between the present work and~\cite{Hoyos:2022ptd}
pertaining to the solution methods of the higher order equations. In~\cite{Hoyos:2022ptd}
the pion scattering amplitudes were computed from the pion correlators which requires
the leading term in the zero momentum expansion only of the solutions of the EOMs. For
this reason, in~\cite{Hoyos:2022ptd} it was sufficient to retain the leading term in the
zero momentum expansion of the transverse Green function only,
\begin{equation}
  G_t(z,z',q)\underset{q^2\to0}{\approx}G_l(z,z').
\end{equation}
One of the major objectives of our work is to go beyond the local approximation (\(q\to0\))
and to obtain the action valid to all orders in the momentum expansion. Thus, the higher
order solutions in the form of truncated expansions are insufficient for our purposes.

\section{Effective Action and Its Free Part}\label{sect:eff}

The solutions of EOMs are substituted into the DBI and CS actions (expanded to the fourth
order in the fields). After integrating over the bulk coordinate \(z\), the effective
action of the boundary theory is obtained. As mentioned in the Introduction, the
effective degrees of freedom are \(\Sigma\) matrix and \(\eta'\)-meson, compactly written
in terms of the \(B\)-fields~\eqref{eq:lo_sols_amp}, and the vector meson multiplets
\(V_{\mu,n}\). Since the higher order solutions are found in the momentum space, the
effective action will also be derived in momentum space.

The action in the bulk is integrated by parts, which separates the terms into the
\emph{surface} and \emph{bulk} actions,
\begin{equation}
  S_\text{eff}=S_\text{surf}+S_\text{bulk}.
\end{equation}
The surface action originates from the \(\mathcal{O}\lb F^2\rb\) terms in the DBI action
only,
\begin{multline}
  S_\text{surf}=-N_f\kappa\int d^4q\left.u(z)^3\eta^{\nu\sigma}\lb
  \pz a^{(1)}_\sigma(q)\,a^{(1)}_\nu(-q)+\pz a^{(1)}_\sigma(q)\,a^{(2)}_\nu(-q)\right.\right.\\\left.\left.+
  \pz a^{(1)}_\sigma(q)\,a^{(3)}_\nu(-q)+\pz a^{(2)}_\sigma(q)\,a^{(1)}_\nu(-q)+
  \pz a^{(2)}_\sigma(q)\,a^{(2)}_\nu(-q)+\pz a^{(3)}_\sigma(q)\,a^{(1)}_\nu(-q)\rb\vphfrac\right|_{z\to-\infty}^{z\to+\infty}\\-
  \frac{\kappa}{2}\int d^4q\left.u(z)^3\eta^{\nu\sigma}\lb
  \pz A^{a,(1)}_\sigma(q)\,A^{a,(1)}_\nu(-q)+\pz A^{a,(1)}_\sigma(q)\,A^{a,(2)}_\nu(-q)\right.\right.\\\left.\left.+
  \pz A^{a,(1)}_\sigma(q)\,A^{a,(3)}_\nu(-q)+\pz A^{a,(2)}_\sigma(q)\,A^{a,(1)}_\nu(-q)\right.\right.\\\left.\left.+
  \pz A^{a,(2)}_\sigma(q)\,A^{a,(2)}_\nu(-q)+\pz A^{a,(3)}_\sigma(q)\,A^{a,(1)}_\nu(-q)\rb\vphfrac\right|_{z\to-\infty}^{z\to+\infty}.
\end{multline}
On the EOMs~\eqref{eq:eoms}, the \emph{bulk} action can be written as
\begin{equation}
  S_\text{bulk}=S_\text{bulk,DBI}+S_\text{bulk,CS},
\end{equation}
where we split the bulk terms into DBI-induced,
\begin{multline}\label{eq:dbi_orig}
  S_\text{int,DBI}\equiv\frac{\kappa}{2}\int\frac{dz}{u(z)}d^4q\,i\eta^{\mu\rho}\eta^{\nu\sigma}q_\mu\,
  \text{STr}\lsb\lb
  2\tilde{F}^{(1)}_{\rho\sigma}(q)+\tilde{F}^{(2)}_{\rho\sigma}(q)+\tilde{F}^{(3)}_{\rho\sigma}(q)\rb
  A^{(1)}_\nu(-q)\rsb\\+
  \frac{\kappa}{2}\int\frac{dz}{u(z)}d^4q\,i\eta^{\mu\rho}\eta^{\nu\sigma}q_\mu\,
  \text{STr}\lsb\lb
  2\tilde{F}^{(1)}_{\rho\sigma}(q)+\tilde{F}^{(2)}_{\rho\sigma}(q)\rb
  A^{(2)}_\nu(-q)\rsb\\+
  \kappa\int\frac{dz}{u(z)}d^4q\,i\eta^{\mu\rho}\eta^{\nu\sigma}q_\mu\,
  \text{STr}\lsb\tilde{F}^{(1)}_{\rho\sigma}(q)\,A^{(3)}_\nu(-q)\rsb\\+
  \frac{\kappa}{2}\int d^4q\,dz\,\eta^{\nu\sigma}\,
  \text{STr}\lsb
  \pz\lsb u(z)^3\lb2\tilde{F}^{(1)}_{z\sigma}(q)+\tilde{F}^{(2)}_{z\sigma}(q)+\tilde{F}^{(3)}_{z\sigma}(q)\rb\rsb
  A^{(1)}_\nu(-q)\rsb\\+
  \frac{\kappa}{2}\int d^4q\,dz\,\eta^{\nu\sigma}\,
  \text{STr}\lsb
  \pz\lsb u(z)^3\lb2\tilde{F}^{(1)}_{z\sigma}(q)+\tilde{F}^{(2)}_{z\sigma}(q)\rb\rsb
  A^{(2)}_\nu(-q)\rsb\\+
  \kappa\int d^4q\,dz\,\eta^{\nu\sigma}\,
  \text{STr}\lsb\pz\lb u(z)^3\tilde{F}^{(1)}_{z\sigma}(q)\rb A^{(3)}_\nu(-q)\rsb+
  \frac{if_{abc}\kappa}{4(2\pi)^2}\int\frac{dz}{u(z)}d^4q\,d^4p\,\eta^{\mu\rho}\eta^{\nu\sigma}p_\rho\\\times\lsb
  \lb A^{b,(1)}_\mu(q-p)\,A^{c,(1)}_\nu(-q)+
  A^{b,(1)}_\mu(q-p)\,A^{c,(2)}_\nu(-q)+A^{b,(2)}_\mu(q-p)\,A^{c,(1)}_\nu(-q)\rb A^{a,(1)}_\sigma(p)\right.\\\left.+
  A^{b,(1)}_\mu(q-p)\,A^{c,(1)}_\nu(-q)\,A^{a,(2)}_\sigma(p)\rsb,
\end{multline}
and CS-induced,
\begin{multline}
  S_\text{bulk,CS}\equiv\frac{1}{12}\int d^4q\,dz\,\veps^{MPRQ\nu}\lsb
  \lb\delta\omega^{(2)}_{MPRQ,\text{ab}}(q)+\delta\omega^{(3)}_{MPRQ,\text{ab}}(q)\rb a^{(1)}_\nu(-q)\right.\\\left.+
  \lb\delta\omega^{a,(2)}_{MPRQ,\text{non-ab}}(q)+\delta\omega^{a,(3)}_{MPRQ,\text{non-ab}}(q)\rb A^{a,(1)}_\nu(-q)\right.\\\left.+
  \delta\omega^{(2)}_{MPRQ,\text{ab}}(q)\,a^{(2)}_\nu(-q)+
  \delta\omega^{a,(2)}_{MPRQ,\text{non-ab}}(q)\,A^{a,(2)}_\nu(-q)\rsb.
\end{multline}
The notations used above can be found in Appendix~\ref{app:rhs}. For compactness, in~\eqref{eq:dbi_orig}
the following notation is also utilized,
\begin{equation}
  \tilde{F}^{(i)}_{MN}\equiv\partial_MA^{(i)}_N-\partial_NA^{(i)}_M.
\end{equation}
EOMs are traditionally used to simplify on-shell actions. Yet, we recall that the
transverse equation at first order in \(\veps\) has been left unsolved. Thus, great
care must be taken when manipulating the action. When done correctly this results in the
extra factors of \(2\) in front of \(\tilde{F}^{(1)}_{\rho\sigma}\) and \(\tilde{F}^{(1)}_{z\sigma}\)
in~\eqref{eq:dbi_orig}. No such caveat appears at higher orders since those are uniquely
related to the first order solutions via EOMs.

The resulting effective action is extremely involved. It is non-local beyond the
quadratic order: the projector~\eqref{eq:proj_def} and the transverse Green function~\eqref{eq:trans_gf}
contain \(q^2\) in the denominators which, if converted into the coordinate space,
results in a non-local action. The final result for the effective action valid for any
\(N_f\) can be found in Appendix~\ref{app:eff}. Let us now discuss the ``free'' part of
this action.

\noindent \(\bullet\) \textbf{The free action.}

At the second order in the dynamical variables \(B_\mu\) (which contain \(\Sigma\) and
\(\eta'\)) and \(V_{\mu,n}\) (the first term \(S_{\text{eff},1}\)~\eqref{eq:Seff1}), the
quadratic terms of the Chiral Lagrangian~\eqref{eq:chi_lagr} are recovered, which is a
standard result of the SS model (here the rescaled coordinates~\eqref{eq:resc_vars} were
converted to the physical ones),\footnote{This part of the effective action is referred to
as ``free'', yet the first term in~\eqref{eq:free_act} contains an infinite number of
self-interaction terms between the \(\pi\)-mesons.}
\begin{multline}\label{eq:free_act}
  S_\text{free}=-\frac{f^2_\pi}{4}\int d^4x\,
  \text{Tr}\lb\partial_\mu\Sigma^{-1}\partial^\mu\Sigma\rb-
  \frac{1}{2}\int d^4x\,\partial_\mu\eta'\partial^\mu\eta'\\-
  \int d^4x\sum_n
  \lb\text{STr}\lsb\partial_\mu V_{\nu,n}\partial^\mu V^\nu_n\rsb+
  m^2_n\,\text{STr}\lsb V_{\mu,n}V^\mu_n\rsb\rb,
\end{multline}
where the pion decay constant \(f_\pi\) is identified as
\begin{equation}\label{eq:fpisq}
  f^2_\pi=\frac{4\kappa\mkk^2}{\pi}=\frac{8\tilde{T}_8(\pi\alpha')^2\mkk^2}{\pi L^4}=
  \frac{N_c\lambda_\text{YM}}{54\pi^4}\mkk^2.
\end{equation}
As mentioned in the Introduction the pseudoscalar\footnote{The parity of \(\eta'\) is
determined in Appendix~\ref{app:parity}.} \(\eta'\)-meson remains massless in the
large-\(N_c\) limit adopted in this paper.

\begin{wrapfigure}{R}{0.4\textwidth}
  \vspace{-4mm}
  \centering
  \includegraphics[width=0.38\textwidth]{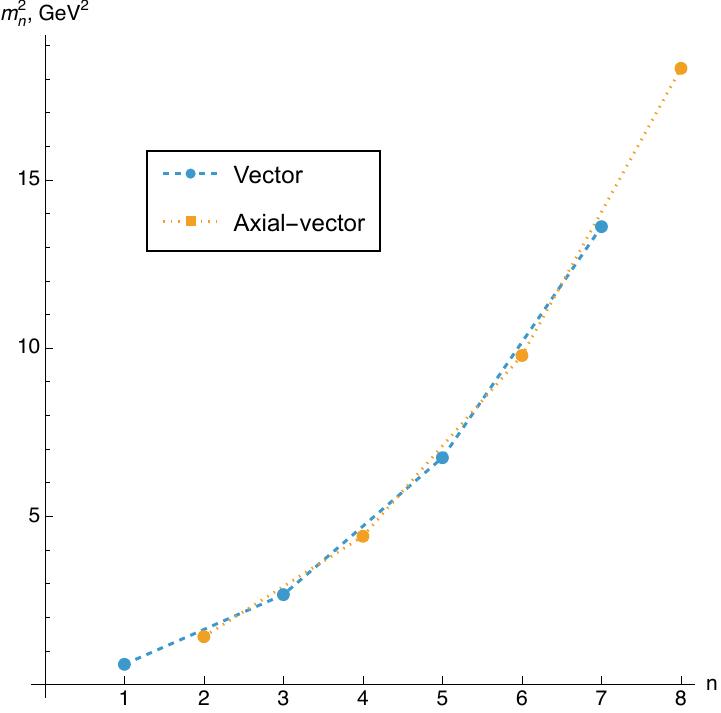}
  \vspace*{-3mm}
  \caption{Meson mass spectra.}
  \vspace*{-6mm}
  \label{fig:spectra}
\end{wrapfigure}

The spin-\(1\) mesons \(V_{\mu,n}\) are either odd (vector) or even (axial-vector) with
respect to the spatial parity transformation (details on the parity can be found in Appendix~\ref{app:parity}).
The parity is alternating with respect to the number \(n\), with the lowest having the
odd parity. The relative normalization of the kinetic and mass terms of the vector mesons
is consistent with identification of the mesons with real physical particles of mass \(m_n\).
The free effective action obtained for mesons could be rewritten in the form of the Proca
action under the condition that the vector mesons are transverse~\eqref{eq:trans_cond}.
The mass squared of the (axial-)vector mesons is given by
\begin{equation}\label{eq:spectra}
  m^2_n\equiv-\lambda_n\mkk^2.
\end{equation}
Fitting the mass of the lowest, \(n=1\) state to that of the \(\rho\)-meson fixes \(\mkk\)
(the experimental value is taken from PDG~\cite{ParticleDataGroup:2024cfk}),
\begin{equation}
  m_\rho=775.26\,\text{MeV},\qquad\lambda_1=-0.67\qrq
  \mkk=946.95\,\text{MeV}.
\end{equation}
The eigenvalue was obtained in~\cite{Sakai:2004cn} from the numerical analysis of the
spectral problem using the shooting method. In Appendix~\ref{app:num} an extended set of
\(39\) eigenvalues computed by us numerically is provided, see Table~\ref{tab:1}. Table~\ref{tab:2}
from the same Appendix~\ref{app:num} presents numerical results for couplings that will
be mentioned in later sections. Figure~\ref{fig:spectra} depicts the mass squared~\eqref{eq:spectra}
of (axial-)vector mesons. As discussed above, the vector meson fields \(V_{\mu,n}\) have
alternating parity with respect to the number \(n\), so the mass spectrum naturally
splits into two trajectories: one for vector mesons and the other for axial-vector. One
can easily notice that the mass spectra are not linear Regge trajectories, which is a
known deficiency of the top-down models of QCD.\footnote{For an example of recent advances
in the study of meson phenomenology within bottom-up AdS/QCD we refer a curious reader
to~\cite{Afonin:2021cwo}.}

The second, dimensionless parameter of the boundary theory, \(\lambda_\text{YM}\), can be
fixed from~\eqref{eq:fpisq} for \(N_c=3\),
\begin{equation}\label{eq:lym}
  f_\pi=92.4\,\text{MeV}\ \Rightarrow\ %
  \lambda_\text{YM}=\frac{54\pi^4f^2_\pi}{N_c\mkk^2}=16.69.
\end{equation}
In some cases (references may be found in~\cite{Hechenberger:2023ljn}) a lower value of
the 't Hooft coupling is chosen which corresponds to \(f_\pi\) going closer to its chiral
limit value of \(87\,\text{MeV}\). With all the parameters of the effective theory fixed,
it can be used to predict values of other interaction couplings such as \(L_{1,2}\).

Beyond the quadratic order the effective action is non-local. To extract local
contributions, expansion of the interaction vertices in powers of \(q\) near \(q\to0\)
(\emph{local expansion}) is performed. In coordinate space this expansion is equivalent
to gradient expansion. In the terms to be discussed in the next two Sections, the local
expansion will be truncated to the leading term. An example of this procedure was
demonstrated in~\eqref{eq:loc_exp} in the Introduction. While the zero momentum limit of
the projector \(D^\vk_\nu(q)\) is actually indeterminate, the expansion procedure is
well-defined.

\section{Low Energy Constants}\label{sect:lecs}

To help with identification of LECs, it should be clarified what terms in the effective
action~\eqref{eq:Seff} can be matched to the LEC terms in the Chiral Lagrangian. We look
for 4-derivative terms in the action. First, note that
\begin{multline}
  B^a_\mu B^a_\nu=-4\,\text{Tr}\lb\Sigma^{-1}\partial_\mu\Sigma T^a\rb
  \text{Tr}\lb\Sigma^{-1}\partial_\nu\Sigma T^a\rb=
  -2\,\text{Tr}\lb\Sigma^{-1}\partial_\mu\Sigma\Sigma^{-1}\partial_\nu\Sigma\rb\\=
  2\,\text{Tr}\lb\partial_\mu\Sigma^{-1}\Sigma\Sigma^{-1}\partial_\nu\Sigma\rb=
  2\,\text{Tr}\lb\partial_\mu\Sigma^{-1}\partial_\nu\Sigma\rb.
\end{multline}
With the help of this identity, the double-trace operators \(L_{1,2}\) of the Chiral
Lagrangian~\eqref{eq:chi_lagr} can be rewritten in terms of \(B^a_\mu\),
\begin{equation}\label{eq:tr_to_B}
  \begin{split}
    \lb\text{Tr}\lb\partial_\mu\Sigma^\dagger\partial^\mu\Sigma\rb\rb^2&=
    \eta^{\mu\rho}\eta^{\nu\sigma}
    \text{Tr}\lb\partial_\mu\Sigma^{-1}\partial_\rho\Sigma\rb
    \text{Tr}\lb\partial_\nu\Sigma^{-1}\partial_\sigma\Sigma\rb=
    \frac{1}{4}\eta^{\mu\rho}\eta^{\nu\sigma}
    B^a_\mu B^a_\rho B^b_\nu B^b_\sigma,\\[0.2cm]
    \lb\text{Tr}\lb\partial_\mu\Sigma^\dagger\partial_\nu\Sigma\rb\rb^2&=
    \eta^{\mu\rho}\eta^{\nu\sigma}
    \text{Tr}\lb\partial_\mu\Sigma^{-1}\partial_\nu\Sigma\rb
    \text{Tr}\lb\partial_\rho\Sigma^{-1}\partial_\sigma\Sigma\rb=
    \frac{1}{4}\eta^{\mu\rho}\eta^{\nu\sigma}
    B^a_\mu B^a_\nu B^b_\rho B^b_\sigma.
  \end{split}
\end{equation}
Similarly, for the single-trace operators~\eqref{eq:L3},
\begin{equation}\label{eq:L3_to_B}
  \begin{split}
    B^a_\mu B^{\mu,b}B^c_\nu B^{\nu,d}\,\text{Tr}\lb T^aT^bT^cT^d\rb&=
    % \lb-2i\rb^4\text{Tr}\lsb\text{Tr}\lb\Sigma^{-1}\partial_\mu\Sigma T^a\rb T^a
    % \text{Tr}\lb\Sigma^{-1}\partial^\mu\Sigma T^b\rb T^b
    % \text{Tr}\lb\Sigma^{-1}\partial_\nu\Sigma T^c\rb T^c
    % \text{Tr}\lb\Sigma^{-1}\partial^\nu\Sigma T^d\rb T^d\rsb=
    %
    % \lb-i\rb^4\text{Tr}\lb\Sigma^{-1}\partial_\mu\Sigma\Sigma^{-1}\partial^\mu\Sigma
    % \Sigma^{-1}\partial_\nu\Sigma\Sigma^{-1}\partial^\nu\Sigma\rb=
    %
    \text{Tr}\lb\partial_\mu\Sigma^{-1}\partial^\mu\Sigma\partial_\nu\Sigma^{-1}\partial^\nu\Sigma\rb,\\[0.2cm]
    %%%
    B^a_\mu B^{\mu,b}B^c_\nu B^{\nu,d}\,\text{Tr}\lb T^aT^cT^bT^d\rb&=
    \text{Tr}\lb\partial_\mu\Sigma^{-1}\partial_\nu\Sigma\partial^\mu\Sigma^{-1}\partial^\nu\Sigma\rb.
  \end{split}
\end{equation}
The equations~\eqref{eq:tr_to_B} and~\eqref{eq:L3_to_B} imply that in the effective
action the terms with LECs must have four \(B^a_\mu\) factors. Obviously, since \(B_\mu\)
contains \(B^a_\mu\)~\eqref{eq:lo_sols_amp}, a term with four factors of \(B_\mu\) is
also a candidate for the role of a term with LECs. However, consider the following
expression,
\begin{equation}
  \partial_\mu B^a_\nu-\partial_\nu B^a_\mu=
  -2i\,\text{Tr}\lsb\lb\partial_\mu\Sigma^{-1}\partial_\nu\Sigma-
  \partial_\nu\Sigma^{-1}\partial_\mu\Sigma\rb T^a\rsb.
\end{equation}
After inserting \(\mathbb{1}=\Sigma\Sigma^{-1}\) between the derivatives the above expression
can be written as a commutator,
\begin{equation}
  \partial_\mu B^a_\nu-\partial_\nu B^a_\mu=
  2i\,\text{Tr}\lb\lsb\Sigma^{-1}\partial_\mu\Sigma,\Sigma^{-1}\partial_\nu\Sigma\rsb T^a\rb.
\end{equation}
On the other hand, commutator of the \(B\)-fields~\eqref{eq:lo_sols_amp} can be expressed
as
\begin{equation}
  \lsb B_\mu,B_\nu\rsb=\lsb B^a_\mu T^a,B^b_\nu T^b\rsb=
  -\lsb\Sigma^{-1}\partial_\mu\Sigma,\Sigma^{-1}\partial_\nu\Sigma\rsb,
\end{equation}
where the abelian parts do not contribute to the commutator. This demonstrates that
the antisymmetrized derivative of \(B^a_\mu\) can be rewritten as the commutator of two
\(B^a_\mu\),
\begin{equation}\label{eq:der_to_B}
  2\partial_{[\mu}B^a_{\nu]}\equiv\partial_\mu B^a_\nu-\partial_\nu B^a_\mu=
  -2i\,\text{Tr}\lb\lsb B_\mu,B_\nu\rsb T^a\rb.
\end{equation}
This means that some terms of the effective action may contain four \(B^a_\mu\) factors
only superficially: if they have an extra derivative and the indices are appropriately
antisymmetrized, they are actually of higher order in \(B^a_\mu\). More generally, recall
that the Chiral Lagrangian is an expansion in derivatives (momenta) of \(\pi\)-mesons.
The terms with LECs are of the order \(\mathcal{O}(p^4)\). Since \(B_\mu\) carries within
it at least one derivative of the pion, the terms with the LECs in the effective action
must have the total number of \(B\)-fields and derivatives acting on them to be equal to
four. Thus, the terms with four \(B\)-fields in \(S_{\text{eff},15}\) and \(S_{\text{eff},16}\),~\eqref{eq:Seff15}
and~\eqref{eq:Seff16}, which might potentially contribute to 4-derivative terms, are in
fact of higher order in momenta and do not contribute to the values of LECs.

There are only two terms in the effective action~\eqref{eq:Seff}, (\(S_{\text{eff},2}\)
and \(S_{\text{eff},3}\)), that satisfy the criteria outlined above and give rise to
terms with the bare LECs of the Chiral Lagrangian. Since \(B\)-fields behave as
derivatives under rescaling of the coordinates, any extra factors arising from converting
the rescaled coordinates~\eqref{eq:resc_vars} into the physical ones are cancelled out by
the factors from the integral measure. As such, for the remainder of this Section all
discussed effective action terms will be treated as if they have already been converted
into physical coordinates.

The first effective action term of interest is \(S_{\text{eff},2}\) which is given by~\eqref{eq:Seff2}.
After certain algebra (see Appendix~\ref{app:lec_terms} for details), in the local
approximation \(S_{\text{eff},2}\) takes the form (see~\eqref{eq:Seff2_4})
\begin{equation}\label{eq:Seff2_res}
  \left.S_{\text{eff},2}\vphfrac\right|_\text{loc}=L^{(l)}\int d^4x\lsb
  \text{Tr}\lb\partial_\mu\Sigma^{-1}\partial^\mu\Sigma\partial_\nu\Sigma^{-1}\partial^\nu\Sigma\rb-
  \text{Tr}\lb\partial_\mu\Sigma^{-1}\partial_\nu\Sigma\partial^\mu\Sigma^{-1}\partial^\nu\Sigma\rb\rsb,
\end{equation}
where \(L^{(l)}\), known since the original work~\cite{Sakai:2004cn}, is
\begin{equation}
  L^{(l)}\equiv-\kappa C=-\frac{N_c\lambda_\text{YM}}{216\pi^3}C.
\end{equation}
The numerical constant \(C\) is computed in~\eqref{eq:C_const},
\begin{equation}
  C=\int\limits_{-\infty}^{+\infty}dz\,\frac{w(z)^2-2w(z)^3+w(z)^4}{u(z)}\approx0.157.
\end{equation}
The second contribution to the bare LECs, which was first obtained in~\cite{Hoyos:2022ptd},
comes from \(S_{\text{eff},3}\). Similarly to \(S_{\text{eff},2}\), taking the local
approximation (details of the computation can be found in Appendix~\ref{app:lec_terms}),
this contribution can be brought to the form (see~\eqref{eq:Seff3_3})
\begin{equation}\label{eq:Seff3_res}
  \left.S_{\text{eff},3}\vphfrac\right|_{BJ,\text{loc}}=L^{(nl)}\int d^4x\lsb
  2\text{Tr}\lb\partial_\mu\Sigma^{-1}\partial^\mu\Sigma\partial_\nu\Sigma^{-1}\partial^\nu\Sigma\rb+
  \text{Tr}\lb\partial_\mu\Sigma^{-1}\partial_\nu\Sigma\partial^\mu\Sigma^{-1}\partial^\nu\Sigma\rb\rsb,
\end{equation}
where
\begin{equation}
  L^{(nl)}\equiv-\frac{\kappa^2}{6\tilde{T}_8}a^{(3)}_{11}=-\frac{9N_c}{128\pi\lambda_\text{YM}}a^{(3)}_{11}.
\end{equation}
The numerical constant \(a^{(3)}_{11}\) is defined in~\eqref{eq:a3_11},
\begin{equation}
  a^{(3)}_{11}=\int\limits_{-\infty}^{+\infty}dz\,w(z)\,\pz\lb u(z)^4(\pz w(z))^3\rb=
  -\frac{7\Gamma\lb\frac{1}{6}\rb}{40\pi^{7/2}\Gamma\lb\frac{2}{3}\rb}\approx-0.013.
\end{equation}
Both effective action terms, \(S_{\text{eff},2}\) and \(S_{\text{eff},3}\), generate the
single-trace operators introduced in~\eqref{eq:L3}. Combining~\eqref{eq:Seff2_res} and~\eqref{eq:Seff3_res}
produces the quartic \(\Sigma\)-field self-interaction terms
\begin{multline}\label{eq:quart_act_Sigma}
  S^{SU(N_f)}_\text{quart}\equiv\int d^4x\lsb
  L^{SU(N_f)}_3\,\text{Tr}\lb\partial_\mu\Sigma^{-1}\partial^\mu\Sigma\partial_\nu\Sigma^{-1}\partial^\nu\Sigma\rb\right.\\\left.+
  \tilde{L}^{SU(N_f)}_3\,\text{Tr}\lb\partial_\mu\Sigma^{-1}\partial_\nu\Sigma\partial^\mu\Sigma^{-1}\partial^\nu\Sigma\rb\rsb,
\end{multline}
where
\begin{equation}\label{eq:LECs_Nf}
  L^{SU(N_f)}_3\equiv 2L^{(nl)}+L^{(l)},\qquad
  \tilde{L}^{SU(N_f)}_3\equiv L^{(nl)}-L^{(l)},
\end{equation}
are the bare LECs for arbitrary \(N_f\). We notice that, contrary to expectations, no
independent double-trace operators labeled by \(L_{1,2}\) emerge here. For \(N_f=2\),
they will appear through relations between the single- and double-trace operators (see
below).

\noindent \(\bullet\) \textbf{Pion elastic scattering}

Consider now the quartic action~\eqref{eq:quart_act_Sigma} expanded to the leading order
in the \(\pi\)-fields
\begin{multline}
  \left.S^{SU(N_f)}_\text{quart}\vphfrac\right|_{\mathcal{O}\lb\pi^4\rb}=\frac{16}{f^4_\pi}\int d^4x\lb
  L^{SU(N_f)}_3\partial_\mu\pi^a\partial^\mu\pi^b\partial_\nu\pi^c\partial^\nu\pi^d\right.\\\left.+
  \tilde{L}^{SU(N_f)}_3\partial_\mu\pi^a\partial_\nu\pi^b\partial^\mu\pi^c\partial^\nu\pi^d\rb
  \text{Tr}\lb T^aT^bT^cT^d\rb.
\end{multline}
For arbitrary \(N_f\) the trace of four generators is given by~\eqref{eq:4gen_tr}, which
simplifies the expression
\begin{multline}\label{eq:SM}
  \left.S^{SU(N_f)}_\text{quart}\vphfrac\right|_{\mathcal{O}\lb\pi^4\rb}=\frac{4L^{SU(N_f)}_3}{f^4_\pi}\int d^4x\,
  \partial_\mu\pi^a\partial^\mu\pi^b\partial_\nu\pi^c\partial^\nu\pi^d
  \lb\frac{1}{N_f}\delta_{ab}\delta_{cd}+\frac{1}{2}d_{abe}d_{cde}\rb\\+
  \frac{4\tilde{L}^{SU(N_f)}_3}{f^4_\pi}\int d^4x\,
  \partial_\mu\pi^a\partial_\nu\pi^b\partial^\mu\pi^c\partial^\nu\pi^d\\\times
  \lsb\frac{1}{N_f}\lb2\delta_{ab}\delta_{cd}-\delta_{ac}\delta_{bd}\rb+
  \frac{1}{2}\lb2d_{abe}d_{cde}-d_{ace}d_{bde}\rb\rsb.
\end{multline}
This action term is responsible for \(2\rightarrow 2\) elastic pion scattering,
\begin{equation}
  \pi^a(p_a)+\pi^b(p_b)\to\pi^c(p_c)+\pi^d(p_d).
\end{equation}
The \(T\)-matrix element for this process:
\begin{equation}
  S=I+iT,\qquad
  T=(2\pi)^4\delta\lb p_a+p_b-p_c-p_d\rb M.
\end{equation}
The tree-level scattering amplitude \(M\) (or rather its low energy limit) can be
straightforwardly read off from~\eqref{eq:SM},
\begin{subequations}\label{eq:amp_Nf}
  \begin{multline}\label{eq:amp_M_Nf}
    M^{SU(N_f)}=\delta_{ab}\delta_{cd}A^{SU(N_f)}(s,t,u)+\delta_{ac}\delta_{bd}A^{SU(N_f)}(t,s,u)+\delta_{ad}\delta_{bc}A^{SU(N_f)}(u,t,s)\\+
    \frac{N_f}{2}\lb A^{SU(N_f)}(s,t,u)d_{abe}d_{cde}+A^{SU(N_f)}(t,s,u)d_{ace}d_{bde}+A^{SU(N_f)}(u,t,s)d_{ade}d_{bce}\rb,
  \end{multline}
  \begin{equation}\label{eq:amp_A_Nf}
    A^{SU(N_f)}(s,t,u)\equiv\frac{8\lb L^{SU(N_f)}_3-\tilde{L}^{SU(N_f)}_3\rb}{N_ff^4_\pi}s^2+
    \frac{8\tilde{L}^{SU(N_f)}_3}{N_ff^4_\pi}\lb t^2+u^2\rb,
  \end{equation}
\end{subequations}
where the Mandelstam variables are defined as
\begin{equation}
  s=-\lb p_a+p_b\rb^2,\qquad
  t=-\lb p_a-p_c\rb^2,\qquad
  u=-\lb p_a-p_d\rb^2.
\end{equation}
Note the appearance of the terms proportional to the product of \(d\)-tensors in the
amplitude~\eqref{eq:amp_M_Nf}.

\noindent \(\bullet\) \(\bm{N_f=3}\)

Let us now project the above results on \(N_f=3\). In this case the \(\tilde{L}_3\)
operator~\eqref{eq:L3} is not independent. With the help of the Cayley-Hamilton relation~\cite{Pich:2018ltt},
\begin{equation}
  \text{Tr}\lb ABAB\rb=-2\,\text{Tr}\lb A^2B^2\rb+\frac{1}{2}\,\text{Tr}\lb A^2\rb\text{Tr}\lb B^2\rb+
  \lb\text{Tr}\lb AB\rb\rb^2,\qquad A,B\in\mathfrak{su}(3),
\end{equation}
the \(\tilde{L_3}\) operator can be rewritten as a linear
combination of \(L_1\), \(L_2\), and \(L_3\) operators. Specifically,
\begin{multline}\label{eq:ch_rel_imp}
  \text{Tr}\lb\partial_\mu\Sigma^{-1}\partial_\nu\Sigma\partial^\mu\Sigma^{-1}\partial^\nu\Sigma\rb=
  \text{Tr}\lb\Sigma^{-1}\partial_\mu\Sigma\Sigma^{-1}\partial_\nu\Sigma\Sigma^{-1}\partial^\mu\Sigma\Sigma^{-1}\partial^\nu\Sigma\rb\\=
  -2\,\text{Tr}\lb\partial_\mu\Sigma^{-1}\partial^\mu\Sigma\partial_\nu\Sigma^{-1}\partial^\nu\Sigma\rb+
  \frac{1}{2}\lb\text{Tr}\lb\partial_\mu\Sigma^{-1}\partial^\mu\Sigma\rb\rb^2+
  \lb\text{Tr}\lb\partial_\mu\Sigma^{-1}\partial_\nu\Sigma\rb\rb^2.
\end{multline}
Substituting~\eqref{eq:ch_rel_imp} into the quartic action~\eqref{eq:quart_act_Sigma} for
the \(\Sigma\)-field gives
\begin{multline}\label{eq:quart_act_Sigma_Nf_3}
  S^{SU(3)}_\text{quart}=\int d^4x\lsb
  L^{SU(3)}_1\lb\text{Tr}\lb\partial_\mu\Sigma^{-1}\partial^\mu\Sigma\rb\rb^2+
  L^{SU(3)}_2\lb\text{Tr}\lb\partial_\mu\Sigma^{-1}\partial_\nu\Sigma\rb\rb^2\right.\\\left.+
  L^{SU(3)}_3\text{Tr}\lb\partial_\mu\Sigma^{-1}\partial^\mu\Sigma\partial_\nu\Sigma^{-1}\partial^\nu\Sigma\rb\rsb,
\end{multline}
where \(SU(3)\) bare LECs are defined as
\begin{equation}\label{eq:LECs_Nf_3}
  L^{SU(3)}_2=2L^{SU(3)}_1=\tilde{L}^{SU(N_f)}_3=L^{(nl)}-L^{(l)},\qquad
  L^{SU(3)}_3=L^{SU(N_f)}_3-2\tilde{L}^{SU(N_f)}_3=3L^{(l)}.
\end{equation}
Here, the definitions~\eqref{eq:LECs_Nf} were used in the last step. The amplitude of the
pion elastic scattering can be computed directly from~\eqref{eq:quart_act_Sigma_Nf_3}
with the result,
\begin{subequations}\label{eq:amp_Nf_3}
  \begin{multline}
    M^{SU(3)}=\delta_{ab}\delta_{cd}A^{SU(3)}(s,t,u)+\delta_{ac}\delta_{bd}A^{SU(3)}(t,s,u)+\delta_{ad}\delta_{bc}A^{SU(3)}(u,t,s)\\+
    \frac{4L^{SU(3)}_3}{f^4_\pi}\lb s^2d_{abe}d_{cde}+t^2d_{ace}d_{bde}+u^2d_{ade}d_{bce}\rb,
  \end{multline}
  \begin{equation}
    A^{SU(3)}(s,t,u)\equiv\frac{8L^{SU(3)}_1}{f^4_\pi}s^2+\frac{8L^{SU(3)}_3}{3f^4_\pi}s^2+\frac{4L^{SU(3)}_2}{f^4_\pi}\lb t^2+u^2\rb.
  \end{equation}
\end{subequations}
The amplitude~\eqref{eq:amp_Nf_3} depends on two coefficients only --- it is either
expressible in terms of \(L^{(l)}\) and \(L^{(nl)}\) or \(L^{SU(N_f)}_3\) and \(\tilde{L}^{SU(N_f)}_3\).
The amplitude~\eqref{eq:amp_Nf_3} can be obtained directly from~\eqref{eq:amp_Nf}
substituting \(N_f=3\) and utilizing the relation~\eqref{eq:d_tens_rel} for the
\(d\)-tensors.

\noindent \(\bullet\) \(\bm{N_f=2}\)

Finally, we separately consider the \(N_f=2\) case. For \(N_f=2\), both single-trace
operators~\eqref{eq:L3} are expressible in terms of the double trace operators \(L_{1,2}\).
In addition to~\eqref{eq:ch_rel_imp}, which is also valid for \(N_f=2\), we use~\cite{Pich:2018ltt},
\begin{equation}\label{eq:Nf_2_op_rel}
  2\,\text{Tr}\lb ABCD\rb=\text{Tr}\lb AB\rb\text{Tr}\lb CD\rb-\text{Tr}\lb AC\rb\text{Tr}\lb BD\rb+
  \text{Tr}\lb AD\rb\text{Tr}\lb BC\rb
\end{equation}
for \(A,B,C,D\in\mathfrak{su}(2)\). Specifically,
\begin{multline}\label{eq:Nf_2_op_rel_imp}
  \text{Tr}\lb\partial_\mu\Sigma^{-1}\partial^\mu\Sigma\partial_\nu\Sigma^{-1}\partial^\nu\Sigma\rb=
  \text{Tr}\lb\Sigma^{-1}\partial_\mu\Sigma\Sigma^{-1}\partial^\mu\Sigma\Sigma^{-1}\partial_\nu\Sigma\Sigma^{-1}\partial^\nu\Sigma\rb\\=
  \frac{1}{2}\lb\text{Tr}\lb\Sigma^{-1}\partial_\mu\Sigma\Sigma^{-1}\partial^\mu\Sigma\rb\rb^2=
  \frac{1}{2}\lb\text{Tr}\lb\partial_\mu\Sigma^{-1}\partial^\mu\Sigma\rb\rb^2.
\end{multline}
The last two terms on the right-hand side of~\eqref{eq:Nf_2_op_rel} cancelled each other
out and the \(L_3\) operator is completely expressible in terms of the \(L_1\) operator.
Substituting~\eqref{eq:ch_rel_imp} and~\eqref{eq:Nf_2_op_rel_imp} into~\eqref{eq:quart_act_Sigma},
the quartic action term for the \(\Sigma\)-field reads
\begin{equation}\label{eq:quart_act_Sigma_Nf_2}
  S^{SU(2)}_\text{quart}=\int d^4x\lsb
  L^{SU(2)}_1\lb\text{Tr}\lb\partial_\mu\Sigma^{-1}\partial^\mu\Sigma\rb\rb^2+
  L^{SU(2)}_2\lb\text{Tr}\lb\partial_\mu\Sigma^{-1}\partial_\nu\Sigma\rb\rb^2\rsb,
\end{equation}
where the \(SU(2)\) bare LECs are defined as
\begin{equation}
  \begin{gathered}
    L^{SU(2)}_1=L^{SU(3)}_1+\frac{L^{SU(3)}_3}{2}=\frac{L^{SU(N_f)}_3-\tilde{L}^{SU(N_f)}_3}{2}=
    \frac{L^{(nl)}}{2}+L^{(l)},\\[0.2cm]
    L^{SU(2)}_2=L^{SU(3)}_2=\tilde{L}^{SU(N_f)}_3=L^{(nl)}-L^{(l)}.
  \end{gathered}
\end{equation}
Here, the relations~\eqref{eq:LECs_Nf} and~\eqref{eq:LECs_Nf_3} were used to connect \(SU(2)\)
couplings to the bare LECs for \(N_f=3\) and arbitrary \(N_f\). The elastic pion
scattering tree level amplitude is again easily computed based on the quartic action~\eqref{eq:quart_act_Sigma_Nf_2}.
The result is well-known in the literature and it is given by~\cite{Ecker:1994gg,Hoyos:2022ptd,
Bijnens:1994qh,Ecker:1996yy,Weinberg:1966kf,Weinberg:1978kz},
\begin{subequations}\label{eq:scat_amp}
  \begin{equation}\label{eq:scat_amp_M}
    M=\delta_{ab}\delta_{cd}A(s,t,u)+\delta_{ac}\delta_{bd}A(t,s,u)+\delta_{ad}\delta_{bc}A(u,t,s),
  \end{equation}
  \begin{equation}\label{eq:scat_amp_A}
    A(s,t,u)=\frac{s}{f^2_\pi}+\frac{8\mathbb{L}^{SU(2)}_1}{f^4_\pi}s^2+\frac{4\mathbb{L}^{SU(2)}_2}{f^4_\pi}(t^2+u^2),
  \end{equation}
\end{subequations}
where the linear in \( s\) contribution in \(A(s,t,u)\) comes from the \(\Sigma\Sigma\)
term of the free action~\eqref{eq:free_act} and is unrelated to the quartic in \(\Sigma\)
terms discussed in this Section (we have omitted this term in~\eqref{eq:amp_A_Nf}). In~\eqref{eq:scat_amp_A},
the effective LECs are defined as
\begin{equation}
  \mathbb{L}^{SU(2)}_1\equiv L^{SU(2)}_1=\frac{L^{(nl)}}{2}+L^{(l)},\qquad
  \mathbb{L}^{SU(2)}_2\equiv L^{SU(2)}_2=L^{(nl)}-L^{(l)}.
\end{equation}
The very same result for the amplitude can be obtained directly from~\eqref{eq:amp_Nf}
substituting \(N_f=2\) and recalling that the \(d\)-tensors vanish in \(SU(2)\).

The coefficient \(L^{{(nl)}}\), which stems from the higher order expansion of the DBI
action was first introduced and discussed in~\cite{Hoyos:2022ptd}. Evidently, \(L^{(l)}\)
and \(L^{{(nl)}}\) have different parametric dependence,
\begin{equation}
  L^{(l)}=\mathcal{O}\lb N_c\lambda_\text{YM}\rb,\qquad
  L^{(nl)}=\mathcal{O}\lb\frac{N_c}{\lambda_\text{YM}}\rb.
\end{equation}
Since the 't Hooft constant was fixed in~\eqref{eq:lym} to a finite value greater than 1,
\(L^{(nl)}\) is suppressed compared to \(L^{(l)}\).

Our results are in full agreement with~\cite{Hoyos:2022ptd}. We would like to further
emphasize that the distinction between the bare and effective LECs is important since the
two sets of LECs are not equivalent. While the effective LECs are physical and measurable,
the bare ones could be altered by vector meson field redefinitions. We demonstrate this
point explicitly in Appendix~\ref{app:ss_redef} on the example of the original works~\cite{Sakai:2004cn,
Sakai:2005yt}.

\clearpage

\noindent \(\bullet\) \textbf{Vector meson exchange contributions to the amplitude.}

The pion scattering amplitude can also acquire contributions from the tree level vector
meson exchange. The effective action~\eqref{eq:Seff} does contain terms that could
potentially generate such contributions. One candidate is \(S_{\text{eff},6}\)~\eqref{eq:Seff6}
in the local approximation (except for the integral in the third line),
\begin{multline}\label{eq:Seff6_local}
  \left.S_{\text{eff},6}\vphfrac\right|_\text{loc}=\sum_n\frac{if_{abc}\kappa a^{(2)}_{6,n}}{2(2\pi)^2}\int d^4q\,d^4p
  \lb2\eta^{\mu\rho}\eta^{\nu\sigma}-\eta^{\mu\sigma}\eta^{\nu\rho}-
  \eta^{\mu\nu}\eta^{\rho\sigma}\rb
  p_\rho B^a_\nu(-q)\,B^b_\sigma(p)\,V^c_{\mu,n}(q-p)\\-
  \sum_n\frac{if_{abc}\kappa a^{(2)}_{6,n}}{2(2\pi)^2}\int d^4q\,d^4p
  \lb2\eta^{\mu\nu}\eta^{\rho\sigma}-\eta^{\mu\sigma}\eta^{\nu\rho}\rb
  p_\rho B^a_\nu(-q)\,B^b_\sigma(q-p)\,V^c_{\mu,n}(p)\\-
  \sum_n\frac{if_{abc}\kappa}{2(2\pi)^2}\int d^4q\,d^4p
  \lb a^{(2)}_{6,n}+\kappa\int dz\lb q^2-\lambda_n\rb\frac{\psi_n(z)}{u(z)}\,\hat{s}^{(2)}_5(q,z)\rb\\\times
  \lb2\eta^{\mu\rho}\eta^{\nu\sigma}-\eta^{\mu\sigma}\eta^{\nu\rho}-
  \eta^{\mu\nu}\eta^{\rho\sigma}\rb
  p_\rho B^a_\sigma(p)\,B^b_\mu(q-p)\,V^c_{\nu,n}(-q)\\-
  \sum_n\frac{if_{abc}\kappa}{(2\pi)^2}\frac{\mathcal{C}_n}{\lambda_n}\int d^4q\,d^4p\,
  \lb\eta^{\mu\rho}\eta^{\nu\sigma}-\eta^{\mu\nu}\eta^{\rho\sigma}\rb
  p_\rho B^a_\sigma(p)\,B^b_\mu(q-p)\,V^c_{\nu,n}(-q).
\end{multline}
Each integral in~\eqref{eq:Seff6_local} contains at least three derivatives (one from \(p_\rho\)
and two more from the \(B\)-fields). Therefore, the resulting contribution to the
amplitude~\eqref{eq:scat_amp_A} is of the order \(\mathcal{O}\lb p^6\rb\), while the
effective LECs emerge at the order \(\mathcal{O}\lb p^4\rb\). Consequently, the effective
action term~\eqref{eq:Seff6_local} does not affect the values of the effective LECs.

Similarly, consider \(S_{\text{eff,5}}\)~\eqref{eq:Seff5}, which, after taking the local
approximation, takes the form (in the coordinate space)
\begin{equation}\label{eq:Seff5_loc}
  \left.S_{\text{eff},5}\vphfrac\right|_\text{loc}=-\sum_n\frac{N_c\veps^{z\mu\nu\rho\sigma}a^{(2)}_{1,n}}{8\pi^2}\int d^4x\,
  \text{STr}\lsb B_\nu B_\rho\partial_\mu V_{\sigma,n}\rsb,
\end{equation}
where the relation~\eqref{eq:ac} was used. The \(\pi\pi V\) vertex read off from~\eqref{eq:Seff5_loc}
is \(\mathcal{O}\lb p^3\rb\) thus contributing \(\mathcal{O}\lb p^6\rb\) to the amplitude.
In fact, the contribution of \(S_{\text{eff,5}}\) is even further suppressed: the
symmetrized trace in~\eqref{eq:Seff5_loc} is symmetric over \(\nu\leftrightarrow\rho\),
when contracted with the antisymmetric Levi-Civita tensor it renders the entire integral
to vanish.

In summary, while tree level exchanges of the vector mesons could potentially contribute
to the effective LECs, they do not. Basically, the vector mesons are irrelevant at this
order in momenta and could be thought as being partially integrated out.

\section{Vector Meson Interactions in the Effective Action}\label{sect:vmes_int}

\noindent \(\bullet\) \textbf{CS-induced.}

As an example of interactions between the vector mesons \(V_{\mu,n}\) and the \(B\)-field
(which contains the pions and \(\eta'\)-meson), we focus on \(S_{\text{eff},7}\)~\eqref{eq:Seff7}.
This term is induced by the CS action. Its distinguishing property (compared to the
DBI-induced terms) is the presence of the Levi-Civita pseudotensor. Applying the
truncated local expansion (which makes it possible to write the expression in coordinate
space),
\begin{equation}\label{eq:Seff7_coord}
  \left.S_{\text{eff},7}\vphfrac\right|_\text{loc}=-\sum_{n,m}\frac{N_c\veps^{z\mu\nu\rho\sigma}a^{(2)}_{2,nm}}{4\pi^2}\int d^4x\,
  \text{STr}\lb B_\nu\partial_\mu V_{\rho,n}V_{\sigma,m}\rb.
\end{equation}
The relation~\eqref{eq:ac} was used here. The coupling is given by the integral
\begin{equation}
  a^{(2)}_{2,nm}=\int\limits_{-\infty}^{+\infty}dz\,w(z)\,\psi_n(z)\,\pz\psi_m(z).
\end{equation}
Appendix~\ref{app:num} contains a set of values for this constant obtained numerically.
Substituting the leading order expansion~\eqref{eq:B_exp} of the \(B\)-field into~\eqref{eq:Seff7_coord}
gives the following interaction terms
\begin{multline}\label{eq:cs_ind_ex}
  \left.S_{\text{eff},7}\vphfrac\right|_{\text{loc},\mathcal{O}\lb\pi,\eta\rb}=\sum_{n,m}\frac{N_c\veps^{z\mu\nu\rho\sigma}a^{(2)}_{2,nm}}{4\pi^2f_\pi}\sqrt{\frac{2}{N_f}}\int d^4x\,
  \partial_\nu\eta'\,\text{STr}\lb\partial_\mu V_{\rho,n}V_{\sigma,m}\rb\\+
  \sum_{n,m}\frac{N_c\veps^{z\mu\nu\rho\sigma}a^{(2)}_{2,nm}}{2\pi^2f_\pi}\int d^4x\,
  \partial_\nu\pi^a\,\text{STr}\lb\partial_\mu V_{\rho,n}V_{\sigma,m}T^a\rb.
\end{multline}
These two terms represent (derivative) interactions of the \(\pi\) and \(\eta'\) mesons
with the vector meson multiplets. For example, taking \(n=m=1\) in the second term, which
corresponds to the lightest vector mesons, produces an interaction resembling the
\(\omega\to\rho\pi\) process, which is known experimentally.

Superficially, it seems that the CS-induced interaction terms~\eqref{eq:cs_ind_ex} might
be parity-violating due to the presence of the Levi-Civita pseudotensor. This is not the
case, however. The terms in~\eqref{eq:cs_ind_ex} contain even number of derivatives. The
sign change from the Levi-Civita pseudotensor under the spatial parity transformation is
compensated by the extra minus sign from the pseudoscalar fields \(\eta'\) or \(\pi^a\).
Since the vector mesons enter quadratically, the entire term conserves parity if both
vector mesons are of the same parity.

Let us now consider the case when the vector mesons have opposite parities. Integrating
the coefficient \(a^{(2)}_{2,nm}\) by parts gives:
\begin{multline}
  a^{(2)}_{2,nm}=\int\limits_{-\infty}^{+\infty}dz\,w(z)\,\psi_n(z)\,\pz\psi_m(z)=
  \left.w(z)\,\psi_n(z)\,\psi_m(z)\right|_{z\to\pm\infty}\\-
  \frac{1}{\pi}\int\limits_{-\infty}^{+\infty}dz\,\frac{\psi_n(z)\,\psi_m(z)}{u(z)^3}-
  \int\limits_{-\infty}^{+\infty}dz\,w(z)\,\pz\psi_n(z)\,\psi_m(z).
\end{multline}
The boundary term is zero due to asymptotics~\eqref{eq:trans_sus_as} of eigenfunctions.
Since the eigenfunctions also have opposite parities, the second, integral term is also
zero since it is an integral of an odd function over a symmetric interval. The last
integral can be identified as \(a^{(2)}_{2,mn}\). Thus, for mesons of opposite parities:
\begin{equation}\label{eq:a2_rel}
  a^{(2)}_{2,nm}=-a^{(2)}_{2,mn},\qquad
  n,m\text{ --- opposite parity}.
\end{equation}
Integrating the second line in the effective action term~\eqref{eq:cs_ind_ex} by parts
gives (the same can be done for the first line as well)
\begin{equation}
  \left.S_{\text{eff},7}\vphfrac\right|_{\substack{\text{loc},\\\mathcal{O}\lb\pi\rb}}=-\sum_{n,m}\frac{N_c\veps^{z\mu\nu\rho\sigma}a^{(2)}_{2,nm}}{2\pi^2f_\pi}\int d^4x\,
  \text{STr}\lb\partial_\mu\partial_\nu\pi^aV_{\rho,n}V_{\sigma,m}T^a+
  \partial_\nu\pi^aV_{\rho,n}\partial_\mu V_{\sigma,m}T^a\rb.
\end{equation}
The first term under the symmetrized trace is symmetric over \(\mu\leftrightarrow\nu\)
and it is contracted with the antisymmetric Levi-Civita pseudotensor, thus, it is zero.
After renaming \(\rho\leftrightarrow\sigma\) and \(n\leftrightarrow m\) in the second
term, utilizing the relation~\eqref{eq:a2_rel}, and reordering factors under the
symmetrized trace,
\begin{equation}
  \left.S_{\text{eff},7}\vphfrac\right|_{\substack{\text{loc},\mathcal{O}\lb\pi\rb,\\n,m\text{ --- opp. par.}}}=-\sum_{n,m}\frac{N_c\veps^{z\mu\nu\rho\sigma}a^{(2)}_{2,nm}}{2\pi^2f_\pi}\int d^4x\,
  \partial_\nu\pi^a\,\text{STr}\lb\partial_\mu V_{\rho,n}V_{\sigma,m}T^a\rb,
\end{equation}
which is the original effective action term (second line of~\eqref{eq:cs_ind_ex}), but
with the opposite sign. This means that this integral must be equal to zero. As such,
there are no CS-induced parity-violating interactions, as expected.

An interaction term similar to~\eqref{eq:cs_ind_ex} was derived in~\cite{Sakai:2005yt}.
In our notations, the numerical constant \(c_{v^nv^m}\) introduced in~\cite{Sakai:2005yt}
for this interaction term can be expressed as
\begin{equation}
  c_{v^nv^m}=-a^{(2)}_{2,2n-1,2m-1}-a^{(2)}_{2,2m-1,2n-1}.
\end{equation}
The numerical results presented in Table~\ref{tab:2} are in agreement with the results
for \(c_{v^1v^n}\) listed in~(4.32) of~\cite{Sakai:2005yt}, up to a sign.

\noindent \(\bullet\) \textbf{DBI-induced.}

Consider \(S_{\text{eff},8}\)~\eqref{eq:Seff8} which is induced by the DBI action. The
leading term of the local expansion can be written in coordinate space (converting the
coefficients with~\eqref{eq:ac}),
\begin{equation}
  \left.S_{\text{eff},8}\vphfrac\right|_\text{loc}=\sum_{n,m}f_{abc}\kappa a^{(2)}_{7,nm}\int d^4x
  \lb2\eta^{\mu\rho}\eta^{\nu\sigma}-\eta^{\mu\sigma}\eta^{\nu\rho}\rb
  B^a_\nu V^b_{\mu,n}\partial_\rho V^c_{\sigma,m},
\end{equation}
where
\begin{equation}
  a^{(2)}_{7,nm}\equiv\int\limits_{-\infty}^{+\infty}dz\,w(z)\,\frac{\psi_n(z)\,\psi_m(z)}{u(z)}.
\end{equation}
The results of the numerical analysis for this coupling for a few lowest \(n\) and \(m\)
can be found in Appendix~\ref{app:num}. Replacing the structure constants with~\eqref{eq:d_tens}
gives\footnote{Despite the appearance, the overall coefficient here and below is actually
real: there is an additional imaginary unit in the commutator, see~\eqref{eq:d_tens}.
There is also a difference in the definition of the structure constants~\eqref{eq:d_tens}
used in this work from the one used in~\cite{Sakai:2004cn,Sakai:2005yt}.}
\begin{equation}
  \left.S_{\text{eff},8}\vphfrac\right|_\text{loc}=-\sum_{n,m}2i\kappa a^{(2)}_{7,nm}\int d^4x
  \lb2\eta^{\mu\rho}\eta^{\nu\sigma}-\eta^{\mu\sigma}\eta^{\nu\rho}\rb
  B^a_\nu\,\text{Tr}\lb\lsb V_{\mu,n},\partial_\rho V_{\sigma,m}\rsb T^a\rb.
\end{equation}
After substituting the leading term of the expansion~\eqref{eq:B_exp} of \(B^a_\nu\),
\begin{equation}\label{eq:dbi_ind_ex}
  \left.S_{\text{eff},8}\vphfrac\right|_{\text{loc},\mathcal{O}\lb\pi\rb}=\sum_{n,m}\frac{4i\kappa a^{(2)}_{7,nm}}{f_\pi}\int d^4x
  \lb2\eta^{\mu\sigma}\eta^{\nu\rho}-\eta^{\mu\nu}\eta^{\rho\sigma}\rb
  \partial_\nu\pi^a\,\text{Tr}\lb\lsb\partial_\mu V_{\rho,n},V_{\sigma,m}\rsb T^a\rb,
\end{equation}
where some indices were renamed to facilitate comparison with~\eqref{eq:cs_ind_ex}. This
interaction term can encode processes such as \(\rho(1450)\to\lb a_1(1260)+\pi\rb_{P\text{-wave}}\),
which was seen experimentally~\cite{ParticleDataGroup:2024cfk}. In fact, in one of the
original works~\cite{Sakai:2005yt} on the SS model a similar interaction term was derived
to describe \(a_1\to\pi\rho\) decay mode. The interaction term contains the numerical
constant \(c_{v^na^m\pi}\) which can be expressed in the notation of this work as
\begin{equation}
  c_{v^na^m\pi}=2\kappa a^{(2)}_{7,2n-1,2m}.
\end{equation}
Our results for this interaction term and the numerical constant are largely\footnote{It
should be noted, however, that there seem to be some sign inconsistencies in the formulas~(3.71)-(3.76)
of~\cite{Sakai:2005yt}. For instance, as defined the experimental value of the numerical
constant \(c_{\rho a_1\pi}\) in~(3.75) should have been negative, since \(m_\rho<m_{a_1}\).}
in agreement with those of~\cite{Sakai:2005yt}. However, it should be again emphasized
the effective action obtained is non-local and beyond the local approximation it contains
additional terms that could contribute to the description of these processes.

The DBI-induced interaction~\eqref{eq:dbi_ind_ex} with vector mesons of the same parities
does not respect the spatial parity symmetry. However, all such terms are, in fact, zero.
Consider the following integral, which is zero, if the eigenfunctions have the same
parity,
\begin{equation}
  \int\limits_{-\infty}^{+\infty}dz\lb2w(z)-1\rb g^{(2)}_{7,nm}(z)=
  \int\limits_{-\infty}^{+\infty}dz\lb2w(z)-1\rb\frac{\psi_n(z)\,\psi_m(z)}{u(z)}=0,
\end{equation}
since \(2w(z)-1\propto\arctan z\) is an odd function, the fraction is an even function
and the integration is over a symmetric interval. The left-hand side can be written as
\begin{equation}
  \int\limits_{-\infty}^{+\infty}dz\lb2w(z)-1\rb g^{(2)}_{7,nm}(z)=
  2a^{(2)}_{7,nm}-\int\limits_{-\infty}^{+\infty}dz\,g^{(2)}_{7,nm}(z),
\end{equation}
which means that
\begin{equation}
  a^{(2)}_{7,nm}=\frac{1}{2}\int\limits_{-\infty}^{+\infty}dz\,\frac{\psi_n(z)\,\psi_m(z)}{u(z)}=\frac{\delta_{nm}}{2\kappa},\qquad
  n,m\text{ --- same parity},
\end{equation}
where we used the defintion~\eqref{eq:aij} of the \(a\)-coefficients and the
orthonormality~\eqref{eq:eigf_ortnorm} of the eigenfunctions. The contribution of the
same parity vector mesons in~\eqref{eq:dbi_ind_ex} can be written as
\begin{equation}\label{eq:t1}
  \left.S_{\text{eff},8}\vphfrac\right|_{\substack{\text{loc},\mathcal{O}\lb\pi\rb,\\n,m\text{ --- same par.}}}=\sum_n\frac{2i}{f_\pi}\int d^4x
  \lb2\eta^{\mu\sigma}\eta^{\nu\rho}-\eta^{\mu\nu}\eta^{\rho\sigma}\rb
  \partial_\nu\pi^a\,\text{Tr}\lb\lsb\partial_\mu V_{\rho,n},V_{\sigma,n}\rsb T^a\rb.
\end{equation}
The integral in~\eqref{eq:t1} can be integrated by parts,
\begin{multline}
  \int d^4x\lb2\eta^{\mu\sigma}\eta^{\nu\rho}-\eta^{\mu\nu}\eta^{\rho\sigma}\rb
  \partial_\nu\pi^a\,\text{Tr}\lb\lsb\partial_\mu V_{\rho,n},V_{\sigma,n}\rsb T^a\rb\\=
  -\int d^4x\lb2\eta^{\mu\sigma}\eta^{\nu\rho}-\eta^{\mu\nu}\eta^{\rho\sigma}\rb
  \pi^a\,\text{Tr}\lb\lsb\partial_\mu\partial_\nu V_{\rho,n},V_{\sigma,n}\rsb T^a\rb\\-
  \int d^4x\lb2\eta^{\mu\sigma}\eta^{\nu\rho}-\eta^{\mu\nu}\eta^{\rho\sigma}\rb
  \pi^a\,\text{Tr}\lb\lsb\partial_\mu V_{\rho,n},\partial_\nu V_{\sigma,n}\rsb T^a\rb.
\end{multline}
In the second integral if the indices \(\mu\leftrightarrow\nu\) and \(\rho\leftrightarrow\sigma\)
are simultaneously swapped, the same integral is obtained, but with an opposite sign.
Thus, the entire integral is equal to zero. In the first integral the first term in the
brackets, \(2\eta^{\mu\sigma}\eta^{\nu\rho}\), can be dropped because of the
transversality~\eqref{eq:trans_cond} of the mesons. For the last remaining term we
explicitly write down the sums over the Lorentz indices (recall, that the signature is
mostly minus),
\begin{multline}
  \int d^4x\,\eta^{\mu\nu}\eta^{\rho\sigma}
  \pi^a\,\text{Tr}\lb\lsb\partial_\mu\partial_\nu V_{\rho,n},V_{\sigma,n}\rsb T^a\rb=
  \int d^4x\,\pi^a\,\text{Tr}\lb\lsb\partial_0\partial_0 V_{0,n},V_{0,n}\rsb T^a\rb\\-
  \int d^4x\,\pi^a\,\text{Tr}\lb\lsb\partial_0\partial_0 V_{i,n},V_{i,n}\rsb T^a\rb-
  \int d^4x\,\pi^a\,\text{Tr}\lb\lsb\partial_i\partial_i V_{0,n},V_{0,n}\rsb T^a\rb\\+
  \int d^4x\,\pi^a\,\text{Tr}\lb\lsb\partial_i\partial_i V_{j,n},V_{j,n}\rsb T^a\rb.
\end{multline}
Changing the integration variable \(x^i\to-x^i\) does not change integration measure
(minus signs from \(d^3x\) compensate changes in the direction of integration). Since the
integrands in each of the integrals are quadratic in the vector meson components and
derivatives, the behavior of the integrand under \(x^i\to-x^i\) is determined solely by
the properties of \(\pi\)-meson which is pseudoscalar. As such, every integral attains
an extra minus sign from such variable change, which means that
\begin{equation}
  \int d^4x\,\eta^{\mu\nu}\eta^{\rho\sigma}
  \pi^a\,\text{Tr}\lb\lsb\partial_\mu\partial_\nu V_{\rho,n},V_{\sigma,n}\rsb T^a\rb=0.
\end{equation}
Thus, we conclude that all the DBI-induced parity-violating terms vanish as expected.

The difference between the CS-induced and DBI-induced interactions terms is clear: the
former only encodes the interactions involving the vector mesons of the same parities,
while the latter describes reactions of vector mesons of opposite parities. Furthermore,
there is no interaction between \(\eta'\) and the vector mesons in the DBI induced
interaction term~\eqref{eq:dbi_ind_ex}. In the \(\pi\)-sector this term is relatively
suppressed by the 't Hooft coupling compared to~\eqref{eq:cs_ind_ex},
\begin{equation}
  \frac{\left.S_{\text{eff},7}\vphfrac\right|_{\text{loc},\mathcal{O}\lb\pi\rb}}{\left.S_{\text{eff},8}\vphfrac\right|_{\text{loc},\mathcal{O}\lb\pi\rb}}=\mathcal{O}\lb\frac{N_c}{\kappa}\rb=\mathcal{O}\lb\frac{1}{\lambda_\text{YM}}\rb.
\end{equation}

\section{Concluding Discussions}\label{sect:conc}

In this work we have derived the effective action for light pseudoscalar and heavy vector
mesons stemming from the SS model in the holographic off-shell formalism. The DBI action
of the SS model was expanded up to fourth order in the bulk field strength. The degrees
of freedom of the effective action are the multiplets of massless pions, \(\eta'\)-meson,
and towers of heavy (axial-)vector meson multiplets. The latter can be thought of as being
integrated into the action. The resulting action (see~\eqref{eq:Seff} in Appendix~\ref{app:eff})
is non-linear, non-local, and rich in physical phenomena.

For arbitrary \(N_f\), we determined the values of bare and effective LECs as well as the
amplitudes of elastic pion scattering induced by the effective action. For \(N_f=2\) our
results are in complete agreement with~\cite{Hoyos:2022ptd}. For larger \(N_f\) we
clarify the difference between the bare and effective LECs. The obtained scattering
amplitude, particularly for \(SU(3)\), is of phenomenological relevance as it can be used
to analyze decays of kaons into pions.

We also discussed the interactions between the pseudoscalar and vector meson sectors that
can be extracted from the obtained effective action. This direction is particularly
interesting in light of the recent work on the radiative vector meson decays~\cite{Hechenberger:2023ljn}.
The CS and DBI actions are responsible for different types of interactions. We observed
that the CS-induced terms describe interactions involving vector mesons with identical
parities while the DBI-induced terms pertain to mesons with opposite parities. At the
same time, the DBI-induced interactions are parametrically suppressed compared to the
CS-induced ones.

The SS model possesses a natural energy scale, \(\mkk\), which corresponds to the energy
cutoff of the effective theory on the boundary. As such, the degrees of freedom of the
effective theory should have energies not exceeding this scale. Indeed, if the off-shell
vector mesons \(V_{\mu,n}\) were not introduced in Section~\ref{sect:sols}, the only
remaining degrees of freedom would be \(\pi\)- and \(\eta'\)-mesons. When vector mesons
are integrated in, the cutoff moves up and may constitute the extension of the energy
region of the effective theory. If a finite tower of the vector mesons is introduced,
this can be viewed as extending the region of the theory's applicability up to the mass
of the heaviest meson.

The inclusion of all orders in the gradient expansion, which manifests in the
non-locality of the effective action, makes the effective theory relativistically causal.
Since the tower of heavy vector mesons is infinite (and thus effective energy region may
be also thought to be infinite), there is space to speculate that the resulting effective
action might be UV complete.\footnote{It should be pointed out that the UV limit of the
SS model, which effectively corresponds to the \(\mkk\to0\) limit, is a five-dimensional
superconformal theory.} Quite obviously, vector meson loops generated by the terms such
as~\eqref{eq:Seff12} and~\eqref{eq:Seff14} (schematically, \(B^2V^2\) and \(V^4\)) would
produce UV divergences. There exists a hypothetical scenario that the divergences might
eventually cancel due to summation over the vector meson towers. This is hard to check,
however, since such a check would require knowledge of the entire meson spectra and
couplings. These couplings are given by integrals involving the eigenfunctions \(\psi_n\),
which are not known at the moment. Thus, exploring this spectral problem might be worth
pursuing. Applying the spectral parameter power series method~\cite{kravchenko2010spectral}
looks like a promising direction for the future.
\\ \ \\
\noindent
\textbf{Acknowledgements.}
We thank Sergey Afonin, Yanyan Bu, Maxim Nefedov, Jacob Sonnenschein, and Ismail Zahed
for useful and illuminating discussions. This research was funded by Binational Science
Foundation grants \#2021789 and \#2022132, by the ISF grant \#910/23, and by VATAT
(Israel planning and budgeting committee) grant for supporting theoretical high energy
physics. The work of TS was additionally supported by the Kreitman School of Advanced
Graduate Studies of Ben-Gurion University of the Negev.

\appendix

\section{The Effective Action --- Complete Expressions}\label{app:eff}

This Appendix contains the explicit expressions for the main result, the effective action.
Here, in addition to the \(c\)-coefficients~\eqref{eq:cij}, \emph{\(a\)-coefficients} are
introduced
\begin{equation}\label{eq:aij}
  a^{(j)}_{i,\dots}\equiv\int\limits_{-\infty}^{+\infty}dz\,w(z)\,g^{(j)}_{i,\dots}(z),
\end{equation}
where the function \(w(z)\) is given in~\eqref{eq:w_def}. The functions \(g^{(j)}_{i,\dots}(z)\)
are given by~\eqref{eq:nlo_g} and~\eqref{eq:nnlo_g} for the second (\(j=2\)) and the
third (\(j=3\)) order. The two types of coefficients are related,\footnote{Except for
\(a^{(2)}_5\) and \(a^{(3)}_{31}\), where the summation and integration operations in the
full higher order transverse functions~\eqref{eq:full_ho_s} are not interchangeable.}
\begin{equation}\label{eq:ac}
  a^{(j)}_{i,\dots}=\sum_k\kappa\frac{\mathcal{C}_k}{\lambda_k}c^{(j)}_{i,\dots,k}.
\end{equation}
Here, \(\lambda_k\) are the eigenvalues of the spectral problem~\eqref{eq:spec_prob} and
\(\mathcal{C}_k\) are the asymptotic values of the corresponding eigenfunctions,
\begin{equation}\label{eq:Cn}
  \mathcal{C}_k\equiv\lim_{z\to+\infty}z^2\pz\psi_k(z)=-\lim_{z\to+\infty}z\psi_k(z).
\end{equation}
The effective action involves the following integrals:
\begin{equation}\label{eq:Ik}
  \mathcal{I}_k\equiv\int\limits_{-\infty}^{+\infty}dz\,\frac{w(z)^{k+1}}{u(z)}.
\end{equation}
The expressions listed below will also involve the longitudinal~\eqref{eq:long_gf} and
transverse~\eqref{eq:trans_gf} Green functions \(G_l\) and \(G_t\). Special functions
\(k(q,z)\), \(l(q,z)\), \(K^a(q,z)\), and \(L^a(q,z)\), defined in~\eqref{eq:KL_not} for
the third order solutions, are used as well.

The effective action is split into 21 terms,
\begin{equation}\label{eq:Seff}
  S_\text{eff}=\sum_{i=1}^{21}S_{\text{eff},i}.
\end{equation}
Below are the expressions for these terms.
\begin{equation}\label{eq:Seff1}
  S_{\text{eff},1}\equiv-\frac{\kappa}{\pi}\int d^4q\,
  \text{STr}\lsb B_\nu(q)\,B^\nu(-q)\rsb-
  \sum_n\int d^4q\lb q^2-\lambda_n\rb
  \text{STr}\lsb V_{\nu,n}(-q)\,V^\nu_n(q)\rsb,
\end{equation}
\begin{multline}\label{eq:Seff2}
  S_{\text{eff},2}\equiv\frac{if_{abc}\kappa}{2(2\pi)^2L^4}\lim_{z\to+\infty}u(z)^3w(z)\int d^4q\,d^4p\,dz'
  \lb2\eta^{\mu\rho}\eta^{\nu\sigma}-\eta^{\mu\sigma}\eta^{\nu\rho}-
  \eta^{\mu\nu}\eta^{\rho\sigma}\rb\\\times
  p_\rho B^a_\sigma(p)\,B^b_\mu(q-p)\,B^c_\vk(-q)\,
  \pz\lb D^\vk_\nu(q)\,G_l(z,z')+\lb\delta^\vk_\nu-D^\vk_\nu(q)\rb G_t(z,z',q)\rb
  g^{(2)}_5(z')\\+
  \frac{f_{abc}f_{cde}\kappa}{2(2\pi)^4L^4}\lim_{z\to+\infty}u(z)^3w(z)\int d^4q\,d^4p_1\,d^4p_2\,dz'\,
  \eta^{\mu\rho}\eta^{\nu\sigma}
  B^a_\vk(-q)\,B^b_\mu(p_2)\,B^d_\rho(p_1-p_2)\,B^e_\sigma(q-p_1)\\\times
  \pz\lb D^\vk_\nu(q)\,G_l(z,z')+\lb\delta^\vk_\nu-D^\vk_\nu(q)\rb G_t(z,z',q)\rb
  g^{(3)}_{31}(z')\\+
  \frac{if_{abc}\kappa\mathcal{I}_1}{2(2\pi)^2}\int d^4q\,d^4p\,
  \eta^{\mu\sigma}\eta^{\nu\rho}
  p_\rho B^a_\sigma(p)\,B^b_\mu(q-p)\,B^c_\nu(-q)\\+
  \frac{if_{abc}\kappa\mathcal{I}_2}{2(2\pi)^2}\int d^4q\,d^4p
  \eta^{\mu\sigma}\eta^{\nu\rho}
  p_\rho B^a_\sigma(p)\,B^b_\mu(q-p)\,B^c_\nu(-q)\\-
  \frac{f_{abc}f_{cde}\kappa\mathcal{I}_3}{4(2\pi)^4}\int d^4q\,d^4p_1\,d^4p_2\,
  \eta^{\mu\sigma}\eta^{\nu\rho}
  B^a_\mu(-q)\,B^b_\nu(p_2)\,B^d_\rho(p_1-p_2)\,B^e_\sigma(q-p_1),
\end{multline}
\begin{multline}\label{eq:Seff3}
  S_{\text{eff},3}\equiv\sum_{i=1}^{20}\sum_k\kappa^2L^4\mathcal{C}_kc^{(3)}_{i,\dots,k}\int d^4q\,
  \lb\frac{\delta^\vk_\nu}{2\lambda_k}-\frac{D^\vk_\nu(q)}{\lambda_k}+
  \frac{\delta^\vk_\nu-D^\vk_\nu(q)}{q^2-\lambda_k}\rb
  \text{STr}\lsb B_\vk(-q)\,J^{\nu,(3)}_{i,\dots}(q)\rsb\\+
  \sum_{i=1}^{20}\sum_n\frac{3\kappa L^4c^{(3)}_{i,\dots,n}}{2}\int d^4q\,
  \text{STr}\lsb V_{\nu,n}(-q)\,J^{\nu,(3)}_{i,\dots}(q)\rsb,
\end{multline}
\begin{multline}
  S_{\text{eff},4}\equiv\sum_k\frac{iN_c\veps^{z\mu\nu\rho\sigma}\kappa\mathcal{C}_kc^{(2)}_{3,k}}{8(2\pi)^2\pi^2}\int d^4q\,d^4p\,
  \lb\frac{D^\vk_\nu(q)}{\lambda_k}-\frac{\delta^\vk_\nu-D^\vk_\nu(q)}{q^2-\lambda_k}\rb\\\times
  p_\mu\,\text{STr}\lsb B_\vk(-q)\,B_\rho(p)\,B_\sigma(q-p)\rsb,
\end{multline}
\begin{multline}\label{eq:Seff5}
  S_{\text{eff},5}\equiv\sum_{n,k}\frac{iN_c\veps^{z\mu\nu\rho\sigma}\kappa\mathcal{C}_kc^{(2)}_{1,n,k}}{8(2\pi)^2\pi^2}\int d^4q\,d^4p\,
  \lb\frac{\delta^\vk_\nu-D^\vk_\nu(q)}{q^2-\lambda_k}-\frac{D^\vk_\nu(q)}{\lambda_k}\rb\\\times
  p_\mu\,\text{STr}\lsb B_\vk(-q)\,B_\rho(q-p)\,V_{\sigma,n}(p)\rsb\\+
  \sum_{n,k}\frac{iN_c\veps^{z\mu\nu\rho\sigma}\kappa\mathcal{C}_kc^{(2)}_{4,n,k}}{8(2\pi)^2\pi^2}\int d^4q\,d^4p\,
  \lb\frac{D^\vk_\nu(q)}{\lambda_k}-\frac{\delta^\vk_\nu-D^\vk_\nu(q)}{q^2-\lambda_k}-
  \frac{\delta^\vk_\nu}{\lambda_k}\rb\\\times
  p_\mu\,\text{STr}\lsb B_\vk(-q)\,B_\rho(p)\,V_{\sigma,n}(q-p)\rsb,
\end{multline}
\begin{multline}\label{eq:Seff6}
  S_{\text{eff},6}\equiv\sum_{n,k}\frac{if_{abc}\kappa^2\mathcal{C}_kc^{(2)}_{6,n,k}}{2(2\pi)^2}\int d^4q\,d^4p
  \lb2\eta^{\mu\rho}\eta^{\nu\sigma}-\eta^{\mu\sigma}\eta^{\nu\rho}-
  \eta^{\mu\nu}\eta^{\rho\sigma}\rb
  \lb\frac{D^\vk_\nu(q)}{\lambda_k}-\frac{\delta^\vk_\nu-D^\vk_\nu(q)}{q^2-\lambda_k}\rb\\\times
  p_\rho B^a_\vk(-q)\,B^b_\sigma(p)\,V^c_{\mu,n}(q-p)\\+
  \sum_{n,k}\frac{if_{abc}\kappa^2\mathcal{C}_kc^{(2)}_{6,n,k}}{2(2\pi)^2}\int d^4q\,d^4p
  \lb2\eta^{\mu\nu}\eta^{\rho\sigma}-\eta^{\mu\sigma}\eta^{\nu\rho}\rb
  \lb\frac{\delta^\vk_\nu-D^\vk_\nu(q)}{q^2-\lambda_k}-\frac{D^\vk_\nu(q)}{\lambda_k}\rb\\\times
  p_\rho B^a_\vk(-q)\,B^b_\sigma(q-p)\,V^c_{\mu,n}(p)\\-
  \sum_n\frac{if_{abc}\kappa}{2(2\pi)^2}\int d^4q\,d^4p
  \lb a^{(2)}_{6,n}+\kappa\int dz\lb q^2-\lambda_n\rb\frac{\psi_n(z)}{u(z)}\,\hat{s}^{(2)}_5(q,z)\rb\\\times
  \lb2\eta^{\mu\rho}\eta^{\nu\sigma}-\eta^{\mu\sigma}\eta^{\nu\rho}-
  \eta^{\mu\nu}\eta^{\rho\sigma}\rb
  p_\rho B^a_\sigma(p)\,B^b_\mu(q-p)\,V^c_{\nu,n}(-q)\\-
  \sum_n\frac{if_{abc}\kappa}{(2\pi)^2}\frac{\mathcal{C}_n}{\lambda_n}\int d^4q\,d^4p\,
  \lb\eta^{\mu\rho}\eta^{\nu\sigma}-\eta^{\mu\nu}\eta^{\rho\sigma}\rb
  p_\rho B^a_\sigma(p)\,B^b_\mu(q-p)\,V^c_{\nu,n}(-q),
\end{multline}
\begin{multline}\label{eq:Seff7}
  S_{\text{eff},7}\equiv\sum_{n,m,k}\frac{iN_c\veps^{z\mu\nu\rho\sigma}\kappa\mathcal{C}_kc^{(2)}_{2,nm,k}}{8(2\pi)^2\pi^2}\int d^4q\,d^4p\,
  \lb\frac{D^\vk_\nu(q)}{\lambda_k}-\frac{\delta^\vk_\nu-D^\vk_\nu(q)}{q^2-\lambda_k}-
  \frac{3\delta^\vk_\nu}{\lambda_k}\rb\\\times
  p_\mu\,\text{STr}\lsb B_\vk(-q)\,V_{\rho,n}(p)\,V_{\sigma,m}(q-p)\rsb,
\end{multline}
\begin{multline}\label{eq:Seff8}
  S_{\text{eff},8}\equiv\sum_{n,m,k}\frac{if_{abc}\kappa^2\mathcal{C}_kc^{(2)}_{7,nm,k}}{2(2\pi)^2}\int d^4q\,d^4p
  \lb2\eta^{\mu\rho}\eta^{\nu\sigma}-\eta^{\mu\sigma}\eta^{\nu\rho}\rb\\\times
  \lb\frac{\delta^\vk_\nu-D^\vk_\nu(q)}{q^2-\lambda_k}-\frac{D^\vk_\nu(q)}{\lambda_k}+
  \frac{3\delta^\vk_\nu}{\lambda_k}\rb
  p_\rho B^a_\vk(-q)\,V^b_{\mu,n}(q-p)\,V^c_{\sigma,m}(p),
\end{multline}
\begin{equation}
  S_{\text{eff},9}\equiv-\sum_{n,m,l}\frac{iN_c\veps^{z\mu\nu\rho\sigma}c^{(2)}_{2,nm,l}}{6(2\pi)^2\pi^2}\int d^4q\,d^4p\,
  p_\mu\,\text{STr}\lsb V_{\nu,l}(q-p)\,V_{\rho,n}(p)\,V_{\sigma,m}(-q)\rsb,
\end{equation}
\begin{equation}
  S_{\text{eff},10}\equiv-\sum_{n,m,l}\frac{2if_{abc}\kappa c^{(2)}_{7,nm,l}}{(2\pi)^2}\int d^4q\,d^4p
  \eta^{\mu\rho}\eta^{\nu\sigma}
  p_\rho V^a_{\sigma,l}(p)\,V^b_{\mu,m}(q-p)\,V^c_{\nu,n}(-q),
\end{equation}
\begin{multline}
  S_{\text{eff},11}\equiv\sum_{k,l}\frac{f_{abc}f_{cde}\kappa^2\mathcal{C}_kc^{(3)}_{32,l,k}}{2(2\pi)^4}\int d^4q\,d^4p_1\,d^4p_2\,
  \lb\eta^{\mu\rho}\eta^{\nu\sigma}-\eta^{\mu\sigma}\eta^{\nu\rho}\rb\\\times
  \lb\frac{D^\vk_\nu(q)}{\lambda_k}-\frac{\delta^\vk_\nu-D^\vk_\nu(q)}{q^2-\lambda_k}\rb
  B^a_\vk(-q)\,B^b_\mu(p_2)\,B^d_\rho(q-p_1)\,V^e_{\sigma,l}(p_1-p_2)\\+
  \sum_{k,l}\frac{f_{abc}f_{cde}\kappa^2\mathcal{C}_kc^{(3)}_{32,l,k}}{2(2\pi)^4}\int d^4q\,d^4p_1\,d^4p_2\,
  \eta^{\mu\sigma}\eta^{\nu\rho}
  \lb\frac{D^\vk_\nu(q)}{\lambda_k}-\frac{\delta^\vk_\nu-D^\vk_\nu(q)}{q^2-\lambda_k}\rb\\\times
  B^a_\vk(-q)\,B^d_\mu(p_2)\,B^e_\rho(q-p_1)\,V^b_{\sigma,l}(p_1-p_2)\\+
  \sum_l\frac{f_{abc}f_{cde}\kappa}{(2\pi)^4}\int d^4q\,d^4p_1\,d^4p_2\,
  \lb a^{(3)}_{32,l}+\frac{\kappa}{2}\int dz\lb q^2-\lambda_l\rb\frac{\psi_l(z)}{u(z)}\,\hat{s}^{(3)}_{31}(q,z)\rb\\\times
  \eta^{\mu\rho}\eta^{\nu\sigma}
  B^a_\mu(q-p_1)\,B^b_\nu(p_2)\,B^d_\rho(p_1-p_2)\,V^e_{\sigma,l}(-q),
\end{multline}
\begin{multline}\label{eq:Seff12}
  S_{\text{eff},12}\equiv\sum_{k,l,m}\frac{f_{abc}f_{cde}\kappa^2\mathcal{C}_kc^{(3)}_{33,lm,k}}{2(2\pi)^4}\int d^4q\,d^4p_1\,d^4p_2\,
  \lb\eta^{\mu\rho}\eta^{\nu\sigma}-\eta^{\mu\nu}\eta^{\rho\sigma}\rb\\\times
  \lb\frac{D^\vk_\nu(q)}{\lambda_k}-\frac{2\delta^\vk_\nu}{\lambda_k}-
  \frac{\delta^\vk_\nu-D^\vk_\nu(q)}{q^2-\lambda_k}\rb
  B^a_\vk(-q)\,B^d_\mu(p_1-p_2)\,V^b_{\rho,l}(p_2)\,V^e_{\sigma,m}(q-p_1)\\+
  \sum_{k,l,m}\frac{f_{abc}f_{cde}\kappa^2\mathcal{C}_kc^{(3)}_{33,lm,k}}{2(2\pi)^4}\int d^4q\,d^4p_1\,d^4p_2\,
  \eta^{\mu\rho}\eta^{\nu\sigma}
  \lb\frac{D^\vk_\nu(q)}{\lambda_k}-\frac{2\delta^\vk_\nu}{\lambda_k}-
  \frac{\delta^\vk_\nu-D^\vk_\nu(q)}{q^2-\lambda_k}\rb\\\times
  B^a_\vk(-q)\,B^b_\mu(p_2)\,V^d_{\rho,l}(p_1-p_2)\,V^e_{\sigma,m}(q-p_1),
\end{multline}
\begin{multline}
  S_{\text{eff},13}\equiv\sum_{k,l,m,n}\frac{f_{abc}f_{cde}\kappa^2\mathcal{C}_kc^{(3)}_{34,lmn,k}}{2(2\pi)^4}\int d^4q\,d^4p_1\,d^4p_2\,
  \eta^{\mu\rho}\eta^{\nu\sigma}\\\times
  \lb\frac{D^\vk_\nu(q)}{\lambda_k}-\frac{5\delta^\vk_\nu}{\lambda_k}-
  \frac{\delta^\vk_\nu-D^\vk_\nu(q)}{q^2-\lambda_k}\rb
  B^a_\vk(-q)\,V^b_{\mu,l}(p_2)\,V^d_{\rho,m}(p_1-p_2)\,V^e_{\sigma,n}(q-p_1),
\end{multline}
\begin{multline}\label{eq:Seff14}
  S_{\text{eff},14}\equiv\sum_{l,m,n,k}\frac{3f_{abc}f_{cde}\kappa c^{(3)}_{34,lmn,k}}{4(2\pi)^4}\int d^4q\,d^4p_1\,d^4p_2\,
  \eta^{\mu\rho}\eta^{\nu\sigma}\\\times
  V^a_{\mu,l}(-q)\,V^b_{\nu,m}(p_2)\,V^d_{\rho,n}(p_1-p_2)\,V^e_{\sigma,k}(q-p_1),
\end{multline}
\begin{multline}\label{eq:Seff15}
  S_{\text{eff},15}\equiv\sum_{\substack{k,l,i,\dots\\i\neq5}}\frac{iN_c\veps^{z\mu\nu\rho\sigma}\kappa^2L^4\mathcal{C}_lc^{(3)}_{23,k,i,\dots,l}}{8(2\pi)^2\pi^2}\int d^4q\,d^4p\,
  \frac{1}{p^2-\lambda_k}\lb\frac{D^\vk_\nu(q)}{\lambda_l}-\frac{\delta^\vk_\nu-D^\vk_\nu(q)}{q^2-\lambda_l}\rb\\\times
  p_\mu\,\text{STr}\lsb B_\vk(-q)\,B_\rho(q-p)\,J^{(2)}_{\sigma,i,\dots}(p)\rsb\\-
  \sum_{k,i,\dots}\frac{iN_c\veps^{z\mu\nu\rho\sigma}\kappa\mathcal{C}_kc^{(3)}_{25,i,\dots,k}}{8(2\pi)^2\pi^2}\int d^4q\,d^4p\,
  D^\lambda_\sigma(q-p)\lb\frac{1}{q^2-\lambda_k}+\frac{1}{2\lambda_k}\rb\\\times
  p_\mu \text{STr}\lsb B_\nu(-q)\,B_\rho(p)\,J^{(2)}_{\lambda,i,\dots}(q-p)\rsb\\+
  \sum_{\substack{k,l,i,\dots\\i\neq5}}\frac{iN_c\veps^{z\mu\nu\rho\sigma}\kappa^2L^4\mathcal{C}_lc^{(3)}_{26,k,i,\dots,l}}{8(2\pi)^2\pi^2}\int d^4q\,d^4p\,
  \frac{\delta^\lambda_\sigma-D^\lambda_\sigma(q-p)}{(q-p)^2-\lambda_k}\\\times
  \lb\frac{\delta^\vk_\nu}{2\lambda_l}-\frac{D^\vk_\nu(q)}{\lambda_l}+
  \frac{\delta^\vk_\nu-D^\vk_\nu(q)}{q^2-\lambda_l}\rb
  p_\mu\,\text{STr}\lsb B_\vk(-q)\,B_\rho(p)\,J^{(2)}_{\lambda,i,\dots}(q-p)\rsb\\+
  \frac{iN_c\veps^{z\mu\nu\rho\sigma}\kappa}{8(2\pi)^2\pi^2}\lim_{z\to+\infty}u(z)^3w(z)\int dz'\,d^4q\,d^4p\,
  p_\mu\,\text{STr}\lsb B_\vk(-q)\,B_\rho(q-p)\,J^{(2)}_{\sigma,5}(p)\rsb\\\times
  \pz\lb D^\vk_\nu(q)\,G_l(z,z')+\lb\delta^\vk_\nu-D^\vk_\nu(q)\rb G_t(z,z',q)\rb
  \partial_{z'}w(z')\,\hat{s}^{(2)}_5(p,z')\\-
  \frac{iN_c\veps^{z\mu\nu\rho\sigma}\kappa}{8(2\pi)^2\pi^2}\lim_{z\to+\infty}u(z)^3w(z)\int dz'\,d^4q\,d^4p\,
  \lb\delta^\lambda_\sigma-D^\lambda_\sigma(q-p)\rb\\\times
  p_\mu\,\text{STr}\lsb B_\vk(-q)\,B_\rho(p)\,J^{(2)}_{\lambda,5}(q-p)\rsb\\\times
  \pz\lb D^\vk_\nu(q)\,G_l(z,z')+\lb\delta^\vk_\nu-D^\vk_\nu(q)\rb G_t(z,z',q)\rb
  w(z')\,\partial_{z'}\hat{s}^{(2)}_5(q-p,z')\\-
  \frac{3iN_c\veps^{z\mu\nu\rho\sigma}\kappa L^4}{16(2\pi)^2\pi^2}\int d^4q\,d^4p\,
  dz\,w(z)\,\pz w(z)\,\hat{s}^{(2)}_5(q,z)\\\times
  \lb\delta^\lambda_\sigma-D^\lambda_\sigma(q)\rb
  p_\mu\,\text{STr}\lsb B_\nu(q-p)\,B_\rho(p)\,J^{(2)}_{\lambda,5}(-q)\rsb,
\end{multline}
\begin{multline}\label{eq:Seff16}
  S_{\text{eff},16}\equiv\sum_{k,i,\dots}\frac{if_{abc}\kappa^2\mathcal{C}_kc^{(3)}_{27,i,\dots,k}}{2(2\pi)^2}\int d^4q\,d^4p
  \lb2\eta^{\mu\rho}\eta^{\nu\sigma}-\eta^{\mu\nu}\eta^{\rho\sigma}-
  \eta^{\mu\sigma}\eta^{\nu\rho}\rb\\\times
  \lb\frac{D^\lambda_\sigma(p)}{\lambda_k}-\frac{\delta^\lambda_\sigma}{\lambda_k}-
  \frac{\delta^\lambda_\sigma-D^\lambda_\sigma(p)}{p^2-\lambda_k}\rb
  p_\rho B^a_\lambda(p)\,B^b_\mu(-q)\,D^\vk_\nu(q-p)\,J^{c,(2)}_{\vk,i,\dots}(q-p)\\+
  \sum_{k,i,\dots}\frac{if_{abc}\kappa^2\mathcal{C}_kc^{(3)}_{27,i,\dots,k}}{2(2\pi)^2}\int d^4q\,d^4p
  \lb2\eta^{\mu\nu}\eta^{\rho\sigma}-\eta^{\mu\rho}\eta^{\nu\sigma}-
  \eta^{\mu\sigma}\eta^{\nu\rho}\rb\\\times
  \lb\frac{\delta^\lambda_\sigma-D^\lambda_\sigma(q)}{q^2-\lambda_k}-\frac{D^\lambda_\sigma(q)}{\lambda_k}\rb
  p_\rho B^a_\lambda(-q)\,B^b_\mu(p)\,D^\vk_\nu(q-p)\,J^{c,(2)}_{\vk,i,\dots}(q-p)\\+
  \sum_{\substack{k,l,i,\dots\\i\neq5}}\frac{if_{abc}\kappa^3L^4\mathcal{C}_lc^{(3)}_{28,k,i,\dots,l}}{2(2\pi)^2}\int d^4q\,d^4p
  \lb2\eta^{\mu\rho}\eta^{\nu\sigma}-\eta^{\mu\nu}\eta^{\rho\sigma}-
  \eta^{\mu\sigma}\eta^{\nu\rho}\rb\\\times
  \frac{\delta^\vk_\nu-D^\vk_\nu(q-p)}{(q-p)^2-\lambda_k}
  \lb\frac{\delta^\lambda_\sigma-D^\lambda_\sigma(p)}{p^2-\lambda_l}-\frac{D^\lambda_\sigma(p)}{\lambda_l}+
  \frac{\delta^\lambda_\sigma}{\lambda_l}\rb
  p_\rho B^a_\lambda(p)\,B^b_\mu(-q)\,J^{c,(2)}_{\vk,i,\dots}(q-p)\\+
  \sum_{\substack{k,l,i,\dots\\i\neq5}}\frac{if_{abc}\kappa^3L^4\mathcal{C}_lc^{(3)}_{28,k,i,\dots,l}}{2(2\pi)^2}\int d^4q\,d^4p
  \lb2\eta^{\mu\nu}\eta^{\rho\sigma}-\eta^{\mu\rho}\eta^{\nu\sigma}-
  \eta^{\mu\sigma}\eta^{\nu\rho}\rb\\\times
  \frac{\delta^\vk_\nu-D^\vk_\nu(q-p)}{(q-p)^2-\lambda_k}
  \lb\frac{D^\lambda_\sigma(q)}{\lambda_l}-\frac{\delta^\lambda_\sigma-D^\lambda_\sigma(q)}{q^2-\lambda_l}\rb
  p_\rho B^a_\lambda(-q)\,B^b_\mu(p)\,J^{c,(2)}_{\vk,i,\dots}(q-p)\\+
  \frac{if_{abc}\kappa^3L^4}{2(2\pi)^2}\lb\lim_{z\to+\infty}-\lim_{z\to-\infty}\rb u(z)^3\int d^4q\,d^4p\,
  \hat{s}^{(2)}_5(q,z)\pz\hat{s}^{(2)}_5(q,z)\\\times
  \lb2\eta^{\mu\rho}\eta^{\nu\sigma}-\eta^{\mu\sigma}\eta^{\nu\rho}-
  \eta^{\mu\nu}\eta^{\rho\sigma}\rb\\\times
  \lb\delta^\vk_\nu-D^\vk_\nu(q)\rb
  p_\rho B^a_\sigma(p)\,B^b_\mu(q-p)\,J^{c,(2)}_{\vk,5}(-q)\\+
  \frac{if_{abc}\kappa^2}{2(2\pi)^2}\lim_{z\to+\infty}u(z)^3w(z)\int dz'\,d^4q\,d^4p
  \lb2\eta^{\mu\rho}\eta^{\nu\sigma}-\eta^{\mu\nu}\eta^{\rho\sigma}-
  \eta^{\mu\sigma}\eta^{\nu\rho}\rb\\\times
  \lb\delta^\vk_\nu-D^\vk_\nu(p)\rb
  p_\rho B^a_\lambda(-q)\,B^b_\mu(q-p)\,J^{c,(2)}_{\vk,5}(p)\\\times
  \pz\lb D^\lambda_\sigma(q)\,G_l(z,z')+\lb\delta^\lambda_\sigma-D^\lambda_\sigma(q)\rb G_t(z,z',q)\rb
  \frac{w(z')}{u(z')}\hat{s}^{(2)}_5(p,z')\\-
  \frac{if_{abc}\kappa^2}{2(2\pi)^2}\lim_{z\to+\infty}u(z)^3w(z)\int dz'\,d^4q\,d^4p
  \lb2\eta^{\mu\sigma}\eta^{\nu\rho}-\eta^{\mu\nu}\eta^{\rho\sigma}-
  \eta^{\mu\rho}\eta^{\nu\sigma}\rb\\\times
  \lb\delta^\vk_\nu-D^\vk_\nu(q-p)\rb
  p_\rho B^a_\lambda(-q)\,B^b_\mu(p)\,J^{c,(2)}_{\vk,5}(q-p)\\\times
  \pz\lb D^\lambda_\sigma(q)\,G_l(z,z')+\lb\delta^\lambda_\sigma-D^\lambda_\sigma(q)\rb G_t(z,z',q)\rb
  \frac{w(z')}{u(z')}\hat{s}^{(2)}_5(q-p,z')\\+
  \frac{if_{abc}\kappa^2L^4}{2(2\pi)^2}\int d^4q\,d^4p\,
  dz\,\frac{w(z)^2}{u(z)}\,\hat{s}^{(2)}_5(q,z)
  \lb2\eta^{\mu\rho}\eta^{\nu\sigma}-\eta^{\mu\nu}\eta^{\rho\sigma}-
  \eta^{\mu\sigma}\eta^{\nu\rho}\rb\\\times
  \lb\delta^\vk_\nu-D^\vk_\nu(q)\rb
  p_\rho B^a_\sigma(p)\,B^b_\mu(q-p)\,J^{c,(2)}_{\vk,5}(-q)\\+
  \frac{if_{abc}\kappa^2L^4}{2(2\pi)^2}\int d^4q\,d^4p\,
  dz\,\frac{w(z)}{u(z)}\,\hat{s}^{(2)}_5(q,z)
  \lb\eta^{\mu\rho}\eta^{\nu\sigma}-\eta^{\mu\nu}\eta^{\rho\sigma}\rb
  p_\rho B^a_\sigma(p)\,B^b_\mu(q-p)\,J^{c(2)}_{\nu,5}(-q)\\+
  \sum_{\substack{k,i,\dots\\i\neq5}}\frac{if_{abc}\kappa^2L^4}{2(2\pi)^2}\frac{\mathcal{C}_k}{\lambda_k}c^{(2)}_{i,\dots,k}\int d^4q\,d^4p\,
  \lb\eta^{\mu\rho}\eta^{\nu\sigma}-\eta^{\mu\nu}\eta^{\rho\sigma}\rb
  \frac{\delta^\vk_\nu-D^\vk_\nu(q)}{q^2-\lambda_k}\\\times
  p_\rho B^a_\sigma(p)\,B^b_\mu(q-p)\,J^{c,(2)}_{\vk,i,\dots}(-q),
\end{multline}
\begin{multline}
  S_{\text{eff},17}\equiv-\sum_{n,k,i,\dots}\frac{iN_c\veps^{z\mu\nu\rho\sigma}\kappa\mathcal{C}_kc^{(3)}_{21,n,i,\dots,k}}{8(2\pi)^2\pi^2}\int d^4q\,d^4p\,
  D^\lambda_\sigma(q-p)\lb\frac{2}{\lambda_k}+\frac{1}{q^2-\lambda_k}\rb\\\times
  p_\mu\,\text{STr}\lsb B_\nu(-q)\,V_{\rho,n}(p)\,J^{(2)}_{\lambda,i,\dots}(q-p)\rsb\\+
  \sum_{\substack{n,k,l,i,\dots\\i\neq5}}\frac{iN_c\veps^{z\mu\nu\rho\sigma}\kappa^2L^4\mathcal{C}_lc^{(3)}_{22,n,k,i,\dots,l}}{8(2\pi)^2\pi^2}\int d^4q\,d^4p\,
  \frac{\delta^\lambda_\sigma-D^\lambda_\sigma(q-p)}{(q-p)^2-\lambda_k}\\\times
  \lb\frac{\delta^\vk_\nu-D^\vk_\nu(q)}{q^2-\lambda_l}-\frac{D^\vk_\nu(q)}{\lambda_l}+
  \frac{2\delta^\vk_\nu}{\lambda_l}\rb
  p_\mu\,\text{STr}\lsb B_\vk(-q)\,V_{\rho,n}(p)\,J^{(2)}_{\lambda,i,\dots}(q-p)\rsb\\+
  \sum_{\substack{n,k,l,i,\dots\\i\neq5}}\frac{iN_c\veps^{z\mu\nu\rho\sigma}\kappa^2L^4\mathcal{C}_lc^{(3)}_{24,n,k,i,\dots,l}}{8(2\pi)^2\pi^2}\int d^4q\,d^4p\,
  \frac{\delta^\lambda_\sigma-D^\lambda_\sigma(q-p)}{(q-p)^2-\lambda_k}\\\times
  \lb\frac{2\delta^\vk_\nu}{\lambda_l}-\frac{D^\vk_\nu(q)}{\lambda_l}+
  \frac{\delta^\vk_\nu-D^\vk_\nu(q)}{q^2-\lambda_l}\rb
  p_\mu\,\text{STr}\lsb B_\vk(-q)\,V_{\rho,n}(p)\,J^{(2)}_{\lambda,i,\dots}(q-p)\rsb\\+
  \sum_{\substack{n,k,l,i,\dots\\i\neq5}}\frac{iN_c\veps^{z\mu\nu\rho\sigma}\kappa^2L^4\mathcal{C}_lc^{(3)}_{24,n,k,i,\dots,l}}{8(2\pi)^2\pi^2}\int d^4q\,d^4p\,
  \frac{\delta^\lambda_\sigma-D^\lambda_\sigma(q)}{q^2-\lambda_k}
  \lb\frac{1}{p^2-\lambda_l}+\frac{2}{\lambda_l}\rb\\\times
  p_\mu\,\text{STr}\lsb B_\nu(p)\,V_{\rho,n}(q-p)\,J^{(2)}_{\lambda,i,\dots}(-q)\rsb\\+
  \frac{iN_c\veps^{z\mu\nu\rho\sigma}\kappa}{8(2\pi)^2\pi^2}\lim_{z\to+\infty}u(z)^3w(z)\int dz'\,d^4q\,d^4p\,
  p_\mu\,\text{STr}\lsb B_\vk(-q)\,V_{\rho,n}(q-p)\,J^{(2)}_{\sigma,5}(p)\rsb\\\times
  \pz\lb D^\vk_\nu(q)\,G_l(z,z')+\lb\delta^\vk_\nu-D^\vk_\nu(q)\rb G_t(z,z',q)\rb
  \partial_{z'}\psi_n(z')\,\hat{s}^{(2)}_5(p,z')\\-
  \frac{iN_c\veps^{z\mu\nu\rho\sigma}\kappa}{8(2\pi)^2\pi^2}\lim_{z\to+\infty}u(z)^3w(z)\int dz'\,d^4q\,d^4p\,
  \lb\delta^\lambda_\sigma-D^\lambda_\sigma(q-p)\rb\\\times
  p_\mu\,\text{STr}\lsb B_\vk(-q)\,V_{\rho,n}(p)\,J^{(2)}_{\lambda,5}(q-p)\rsb\\\times
  \pz\lb D^\vk_\nu(q)\,G_l(z,z')+\lb\delta^\vk_\nu-D^\vk_\nu(q)\rb G_t(z,z',q)\rb
  \psi_n(z')\,\partial_{z'}\hat{s}^{(2)}_5(q-p,z')\\-
  \sum_n\frac{iN_c\veps^{z\mu\nu\rho\sigma}\kappa L^4}{24(2\pi)^2\pi^2}\int d^4q\,d^4p\,
  dz\,\pz w(z)\,\psi_n(z)\,\hat{s}^{(2)}_5(q-p,z)\\\times
  \lb3\delta^\lambda_\sigma-2D^\lambda_\sigma(q-p)\rb
  p_\mu\,\text{STr}\lsb B_\nu(-q)\,V_{\rho,n}(p)\,J^{(2)}_{\lambda,5}(q-p)\rsb\\+
  \sum_n\frac{iN_c\veps^{z\mu\nu\rho\sigma}\kappa L^4}{24(2\pi)^2\pi^2}\int d^4q\,d^4p\,
  dz\,w(z)\,\pz \psi_n(z)\,\hat{s}^{(2)}_5(q-p,z)\\\times
  D^\lambda_\sigma(q-p)\,
  p_\mu\,\text{STr}\lsb B_\nu(-q)\,V_{\rho,n}(p)\,J^{(2)}_{\lambda,5}(q-p)\rsb\\+
  \sum_n\frac{iN_c\veps^{z\mu\nu\rho\sigma}\kappa L^4}{24(2\pi)^2\pi^2}\int d^4q\,d^4p\,
  dz\,w(z)\,\pz \psi_n(z)\,\hat{s}^{(2)}_5(q-p,z)\\\times
  \lb3\delta^\lambda_\sigma-2D^\lambda_\sigma(q-p)\rb
  p_\mu\,\text{STr}\lsb B_\nu(p)\,V_{\rho,n}(-q)\,J^{(2)}_{\lambda,5}(q-p)\rsb\\-
  \sum_n\frac{iN_c\veps^{z\mu\nu\rho\sigma}\kappa L^4}{24(2\pi)^2\pi^2}\int d^4q\,d^4p\,
  dz\,\pz w(z)\,\psi_n(z)\,\hat{s}^{(2)}_5(q-p,z)\\\times
  D^\lambda_\sigma(q-p)
  p_\mu\,\text{STr}\lsb B_\nu(p)\,V_{\rho,n}(-q)\,J^{(2)}_{\lambda,5}(q-p)\rsb\\+
  \sum_n\frac{iN_c\veps^{z\mu\nu\rho\sigma}\kappa}{8(2\pi)^2\pi^2}\int d^4q\,d^4p\,dz\,
  \frac{dz'}{u(z')}\,\psi_n(z')\,G_t(z',z,q)
  \lb\delta^\lambda_\sigma-D^\lambda_\sigma(q-p)\rb
  \lb q^2-\lambda_n\rb\\\times
  p_\mu\,\text{STr}\lsb B_\nu(p)\,V_{\rho,n}(-q)\,J^{(2)}_{\lambda,5}(q-p)\rsb
  w(z)\,\pz\hat{s}^{(2)}_5(q-p,z)\\-
  \sum_n\frac{iN_c\veps^{z\mu\nu\rho\sigma}\kappa}{8(2\pi)^2\pi^2}\int d^4q\,d^4p\,dz\,
  \frac{dz'}{u(z')}\,\psi_n(z')\,G_t(z',z,q)
  \lb q^2-\lambda_n\rb\\\times
  p_\mu\,\text{STr}\lsb B_\nu(q-p)\,V_{\rho,n}(-q)\,J^{(2)}_{\sigma,5}(p)\rsb
  \pz w(z)\,\hat{s}^{(2)}_5(p,z),
\end{multline}
\begin{multline}
  S_{\text{eff},18}\equiv\sum_{n,k,i,\dots}\frac{if_{abc}\kappa^2\mathcal{C}_kc^{(3)}_{29,n,i,\dots,k}}{2(2\pi)^2}\int d^4q\,d^4p
  \lb2\eta^{\mu\rho}\eta^{\nu\sigma}-\eta^{\mu\nu}\eta^{\rho\sigma}-
  \eta^{\mu\sigma}\eta^{\nu\rho}\rb\\\times
  \lb\frac{D^\lambda_\sigma(p)}{\lambda_k}-\frac{\delta^\lambda_\sigma-D^\lambda_\sigma(p)}{p^2-\lambda_k}-
  \frac{2\delta^\lambda_\sigma}{\lambda_k}\rb
  p_\rho B^a_\lambda(p)\,V^b_{\mu,n}(-q)\,D^\vk_\nu(q-p)\,J^{c,(2)}_{\vk,i,\dots}(q-p)\\+
  \sum_{n,k,i,\dots}\frac{if_{abc}\kappa^2\mathcal{C}_kc^{(3)}_{29,n,i,\dots,k}}{2(2\pi)^2}\int d^4q\,d^4p
  \lb2\eta^{\mu\nu}\eta^{\rho\sigma}-\eta^{\mu\sigma}\eta^{\nu\rho}\rb\\\times
  \lb\frac{2\delta^\lambda_\sigma}{\lambda_k}-\frac{D^\lambda_\sigma(q)}{\lambda_k}+
  \frac{\delta^\lambda_\sigma-D^\lambda_\sigma(q)}{q^2-\lambda_k}\rb
  p_\rho B^a_\lambda(-q)\,V^b_{\mu,n}(p)\,D^\vk_\nu(q-p)\,J^{c,(2)}_{\vk,i,\dots}(q-p)\\+
  \sum_{\substack{n,k,l,i,\dots\\i\neq5}}\frac{if_{abc}\kappa^3L^4\mathcal{C}_lc^{(3)}_{30,n,k,i,\dots,l}}{2(2\pi)^2}\int d^4q\,d^4p
  \lb2\eta^{\mu\rho}\eta^{\nu\sigma}-\eta^{\mu\nu}\eta^{\rho\sigma}-
  \eta^{\mu\sigma}\eta^{\nu\rho}\rb\\\times
  \frac{\delta^\vk_\nu-D^\vk_\nu(q-p)}{(q-p)^2-\lambda_k}
  \lb\frac{2\delta^\lambda_\sigma}{\lambda_l}-\frac{D^\lambda_\sigma(p)}{\lambda_l}+
  \frac{\delta^\lambda_\sigma-D^\lambda_\sigma(p)}{p^2-\lambda_l}\rb
  p_\rho B^a_\lambda(p)\,V^b_{\mu,n}(-q)\,J^{c,(2)}_{\vk,i,\dots}(q-p)\\+
  \sum_{\substack{n,k,l,i,\dots\\i\neq5}}\frac{if_{abc}\kappa^3L^4\mathcal{C}_lc^{(3)}_{30,n,k,i,\dots,l}}{2(2\pi)^2}\int d^4q\,d^4p
  \lb2\eta^{\mu\nu}\eta^{\rho\sigma}-\eta^{\mu\sigma}\eta^{\nu\rho}\rb\\\times
  \frac{\delta^\vk_\nu-D^\vk_\nu(q-p)}{(q-p)^2-\lambda_k}
  \lb\frac{D^\lambda_\sigma(q)}{\lambda_l}-\frac{\delta^\lambda_\sigma-D^\lambda_\sigma(q)}{q^2-\lambda_l}-
  \frac{2\delta^\lambda_\sigma}{\lambda_l}\rb
  p_\rho B^a_\lambda(-q)\,V^b_{\mu,n}(p)\,J^{c,(2)}_{\vk,i,\dots}(q-p)\\+
  \frac{if_{abc}\kappa^2}{2(2\pi)^2}\lim_{z\to+\infty}u(z)^3w(z)\int dz'\,d^4q\,d^4p\,
  \lb2\eta^{\mu\rho}\eta^{\nu\sigma}-\eta^{\mu\nu}\eta^{\rho\sigma}-
  \eta^{\mu\sigma}\eta^{\nu\rho}\rb\\\times
  \lb\delta^\vk_\nu-D^\vk_\nu(p)\rb
  p_\rho B^a_\lambda(-q)\,V^b_{\mu,n}(q-p)\,J^{c,(2)}_{\vk,5}(p)\\\times
  \pz\lb D^\lambda_\sigma(q)\,G_l(z,z')+\lb\delta^\lambda_\sigma-D^\lambda_\sigma(q)\rb G_t(z,z',q)\rb
  \frac{\psi_n(z')}{u(z')}\,\hat{s}^{(2)}_5(p,z')\\+
  \frac{if_{abc}\kappa^2}{2(2\pi)^2}\lim_{z\to+\infty}u(z)^3w(z)\int dz'\,d^4q\,d^4p\,
  \lb2\eta^{\mu\sigma}\eta^{\nu\rho}-\eta^{\mu\nu}\eta^{\rho\sigma}\rb
  \lb\delta^\vk_\nu-D^\vk_\nu(q-p)\rb\\\times
  p_\rho B^a_\lambda(-q)\,V^c_{\mu,n}(p)\,J^{b,(2)}_{\vk,5}(q-p)\\\times
  \pz\lb D^\lambda_\sigma(q)\,G_l(z,z')+\lb\delta^\lambda_\sigma-D^\lambda_\sigma(q)\rb G_t(z,z',q)\rb
  \frac{\psi_n(z')}{u(z')}\,\hat{s}^{(2)}_5(q-p,z')\\+
  \sum_n\frac{if_{abc}\kappa^2L^4}{2(2\pi)^2}\int d^4q\,d^4p\,
  dz\,\frac{w(z)\,\psi_n(z)}{u(z)}\,\hat{s}^{(2)}_5(q,z)
  \lb2\eta^{\mu\rho}\eta^{\nu\sigma}-\eta^{\mu\nu}\eta^{\rho\sigma}-
  \eta^{\mu\sigma}\eta^{\nu\rho}\rb\\\times
  \lb\delta^\vk_\nu-D^\vk_\nu(q)\rb
  p_\rho B^a_\sigma(p)\,V^b_{\mu,n}(q-p)\,J^{c,(2)}_{\vk,5}(-q)\\-
  \sum_n\frac{if_{abc}\kappa^2L^4}{2(2\pi)^2}\int d^4q\,d^4p\,
  dz\,\frac{w(z)\,\psi_n(z)}{u(z)}\,\hat{s}^{(2)}_5(q,z)
  \lb2\eta^{\mu\nu}\eta^{\rho\sigma}-\eta^{\mu\sigma}\eta^{\nu\rho}\rb\\\times
  \lb\delta^\vk_\nu-D^\vk_\nu(q)\rb
  p_\rho B^a_\sigma(q-p)\,V^b_{\mu,n}(p)\,J^{c,(2)}_{\vk,5}(-q)\\-
  \sum_n\frac{if_{abc}\kappa^2}{2(2\pi)^2}\int d^4q\,d^4p\,dz\,dz'\,
  \frac{\psi_n(z')}{u(z')}\,G_t(z',z,q)
  \lb2\eta^{\mu\nu}\eta^{\rho\sigma}-\eta^{\mu\rho}\eta^{\nu\sigma}-
  \eta^{\mu\sigma}\eta^{\nu\rho}\rb\\\times
  \lb\delta^\vk_\nu-D^\vk_\nu(p)\rb
  \lb q^2-\lambda_n\rb
  p_\rho B^a_\sigma(q-p)\,V^b_{\mu,n}(-q)\,J^{c,(2)}_{\vk,5}(p)\,
  \frac{w(z)}{u(z)}\,\hat{s}^{(2)}_5(p,z)\\+
  \sum_n\frac{if_{abc}\kappa^2}{2(2\pi)^2}\int d^4q\,d^4p\,dz\,dz'\,
  \frac{\psi_n(z')}{u(z')}\,G_t(z',z,q)
  \lb2\eta^{\mu\sigma}\eta^{\nu\rho}-\eta^{\mu\rho}\eta^{\nu\sigma}-
  \eta^{\mu\nu}\eta^{\rho\sigma}\rb\\\times
  \lb\delta^\vk_\nu-D^\vk_\nu(q-p)\rb
  \lb q^2-\lambda_n\rb
  p_\rho B^a_\sigma(p)\,V^b_{\mu,n}(-q)\,J^{c,(2)}_{\vk,5}(q-p)\,
  \frac{w(z)}{u(z)}\,\hat{s}^{(2)}_5(q-p,z),
\end{multline}
\begin{multline}
  S_{\text{eff},19}\equiv-\sum_{n,m,i,\dots}\frac{7iN_c\veps^{z\mu\nu\rho\sigma}c^{(3)}_{21,n,i,\dots,m}}{32(2\pi)^2\pi^2}\int d^4q\,d^4p\,
  D^\lambda_\sigma(q)\\\times
  p_\mu\,\text{STr}\lsb V_{\nu,n}(q-p)\,V_{\rho,m}(p)\,J^{(2)}_{\lambda,i,\dots}(-q)\rsb\\-
  \sum_{\substack{n,m,k,i,\dots\\i\neq5}}\frac{7iN_c\veps^{z\mu\nu\rho\sigma}\kappa L^4c^{(3)}_{24,n,k,i,\dots,m}}{16(2\pi)^2\pi^2}\int d^4q\,d^4p\,
  \frac{\delta^\lambda_\sigma-D^\lambda_\sigma(q)}{q^2-\lambda_k}\\\times
  p_\mu\,\text{STr}\lsb V_{\nu,n}(q-p)\,V_{\rho,m}(p)\,J^{(2)}_{\lambda,i,\dots}(-q)\rsb\\-
  \sum_{n,m}\frac{iN_c\veps^{z\mu\nu\rho\sigma}\kappa L^4}{8(2\pi)^2\pi^2}\int d^4q\,d^4p\,
  dz\,\pz \psi_n(z)\,\psi_m(z)\,\hat{s}^{(2)}_5(q-p,z)\\\times
  \lb\delta^\lambda_\sigma-D^\lambda_\sigma(q-p)\rb
  p_\mu\,\text{STr}\lsb V_{\nu,n}(-q)\,V_{\rho,m}(p)\,J^{(2)}_{\lambda,5}(q-p)\rsb\\-
  \sum_{n,m}\frac{iN_c\veps^{z\mu\nu\rho\sigma}\kappa}{8(2\pi)^2\pi^2}\int d^4q\,d^4p\,dz\,dz'\,
  \frac{\psi_n(z')}{u(z')}\,G_t(z',z,q)
  \lb\delta^\lambda_\sigma-D^\lambda_\sigma(q-p)\rb
  \lb q^2-\lambda_n\rb\\\times
  p_\mu\,\text{STr}\lsb V_{\nu,n}(-q)\,V_{\rho,m}(p)\,J^{(2)}_{\lambda,5}(q-p)\rsb
  \psi_m(z)\,\pz\hat{s}^{(2)}_5(q-p,z)\\+
  \sum_{n,m}\frac{iN_c\veps^{z\mu\nu\rho\sigma}\kappa}{8(2\pi)^2\pi^2}\int d^4q\,d^4p\,dz\,dz'\,
  \frac{\psi_n(z')}{u(z')}\,G_t(z',z,q)
  \lb q^2-\lambda_n\rb\\\times
  p_\mu\,\text{STr}\lsb V_{\nu,n}(-q)\,V_{\rho,m}(q-p)\,J^{(2)}_{\sigma,5}(p)\rsb
  \pz \psi_m(z)\,\hat{s}^{(2)}_5(p,z),
\end{multline}
\begin{multline}
  S_{\text{eff},20}\equiv-\sum_{n,m,i,\dots}\frac{3if_{abc}\kappa c^{(3)}_{29,n,i,\dots,m}}{2(2\pi)^2}\int d^4q\,d^4p
  \lb2\eta^{\mu\rho}\eta^{\nu\sigma}-\eta^{\mu\sigma}\eta^{\nu\rho}\rb\\\times
  p_\rho V^a_{\sigma,n}(p)\,V^b_{\mu,m}(q-p)\,D^\vk_\nu(q)\,J^{c,(2)}_{\vk,i,\dots}(-q)\\-
  \sum_{\substack{n,m,k,i,\dots\\i\neq5}}\frac{2if_{abc}\kappa^2L^4c^{(3)}_{30,n,k,i,\dots,m}}{(2\pi)^2}\int d^4q\,d^4p
  \lb2\eta^{\mu\rho}\eta^{\nu\sigma}-\eta^{\mu\sigma}\eta^{\nu\rho}\rb
  \frac{\delta^\vk_\nu-D^\vk_\nu(q)}{q^2-\lambda_k}\\\times
  p_\rho V^a_{\sigma,n}(p)\,V^b_{\mu,m}(q-p)\,J^{c,(2)}_{\vk,i,\dots}(-q)\\+
  \sum_{n,m}\frac{if_{abc}\kappa^2L^4}{2(2\pi)^2}\int d^4q\,d^4p\,
  dz\,\frac{\psi_n(z)\,\psi_m(z)}{u(z)}\,\hat{s}^{(2)}_5(q,z)
  \lb2\eta^{\mu\rho}\eta^{\nu\sigma}-\eta^{\mu\sigma}\eta^{\nu\rho}\rb\\\times
  \lb\delta^\vk_\nu-D^\vk_\nu(q)\rb
  p_\rho V^a_{\sigma,n}(p)\,V^b_{\mu,m}(q-p)\,J^{c,(2)}_{\vk,5}(-q)\\+
  \sum_{n,m}\frac{if_{abc}\kappa^2}{2(2\pi)^2}\int d^4q\,d^4p\,dz\,dz'\,
  \frac{\psi_n(z')}{u(z')}\,G_t(z',z,q)
  \lb2\eta^{\mu\rho}\eta^{\nu\sigma}-\eta^{\mu\nu}\eta^{\rho\sigma}-
  \eta^{\mu\sigma}\eta^{\nu\rho}\rb\\\times
  \lb\delta^\vk_\nu-D^\vk_\nu(p)\rb
  \lb q^2-\lambda_n\rb
  p_\rho V^a_{\sigma,n}(-q)\,V^b_{\mu,m}(q-p)\,J^{c,(2)}_{\vk,5}(p)\,
  \frac{\psi_m(z)}{u(z)}\,\hat{s}^{(2)}_5(p,z)\\-
  \sum_{n,m}\frac{if_{abc}\kappa^2}{2(2\pi)^2}\int d^4q\,d^4p\,dz\,dz'\,
  \frac{\psi_n(z')}{u(z')}\,G_t(z',z,q)
  \lb2\eta^{\mu\sigma}\eta^{\nu\rho}-\eta^{\mu\nu}\eta^{\rho\sigma}-
  \eta^{\mu\rho}\eta^{\nu\sigma}\rb\\\times
  \lb\delta^\vk_\nu-D^\vk_\nu(q-p)\rb
  \lb q^2-\lambda_n\rb
  p_\rho V^a_{\sigma,n}(-q)\,V^b_{\mu,m}(p)\,J^{c,(2)}_{\vk,5}(q-p)\,
  \frac{\psi_m(z)}{u(z)}\,\hat{s}^{(2)}_5(q-p,z),
\end{multline}
\begin{multline}
  S_{\text{eff},21}\equiv\sum_{\substack{n,i,\dots\\i\neq31}}\frac{if_{abc}\kappa^2L^4c^{(3)}_{i,\dots,n}}{2(2\pi)^2}\frac{\mathcal{C}_n}{\lambda_n}\int d^4q\,d^4p\,
  \lb\eta^{\mu\rho}\eta^{\nu\sigma}-\eta^{\mu\nu}\eta^{\rho\sigma}\rb
  \frac{\delta^\vk_\nu-D^\vk_\nu(q)}{q^2-\lambda_n}\\\times
  p_\rho B^a_\sigma(p)\,B^b_\mu(q-p)\,J^{c,(3)}_{\vk,i,\dots}(-q)\\+
  \frac{if_{abc}\kappa^2L^4}{2(2\pi)^2}\int d^4q\,d^4p\,
  dz\,\frac{w(z)}{u(z)}\,\hat{s}^{(3)}_{31}(q,z)\,
  \lb\eta^{\mu\rho}\eta^{\nu\sigma}-\eta^{\mu\nu}\eta^{\rho\sigma}\rb
  p_\rho B^a_\sigma(p)\,B^b_\mu(q-p)\,J^{c,(3)}_{\nu,31}(-q)\\+
  \sum_{\substack{i,\dots\\i\in\lsb35,42\rsb}}\kappa\int d^4q\,\frac{dz}{u(z)}\,w(z)\,
  \eta^{\mu\rho}\eta^{\nu\sigma}q_\mu\,
  \text{STr}\lsb\lsb B_\rho,B_\sigma\rsb(q)\,K_{\nu,i,\dots}(-q,z)\rsb.
\end{multline}

\section{CS \texorpdfstring{\(\bm{5}\)}{5}-form and CS Action}\label{app:cs5form}

In this Appendix some technical details related to the CS \(5\)-form and CS action~\eqref{eq:cs_term}
are discussed.

First, let us note that there exists another formulation of the CS action~\cite{Sakai:2004cn},
\begin{equation}\label{eq:cs_term2}
  S_\text{CS}=\frac{1}{48\pi^3}\int\limits_\text{D8}C_3\wedge\text{STr}F^3,
\end{equation}
which can be related to~\eqref{eq:cs_term} by the graded Leibniz rule for the exterior
derivative of the exterior product,
\begin{equation}
  d(P\wedge Q)=dP\wedge Q+(-1)^pP\wedge dQ,
\end{equation}
where \(p\) is the degree of the form \(P\). These two formulations of the CS action,~\eqref{eq:cs_term}
and~\eqref{eq:cs_term2}, are not always equal. In this work the former,~\eqref{eq:cs_term}
is used, following~\cite{Sakai:2004cn,Green:1996dd}. \eqref{eq:cs_term2} vanishes under
assumption that the gauge field does not depend on the coordinates on the \(S_4\) sphere
and does not have non-zero components along these coordinates.\footnote{To see this, note
that \(F^3\) in the coordinate form must have at least six indices. Since only five
coordinates are ``transverse'' to the sphere, one of those six indices must be along the
sphere, which eliminates the term.}

The precise definition of the CS \(5\)-form depends on the conventions used to define the
non-abelian algebra, specifically, the definition of its structure constants. With the
definition~\eqref{eq:str_const} used in this work, the appropriate CS 5-form is
\begin{equation}\label{eq:omega5_2}
  \omega_5(A)=\text{STr}\lb AF^2-\frac{i}{2}A^3F-\frac{1}{10}A^5\rb,
\end{equation}
and it obeys
\begin{equation}
  d\omega_5(A)=\text{STr}\,F^3.
\end{equation}
The derivation of the CS \(5\)-form~\eqref{eq:omega5_2} is often omitted in the
literature. An interested reader can find the details and the references to the original
works in~\cite{Bertlmann:1996xk} (specifically, sections 7.1 and 7.2 therein).

The expression for the CS \(5\)-form can be further simplified. \eqref{eq:omega5_2} can
be rewritten as
\begin{equation}
  \omega_5(A)=\text{STr}\lb A(dA)^2+\frac{3i}{2}A^3dA-\frac{3}{5}A^5\rb.
\end{equation}
Then the \(5\)-form can be split into sum over terms with different number of \(\mathfrak{su}(N_f)\)
generators,
\begin{multline}
  \omega_5=N_fW_0+W^{ab}_2\,\text{STr}\lb T^aT^b\rb+
  W^{abc}_3\,\text{STr}\lb T^aT^bT^c\rb\\+
  W^{abcd}_4\,\text{STr}\lb T^aT^bT^cT^d\rb+
  W^{abcde}_5\,\text{STr}\lb T^aT^bT^cT^dT^e\rb,
\end{multline}
where
\begin{equation}\label{eq:Wexp0}
  \begin{split}
    W_0&\equiv a(da)^2+\frac{3i}{2}a^3da-\frac{3}{5}a^5,\\
    W^{ab}_2&\equiv adA^adA^b+2A^adadA^b+\frac{3i}{2}\lb a^2A^adA^b+aA^aA^bda\rb-
    \frac{6}{5} a^3A^aA^b,\\
    W^{abc}_3&\equiv A^adA^bdA^c+\frac{3i}{2}\lb aA^aA^bdA^c+A^aA^bA^cda\rb-
    \frac{6}{5}a^2A^aA^bA^c,\\
    W^{abcd}_4&\equiv\frac{3i}{2}A^aA^bA^cdA^d-\frac{3}{5}aA^aA^bA^cA^d,\qquad
    W^{abcde}_5\equiv-\frac{3}{5}A^aA^bA^cA^dA^e.
  \end{split}
\end{equation}
Some of the terms in these expressions are symmetric over the generator indices (or
rather contracted with fully symmetric tensors). In addition, there are terms which are
at least quadratic in either abelian or non-abelian fields. By transforming to the
coordinate form it can be demonstrated that all such terms vanish. For instance, consider
the following term,
\begin{equation}
  a^3da=a_Ma_Na_P\partial_{[Q}a_{R]}dx^M\wedge dx^N\wedge dx^P\wedge dx^Q\wedge dx^R.
\end{equation}
The symmetric product of the fields \(a_Ma_Na_P\) contracted with the antisymmetric
exterior product \(dx^M\wedge dx^N\wedge dx^P\) makes the entire expression vanish. As a
result, the expressions~\eqref{eq:Wexp0} are simplified,
\begin{equation}
  \begin{gathered}
    W_0=a(da)^2,\qquad
    W^{ab}_2=adA^adA^b+2A^ada\,dA^b,\qquad
    W^{abc}_3=A^adA^bdA^c,\\
    W^{abcd}_4=W^{abcde}_5=0.
  \end{gathered}
\end{equation}
We can further use the results for symmetrized traces of products of two~\eqref{eq:1_2gen_tr}
and three~\eqref{eq:3gen_symtr} generators to write the decomposition of the \(5\)-form
as
\begin{equation}
  \omega_5=N_fW_0+\frac{1}{2}W^{aa}_2+\frac{1}{4}W^{abc}_3d_{abc}.
\end{equation}
To make comparison with~\cite{Hoyos:2022ptd}, note that only the \(W^{aa}_2\) term is
included in~\cite{Hoyos:2022ptd}.\footnote{More precisely, in~\cite{Hoyos:2022ptd} the
second term in \(W^{aa}_2\) was integrated by parts under the integral of the CS action~\eqref{eq:cs_term}.}
The purely abelian contribution \(W_0\) is omitted and the purely non-abelian term
\(W^{abc}_3\) is absent, since the \(d\)-tensor is zero for \(N_f=2\) considered in~\cite{Hoyos:2022ptd}.

\section{Weak Field Expansion of the DBI Action}\label{app:dbi_exp}

In this Appendix we expand the DBI action~\eqref{eq:dbi} in even powers of the field
strength up to the fourth order in the fields. The experession under the determinant:
\begin{multline}
  G_{MN}+2\pi\alpha'F_{MN}=
  G_{ML}\delta^L_N+2\pi\alpha'\delta^K_MF_{KN}\\=
  G_{ML}\delta^L_N+2\pi\alpha'G_{ML}G^{LK}F_{KN}=
  G_{ML}\lb\delta^L_N+{M^L}_N\rb,
\end{multline}
where
\begin{equation}
  {M^L}_N\equiv 2\pi\alpha'G^{LK}F_{KN}.
\end{equation}
Since \(F_{MN}\) is antisymmetric,
\begin{equation}
  \det\lb\delta^L_N+{M^L}_N\rb=
  \det\lb\delta^L_N+{M^L}_N\rb^T=
  \det\lb\delta^L_N-{M^L}_N\rb.
\end{equation}
Consequently,
\begin{equation}
  \det\lb\delta^L_N-\lb{M^L}_N\rb^2\rb=
  \det\lb\delta^L_N-{M^L}_N\rb\det\lb\delta^L_N+{M^L}_N\rb=
  \lb\det\lb\delta^L_N+{M^L}_N\rb\rb^2,
\end{equation}
implying
\begin{equation}
  \det\lb\delta^L_N+{M^L}_N\rb=\sqrt{\det\lb\delta^L_N-\lb{M^L}_N\rb^2\rb}.
\end{equation}
The square root in the DBI action~\eqref{eq:dbi} can be expressed as following:
\begin{equation}\label{eq:dbi_det}
  \sqrt{-\det\lb G_{MN}+2\pi\alpha'F_{MN}\rb}=
  \sqrt{-G}\lsb\det\lb\delta^L_N-\lb{M^L}_N\rb^2\rb\rsb^{1/4},\qquad
  G\equiv\det G_{MN}.
\end{equation}
The square root of the metric determinant,
\begin{equation}
  \sqrt{-G}=\lb\frac{9}{4}\rb^2L^9u(z)^{11/4}
\end{equation}
contributes to the overall factor of the expanded DBI action. The right-hand side
of~\eqref{eq:dbi_det} can be manipulated further
\begin{multline}
  \lb\det\lb\delta^L_N-\lb{M^L}_N\rb^2\rb\rb^{1/4}=
  \exp\lsb\frac{1}{4}\ln\det\lb\delta^L_N-\lb{M^L}_N\rb^2\rb\rsb\\=
  \exp\lsb\frac{1}{4}\text{tr}\ln\lb\delta^L_N-\lb{M^L}_N\rb^2\rb\rsb,
\end{multline}
where \(\text{tr}\) is a trace with respect to the Lorentz indices. In the low energy
limit \(\alpha'=l_s^2\to0\) (equivalently, \({M^L}_N\to0\)):
\begin{multline}\label{eq:dbi_exp_interim}
  \exp\lsb\frac{1}{4}\text{tr}\ln\lb\delta^L_N-\lb{M^L}_N\rb^2\rb\rsb=
  \exp\lsb-\frac{1}{4}\text{tr}\lb\lb{M^L}_N\rb^2+\frac{1}{2}\lb{M^L}_N\rb^4+\mathcal{O}\lb\alpha'^6\rb\rb\rsb\\=
  1-\frac{1}{4}\text{tr}\lb\lb{M^L}_N\rb^2+\frac{1}{2}\lb{M^L}_N\rb^4\rb+
  \frac{1}{32}\lsb\text{tr}\lb\lb{M^L}_N\rb^2+\frac{1}{2}\lb{M^L}_N\rb^4\rb\rsb^2+\mathcal{O}\lb\alpha'^6\rb.
\end{multline}
This expansion contains all the terms up to the fourth order in the field strength. The
third term in~\eqref{eq:dbi_exp_interim} gives one contribution of order \(\alpha'^4\):
\begin{equation}
  \lsb\text{tr}\lb\lb{M^L}_N\rb^2+\frac{1}{2}\lb{M^L}_N\rb^4\rb\rsb^2=
  \lb\text{tr}\lb{M^L}_N\rb^2\rb^2+\mathcal{O}\lb\alpha'^6\rb.
\end{equation}
After taking the symmetrized trace of~\eqref{eq:dbi_exp_interim}, the expansion of the
DBI action is
\begin{equation}
  \text{STr}\exp\lsb\frac{1}{4}\text{tr}\ln\lb\delta^L_N-\lb{M^L}_N\rb^2\rb\rsb=
  N_f+S^{(2)}+S^{(4)}+\mathcal{O}\lb\alpha'^6\rb,
\end{equation}
where \(N_f=\text{STr}\,\mathbb{1}_{N_f}\). The quadratic and quartic terms are
\begin{equation}
  S^{(2)}\equiv-\frac{1}{4}\,\text{STr}\,\text{tr}\lb{M^L}_N\rb^2,\qquad
  S^{(4)}\equiv-\frac{1}{8}\,\text{STr}\,\text{tr}\lb{M^L}_N\rb^4+
  \frac{1}{32}\,\text{STr}\lb\text{tr}\lb{M^L}_N\rb^2\rb^2.
\end{equation}
or, after substituting \(M^L{}_N\),
\begin{equation}
  S^{(2)}=N_f(\pi\alpha')^2f^{MN}f_{MN}+\frac{(\pi\alpha')^2}{2}F^{MN,a}F_{MN}^a,
\end{equation}
\begin{multline}
  S^{(4)}=(\pi\alpha')^4\lsb
  N_f\lb\frac{1}{2}f^{KL}f_{KL}f^{MN}f_{MN}-2f_{LM}f_{NK}f^{KL}f^{MN}\rb\right.\\\left.+
  \frac{1}{2}f_{KL}f^{KL}F^{MN,a}F^a_{MN}+f_{KL}f^{MN}F^{KL,a}F^a_{MN}\right.\\\left.-
  4f_{LM}f^{MN}F^{KL,a}F^a_{NK}-2f_{LM}f_{NK}F^{KL,a}F^{MN,a}\right.\\\left.+
  \lb\frac{1}{2}f_{KL}F^{KL,a}F^{MN,b}F^c_{MN}-2f^{KL}F^a_{LM}F^{MN,b}F^c_{NK}\rb d_{abc}\right.\\\left.+
  \lb\frac{1}{2}F^{KL,a}F^b_{KL}F^{MN,c}F^d_{MN}-2F^{KL,a}F^b_{LM}F^{MN,c}F^d_{NK}\rb\text{STr}\lb T^aT^bT^cT^d\rb\rsb.
\end{multline}

\section{Gauge Fixing and Boundary Conditions}\label{app:bcond}

In this Appendix we discuss the gauge fixing and related boundary conditions for the
bulk gauge field. Abelian and non-abelian fields will be discussed separately.

\noindent \(\bullet\) \textbf{The abelian case.}

The boundary conditions follow from requirement for the 4D effective action to be finite.
To obtain the effective action, the off-shell action in the bulk is integrated over \(z\).
At the lowest order in the expansion~\eqref{eq:dbi_exp} of the DBI action, if the
following integral converges,
\begin{equation}
  \int\limits_{-\infty}^{+\infty}dz\,u(z)^2\lb f_{MN}\rb^2<\infty,
\end{equation}
the effective action is finite. This requirement is satisfied if
\begin{equation}
  f_{MN}\underset{z\to\pm\infty}{\to}0.
\end{equation}
Consequently, the bulk gauge field at the boundary must be a pure gauge:
\begin{equation}
  a_M\underset{z\to\pm\infty}{\to}-\partial_M\alpha_\pm,
\end{equation}
where \(e^{i\alpha_\pm(x^\mu)}\in U(1)\). This boundary condition can be made homogeneous
by performing gauge transformation with \(e^{i\beta\lb x^\mu,z\rb}\in U(1)\),
\begin{equation}\label{eq:ab_bcond2}
  a_M^\beta=a_M-\partial_M\beta,\qquad
  \beta_\pm\lb x^\mu\rb\equiv\lim_{z\to\pm\infty}\beta=-\alpha_\pm+\text{Const}\qrq
  a_M^\beta\underset{z\to\pm\infty}{\to}0.
\end{equation}
Next, the axial gauge \(a_z=0\) is imposed by performing gauge transformation with
\(e^{i\gamma\lb x^\mu,z\rb}\in U(1)\),
\begin{equation}
  a^\gamma_z=0=a^\beta_z-\pz\gamma,\qquad
  \gamma=\int\limits_0^zdz'\,a^\beta_z(x^\mu,z').
\end{equation}
After this transformation, the vanishing boundary condition~\eqref{eq:ab_bcond2} is
modified:
\begin{equation}\label{eq:gamma_pm}
  a^\gamma_\mu\underset{z\to\pm\infty}{\to}-\partial_\mu\gamma_\pm,\qquad
  \gamma_\pm\lb x^\mu\rb\equiv\int\limits_0^{\pm\infty}dz\,a^\beta_z(x^\mu,z).
\end{equation}
Exponents of \(\gamma_\pm\) are identified as elements of the left-right symmetry group
of the boundary theory, \(\lb e^{i\gamma_+},e^{i\gamma_-}\rb\in U(1)_L\times U(1)_R\).
The boundary conditions can be further simplified by performing a residual,
\(z\)-independent gauge transformation, which keeps the axial gauge intact,
\begin{equation}
  a^{\gamma'}_\mu=a_\mu-\partial_\mu\lb\gamma+\gamma'\rb,\qquad
  e^{i\gamma'\lb x^\mu\rb}\in U(1),
\end{equation}
such that
\begin{equation}
  a^{\gamma'}_\mu\underset{z\to-\infty}{\to}0.
\end{equation}
It is easy to see that this condition can be satisfied if
\begin{equation}
  \gamma'=-\gamma_-+\text{Const},
\end{equation}
where the undefined constant can be set to zero. Then the boundary condition at
\(z\to+\infty\) becomes
\begin{equation}
  a^{\gamma'}_\mu\underset{z\to+\infty}{\to}-\partial_\mu\lb\gamma_+-\gamma_-\rb\equiv-\partial_\mu\eta'.
\end{equation}
In the interpretation of \(e^{i\gamma_\pm}\) as elements of the boundary \(U(1)_{L,R}\)
groups, \(e^{i\lb\gamma_+-\gamma_-\rb}\) represents an element of the \(U(1)_A\) axial
group. As such, \(\gamma_+-\gamma_-\) is identified as the \(\eta'\)-meson (up to a
constant factor),
\begin{equation}\label{eq:eta_def}
  \eta'=\gamma_+-\gamma_-=\int\limits_{-\infty}^{+\infty}dz\,a^\beta_z(x^\mu,z),
\end{equation}
where~\eqref{eq:gamma_pm} was used. To summarize, the boundary conditions for the abelian
gauge fixed field take the form
\begin{equation}
  a_z=0,\qquad
  a_\mu\underset{z\to+\infty}{\to}-\sqrt{\frac{2}{N_f}}\frac{1}{f_\pi}\partial_\mu\eta',\qquad
  a_\mu\underset{z\to-\infty}{\to}0,
\end{equation}
where \(f_\pi\) is later identified as the pion decay constant (see related discussions
in the Introduction). Here, the \(\eta'\)-meson field is rescaled, which together with
the normalization choice for the abelian generator made in~\eqref{eq:bf_decomp} ensures
the canonical normalization of its kinetic term in the effective action.

\noindent \(\bullet\) \textbf{The non-abelian case.}

Discussion of the boundary conditions for the non-abelian part of the bulk gauge field is
quite similar to the abelian case. The following notation will be used below:
\begin{equation}
  \mathcal{A}_M\equiv A^a_MT^a,\qquad
  \mathcal{F}_{MN}\equiv F^a_{MN}T^a=\partial_M\mathcal{A}_N-\partial_N\mathcal{A}_M+
  i\lsb\mathcal{A}_M,\mathcal{A}_N\rsb.
\end{equation}
The starting argument is identical to that of the abelian case: for finiteness of the
effective action it is sufficient to require that the non-abelian bulk gauge field at the
boundary is a pure gauge,
\begin{equation}\label{eq:bcond1}
  \int\limits_{-\infty}^{+\infty}dz\,u(z)^2\lb F^a_{MN}\rb^2<\infty\qrq
  F^a_{MN}\underset{z\to\pm\infty}{\to}0\qrq
  \mathcal{A}_M\underset{z\to\pm\infty}{\to}-if_\pm\partial_M f^{-1}_\pm,
\end{equation}
where \(f_\pm(x^\mu)\in SU(N_f)\). Next, a gauge transformation is performed with
\(g\lb x^\mu,z\rb\in SU(N_f)\),
\begin{equation}\label{eq:gtf}
  \mathcal{A}^g_M=g\mathcal{A}_Mg^{-1}-ig\partial_Mg^{-1},
\end{equation}
such that the boundary condition~\eqref{eq:bcond1} becomes homogeneous:
\begin{equation}\label{eq:bcond2}
  \mathcal{A}^g_M\underset{z\to\pm\infty}{\to}0.
\end{equation}
Substituting~\eqref{eq:gtf} into~\eqref{eq:bcond2} gives equations for \(g_\pm\equiv\lim_{z\to\pm\infty}g\),
\begin{equation}
  g_\pm f_\pm\partial_Mf^{-1}_\pm g^{-1}_\pm+g_\pm\partial_Mg^{-1}_\pm=0,
\end{equation}
with solution
\begin{equation}
  g_\pm\lb x^\mu\rb=P\exp\int dx^M\,f_\pm\partial_Mf^{-1}_\pm.
\end{equation}
Here, \(P\exp\) is a path-ordered exponent. This condition does not fully fix the gauge.
Any field configuration \(\mathcal{A}^{g'}_M\), gauge transformed with \(g'\lb x^\mu,z\rb\in SU(N_f)\)
that satisfies the requirement
\begin{equation}\label{eq:gprime}
  \partial_M g'\underset{z\to\pm\infty}{\to}0,
\end{equation}
has the same homogeneous boundary condition~\eqref{eq:bcond2}:
\begin{equation}
  \mathcal{A}^{g'}_M=g'\mathcal{A}^g_Mg'^{-1}-ig'\partial_Mg'^{-1}\underset{z\to\pm\infty}{\to}
  -ig'\partial_Mg'^{-1}=i\partial_Mg'g'^{-1}\to 0.
\end{equation}
Boundary values of these gauge transformations are denoted as
\begin{equation}
  g_L\lb x^\mu\rb\equiv\lim_{z\to+\infty}g'\lb x^\mu,z\rb,\qquad
  g_R\lb x^\mu\rb\equiv\lim_{z\to-\infty}g'\lb x^\mu,z\rb,
\end{equation}
and identified in~\cite{Sakai:2004cn,Sugimoto:2004mh} as elements of the left-right
non-abelian gauge group of the boundary theory, \(\lb g_L,g_R\rb\in SU(N_f)_L\times SU(N_f)_R\).

The axial gauge \(\mathcal{A}_z=0\) is imposed by performing gauge transformation with
\(h\lb x^\mu,z\rb\in SU(N_f)\),
\begin{equation}
  \mathcal{A}^h_z=0=h\mathcal{A}^g_z(x^\mu,z)h^{-1}-ih\pz h^{-1},\qquad
  h=P\exp\lb i\int\limits_0^zdz'\mathcal{A}^g_z(x^\mu,z')\rb.
\end{equation}
After this transformation, the boundary conditions for the bulk field change:
\begin{equation}\label{eq:bcond3}
  \mathcal{A}^h_\mu\underset{z\to\pm\infty}{\to}h_\pm \mathcal{A}^g_\mu(x^\mu,\pm\infty)h^{-1}_\pm-ih_\pm\partial_\mu h^{-1}_\pm=
  -ih_\pm\partial_\mu h^{-1}_\pm,
\end{equation}
where~\eqref{eq:bcond2} was used and
\begin{equation}
  h_\pm\lb x^\mu\rb\equiv P\exp\lb i\int\limits_0^{\pm\infty}dz\,\mathcal{A}^g_z(x^\mu,z)\rb,\qquad
  h^{-1}_\pm\lb x^\mu\rb=P\exp\lb-i\int\limits_0^{\pm\infty}dz\,\mathcal{A}^g_z(x^\mu,z)\rb.
\end{equation}
Under the gauge transformations \(g'\), see~\eqref{eq:gprime}, \(h\) changes as
\begin{equation}
  h\to P\exp\lb i\int\limits_0^zdz'\,g'\mathcal{A}^g_zg'^{-1}+
  \int\limits_0^zdz'\,g'\pz g'^{-1}\rb.
\end{equation}
The second integral is easily computed, so the transformation law for \(h\) is given by
\begin{equation}
  h(x^\mu,z)\to g'(x^\mu,0)\,h(x^\mu,z)\,g'(x^\mu,z)^{-1},
\end{equation}
where the order of factors on the right-hand side is determined by the path-ordered
exponent. Consequently, \(h_\pm\) and \(h^{-1}_\pm\) transform as
\begin{equation}
  h_\pm\to h'h_\pm g_{L,R}^{-1},\qquad
  h^{-1}_\pm\to g_{L,R}h^{-1}_\pm h'^{-1},
\end{equation}
where the residual, \(z\)-independent gauge transformation is denoted
\begin{equation}
  h'(x^\mu)\equiv g'(x^\mu,0).
\end{equation}
As discussed in the Introduction, the \(\pi\)-meson matrix \(\Sigma\) transforms under
\(SU(N_f)_L\times SU(N_f)_R\) as
\begin{equation}\label{eq:sigma_trans}
  \Sigma\to g_L\Sigma g^{-1}_R,
\end{equation}
which makes it possible to construct \(\Sigma\) out of \(h_\pm\),
\begin{equation}\label{eq:Sigma_split}
  \Sigma=h^{-1}_+h_-=P\exp\lb-i\int\limits_{-\infty}^{+\infty}dz\,\mathcal{A}^g_z(x^\mu,z)\rb,
\end{equation}
evidently satisfying the transformation property~\eqref{eq:sigma_trans}. Construction of
\(\Sigma\) in~\eqref{eq:Sigma_split} resembles the HLS approach mentioned in Introduction.
Within this approach \(\Sigma\) is split as in~\eqref{eq:Sigma_split}. This division is
not unique and the freedom left (\(h'\) in our notation) is identified as the gauge
transformation of the HLS. The vector mesons appearing in this work cannot be identified
as the vector gauge fields of the HLS.

\(\Sigma\) of~\eqref{eq:Sigma_split} does not appear directly in the boundary conditions.
Instead it is ``split'' between the \(z\to+\infty\) and \(z\to-\infty\) boundaries~\eqref{eq:bcond3}.
The residual \(z\)-independent gauge symmetry \(h'\), that was mentioned previously, can
be used to collect \(\Sigma\) at one of the boundaries. This transformation preserves the
axial gauge. The boundary conditions~\eqref{eq:bcond3} change into (note that \(\lim_{z\to\pm\infty}h'=h'\)
due to its \(z\)-independence):
\begin{equation}
  \mathcal{A}^{h'}_\mu\underset{z\to\pm\infty}{\to}-ih'_\pm\partial_\mu h'^{-1}_\pm,\qquad
  h'_\pm\equiv h'h_\pm,\qquad
  h'^{-1}_\pm\equiv h^{-1}_\pm h'^{-1}.
\end{equation}
Choosing \(h'=h_-^{-1}\) as the gauge transformation,
\begin{equation}
  h'_-=1,\qquad
  h'^{-1}_+=h^{-1}_+h_-=\Sigma.
\end{equation}
Then the boundary conditions for the bulk gauge field are
\begin{equation}
  \mathcal{A}^{h'}_\mu\underset{z\to+\infty}{\to}-i\Sigma^{-1}\partial_\mu\Sigma,\quad
  \mathcal{A}^{h'}_\mu\underset{z\to-\infty}{\to}0.
\end{equation}
With the help of~\eqref{eq:gen_comp}, the gauge fixing combined with the boundary
conditions for \(A^a_M\) can be written as
\begin{equation}
  A^a_z=0,\quad
  A^a_\mu\underset{z\to+\infty}{\to}-2i\,\text{Tr}\lb\Sigma^{-1}\partial_\mu\Sigma T^a\rb,\quad
  A^a_\mu\underset{z\to-\infty}{\to}0.
\end{equation}

\section{The Sources}\label{app:rhs}

In this Appendix we provide the explicit expressions for the source terms (right-hand
side expressions) of the EOMs~\eqref{eq:pert_eq} for each order in the expansion~\eqref{eq:field_pertr}
considered in this work. The source terms in the abelian case are the remaining
\(\mathcal{T}\)-tensor terms~\eqref{eq:T1}--\eqref{eq:T4} of the \(e\)-tensor~\eqref{eq:E_ab},
and the CS induced terms~\eqref{eq:delta_omega_ab}.

In the non-abelian case there are four types of source terms:
\begin{itemize}
  \item the terms of the non-abelian equation which contain the structure constant (see
  the second term on the left-hand side of the second line of~\eqref{eq:eoms}),
  \item the remaining \(\mathcal{T}\)-tensor terms~\eqref{eq:T1}--\eqref{eq:T4} from the
  \(E\)-tensor~\eqref{eq:E_n_ab},
  \item the non-linear terms of the non-abelian field strength tensors~\eqref{eq:n_ab_fs_pert},
  \item the terms derived from the CS action~\eqref{eq:delta_omega_n_ab}.
\end{itemize}
The expressions for the sources require introduction of several additional notations. The
\(\veps\) expansion:
\begin{equation}\label{eq:more_pertr_exp}
  \begin{split}
    e_{RS}&=\veps e^{(1)}_{RS}+\veps^2e^{(2)}_{RS}+\veps^3e^{(3)}_{RS}+\mathcal{O}(\veps^4),\\[0.2cm]
    E^a_{RS}&=\veps E^{a,(1)}_{RS}+\veps^2E^{a,(2)}_{RS}+\veps^3E^{a,(3)}_{RS}+\mathcal{O}(\veps^4),\\[0.2cm]
    \delta\omega_{MPRQ,\text{ab}}&=\veps^2\delta\omega^{(2)}_{MPRQ,\text{ab}}+
    \veps^3\delta\omega^{(3)}_{MPRQ,\text{ab}}+\mathcal{O}\lb\veps^4\rb,\\[0.2cm]
    \delta\omega^a_{MPRQ,\text{non-ab}}&=\veps^2\delta\omega^{a,(2)}_{MPRQ,\text{non-ab}}+
    \veps^3\delta\omega^{a,(3)}_{MPRQ,\text{non-ab}}+\mathcal{O}\lb\veps^4\rb,
  \end{split}
\end{equation}
The terms induced by variation of the CS action (last two lines in~\eqref{eq:more_pertr_exp})
are \(\mathcal{O}\lb\veps^2\rb\). The \(\veps\)-expansion terms:
\begin{equation}
  \begin{split}
    e^{(1)}_{RS}&=4N_f\tilde{T}_8(\pi\alpha')^2f^{(1)}_{RS},\qquad
    e^{(2)}_{RS}=4N_f\tilde{T}_8(\pi\alpha')^2f^{(2)}_{RS},\\[0.2cm]
    e^{(3)}_{RS}&=\tilde{T}_8(\pi\alpha')^2\lsb4N_ff^{(3)}_{RS}+(\pi\alpha')^2\lb4N_f\mathcal{T}^{(3)}_{RS,1}+2\mathcal{T}^{aa,(3)}_{RS,3}+d_{abc}\mathcal{T}^{abc,(3)}_{RS,4}\rb\rsb,
  \end{split}
\end{equation}
\begin{equation}
  E^{a,(1)}_{RS}=2\tilde{T}_8(\pi\alpha')^2 f^{a,(1)}_{RS},\qquad
  E^{a,(2)}_{RS}=2\tilde{T}_8(\pi\alpha')^2\lsb f^{a,(2)}_{RS}-f_{abc}A^{b,(1)}_RA^{c,(1)}_S\rsb,
\end{equation}
\begin{multline}
  E^{a,(3)}_{RS}=\tilde{T}_8(\pi\alpha')^2\lsb2f^{a,(3)}_{RS}-
  2f_{abc}\lb A^{b,(1)}_RA^{c,(2)}_S+A^{b,(2)}_RA^{c,(1)}_S\rb\right.\\\left.+
  (\pi\alpha')^2\lb2\mathcal{T}^{a,(3)}_{RS,2}+d_{abc}\mathcal{T}^{bc,(3)}_{RS,3}+4\,\text{STr}\lb T^aT^bT^cT^d\rb\mathcal{T}^{bcd,(3)}_{RS,4}\rb\rsb,
\end{multline}
\begin{equation}
  \delta\omega^{(2)}_{MPRQ,\text{ab}}\equiv\frac{N_c}{8\pi^2}\lb
  N_f\partial_Ma^{(1)}_P\partial_Ra^{(1)}_Q+
  \frac{1}{2}\partial_MA^{a,(1)}_P\partial_RA^{a,(1)}_Q\rb,
\end{equation}
\begin{multline}
  \delta\omega^{(3)}_{MPRQ,\text{ab}}\equiv\frac{N_c}{8\pi^2}\lb
  N_f\partial_Ma^{(1)}_P\partial_Ra^{(2)}_Q+N_f\partial_Ma^{(2)}_P\partial_Ra^{(1)}_Q\right.\\\left.+
  \frac{1}{2}\partial_MA^{a,(1)}_P\partial_RA^{a,(2)}_Q+\frac{1}{2}\partial_MA^{a,(2)}_P\partial_RA^{a,(1)}_Q\rb,
\end{multline}
\begin{equation}
  \delta\omega^{a,(2)}_{MPRQ,\text{non-ab}}\equiv\frac{N_c}{8\pi^2}\lb
  \partial_Ma^{(1)}_P\partial_RA^{a,(1)}_Q+\frac{1}{4}d_{abc}\partial_MA^{b,(1)}_P\partial_RA^{c,(1)}_Q\rb,
\end{equation}
\begin{multline}
  \delta\omega^{a,(3)}_{MPRQ,\text{non-ab}}\equiv\frac{N_c}{8\pi^2}\lb
  \partial_Ma^{(1)}_P\partial_RA^{a,(2)}_Q+\partial_Ma^{(2)}_P\partial_RA^{a,(1)}_Q\right.\\\left.+
  \frac{1}{4}d_{abc}\partial_MA^{b,(1)}_P\partial_RA^{c,(2)}_Q+\frac{1}{4}d_{abc}\partial_MA^{b,(2)}_P\partial_RA^{c,(1)}_Q\rb.
\end{multline}
The expansions for the \(\mathcal{T}\)-tensors are \(\mathcal{O}\lb\veps^3\rb\),
\begin{equation}
  \mathcal{T}^{\dots}_{RS,i}=\veps^3\mathcal{T}^{\dots,(3)}_{RS,i}+\mathcal{O}(\veps^4),
\end{equation}
\begin{equation}
  \begin{split}
    \mathcal{T}^{(3)}_{RS,1}&\equiv f^{(1)}_{RS}f^{(1)}_{KL}f^{KL,(1)}-4f^{(1)}_{KR}f^{(1)}_{LS}f^{KL,(1)},\\
    \mathcal{T}^{a,(3)}_{RS,2}&\equiv f^{(1)}_{KL}f^{KL,(1)}F^{a,(1)}_{RS}+2f^{(1)}_{KL}f^{(1)}_{RS}F^{KL,a,(1)}\\&-
    4f^{(1)}_{LS}f^{KL,(1)}F^{a,(1)}_{KR}-4f^{(1)}_{KR}f^{KL,(1)}F^{a,(1)}_{LS}-4f^{(1)}_{KR}f^{(1)}_{LS}F^{KL,a,(1)},\\
    \mathcal{T}^{ab,(3)}_{RS,3}&\equiv f^{(1)}_{RS}F^{KL,a,(1)}F^{b,(1)}_{KL}+2f^{(1)}_{KL}F^{KL,a,(1)}F^{b,(1)}_{RS}\\&-
    4f^{(1)}_{KR}F^{KL,a,(1)}F^{b,(1)}_{LS}-4f^{(1)}_{LS}F^{KL,a,(1)}F^{b,(1)}_{KR}-4f^{KL,(1)}F^{a,(1)}_{KR}F^{b,(1)}_{LS},\\
    \mathcal{T}^{abc,(3)}_{RS,4}&\equiv F^{KL,a,(1)}F^{b,(1)}_{KL}F^{c,(1)}_{RS}-4F^{KL,a,(1)}F^{b,(1)}_{KR}F^{c,(1)}_{LS}.
  \end{split}
\end{equation}
Finally, the sources for perturbative EOMs~\eqref{eq:pert_eq} are
\begin{equation}\label{eq:J2_ab}
  j^{\nu,(1)}=0,\qquad
  j^{\nu,(2)}=-\frac{1}{2N_f\kappa L^4}\veps^{MPRQ\nu}\delta\omega^{(2)}_{MPRQ,\text{ab}},
\end{equation}
\begin{multline}\label{eq:J3_ab}
  j^{\nu,(3)}=-\frac{(\pi\alpha')^2}{4N_f}\partial_M\lsb u(z)^2G^{MR}G^{\nu\sigma}\lb4N_f\mathcal{T}^{(3)}_{R\sigma,1}+2\mathcal{T}^{aa,(3)}_{R\sigma,3}+d_{abc}\mathcal{T}^{abc,(3)}_{R\sigma,4}\rb\rsb\\-
  \frac{1}{2N_f\kappa L^4}\veps^{MPRQ\nu}\delta\omega^{(3)}_{MPRQ,\text{ab}},
\end{multline}
\begin{equation}
  J^{\nu,a,(1)}=0,
\end{equation}
\begin{multline}\label{eq:J2_n_ab}
  J^{\nu,a,(2)}=\frac{1}{\kappa L^4}f_{abe}u(z)^2G^{MR}G^{\nu\sigma}A^{b,(1)}_ME^{e,(1)}_{R\sigma}+
  f_{abc}\partial_M\lb u(z)^2G^{MR}G^{\nu\sigma}A^{b,(1)}_RA^{c,(1)}_\sigma\rb\\-
  \frac{1}{\kappa L^4}\veps^{MPRQ\nu}\delta\omega^{a,(2)}_{MPRQ,\text{non-ab}},
\end{multline}
\begin{multline}\label{eq:J3_n_ab}
  J^{\nu,a,(3)}=\frac{1}{\kappa L^4}f_{abe}u(z)^2G^{MR}G^{\nu\sigma}\lb A^{b,(1)}_ME^{e,(2)}_{R\sigma}+A^{b,(2)}_ME^{e,(1)}_{R\sigma}\rb\\-
  \frac{(\pi\alpha')^2}{2}\partial_M\lsb u(z)^2G^{MR}G^{\nu\sigma}\lb2\mathcal{T}^{a,(3)}_{R\sigma,2}+d_{abc}\mathcal{T}^{bc,(3)}_{R\sigma,3}+4\,\text{STr}\lb T^aT^bT^cT^d\rb\mathcal{T}^{bcd,(3)}_{R\sigma,4}\rb\rsb\\+
  f_{abc}\partial_M\lsb u(z)^2G^{MR}G^{\nu\sigma}\lb A^{b,(1)}_RA^{c,(2)}_\sigma+A^{b,(2)}_RA^{c,(1)}_\sigma\rb\rsb-
  \frac{1}{\kappa L^4}\veps^{MPRQ\nu}\delta\omega^{a,(3)}_{MPRQ,\text{non-ab}}.
\end{multline}
The source terms are non-linear. At quadratic level,~\eqref{eq:J2_ab} and~\eqref{eq:J2_n_ab},
the non-abelian gauge field mixes with the abelian one solely due to the CS-induced terms,
\(\delta\omega^{(2)}_{MPRQ,\text{ab}}\) and \(\delta\omega^{a,(2)}_{MPRQ,\text{non-ab}}\).

\section{Spectral Problem of the Transverse Equation}\label{app:spec_prob}

In this Appendix we discuss various properties of the eigensystem of the spectral problem
of the transverse equation~\eqref{eq:spec_prob} that can be established even in absence
of analytical solutions for the eigensystem.

For the equation~\eqref{eq:spec_prob} (primes denote derivatives with respect to \(z\))
\begin{equation}
  \psi_n''(z)+\frac{2z}{1+z^2}\psi_n'(z)-\frac{\lambda_n}{\lb1+z^2\rb^{4/3}}\psi_n(z)=0,
\end{equation}
at large \(z\), the following approximate equation holds:
\begin{equation}
  \psi_n''(z)+\frac{2}{z}\psi_n'(z)\approx0,\qquad z\to+\infty.
\end{equation}
It has an asymptotic solution
\begin{equation}
  \pz\psi_n(z)\approx\frac{C_1}{z^2}\qrq\psi_n(z)\approx\frac{C_1}{z}+C_2,
\end{equation}
where \(C_1\) and \(C_2\) are integration constants. Homogeneous boundary conditions
imply \(C_2=0\). Hence,
\begin{equation}\label{eq:trans_sus_as}
  \psi_n(z)\underset{z\to\pm\infty}{=}\mathcal{O}\lb\frac{1}{z}\rb,\qquad
  \pz\psi_n(z)\underset{z\to\pm\infty}{=}\mathcal{O}\lb\frac{1}{z^2}\rb.
\end{equation}
Assuming that the eigensystem is discrete, a number of properties can be established.
First, the eigenfunctions are orthogonal since the differential operator of the spectral
problem is symmetric. Introduction of an inner product between the eigenfunctions imposes
an orthonormalization condition:
\begin{equation}\label{eq:eigf_ortnorm}
  \kappa\left\langle\psi_n,\psi_m\right\rangle=
  \kappa\int\limits_{-\infty}^{+\infty}\frac{dz}{u(z)}\psi_n(z)\psi_m(z)=
  \delta_{nm}.
\end{equation}
Convergence of this integral requires that the eigenfunctions fall down at infinity
faster than \(z^{-1/6}\), which is consistent with the asymptotic behavior~\eqref{eq:trans_sus_as}.
The constant prefactor defined in~\eqref{eq:kappa} ensures that the kinetic term for the
vector mesons has canonical normalization consistent with the Proca action. As a
consequence of the orthogonality,
\begin{equation}
  \kappa\int\limits_{-\infty}^{+\infty}dz\,u(z)^3\pz\psi_n(z)\,\pz\psi_m(z)=-\lambda_n\delta_{nm}.
\end{equation}
The completeness relation compatible with the normalization condition~\eqref{eq:eigf_ortnorm}
\begin{equation}\label{eq:comp_rel}
  \kappa\sum_{n=1}^{\infty}\frac{1}{u(z')}\psi_n(z)\,\psi_n(z')=\delta(z-z')
\end{equation}
is used in the derivation of the Green function \(G_t\) in Section~\ref{sect:ho}. Note
there is no non-trivial zero mode corresponding to \(\lambda=0\) satisfying the
homogeneous boundary condition~\eqref{eq:lo_ab_bcond}. There is also an orthonormal
complete basis in the space of square-integrable functions that behave as \(z^{-2}\) at
infinity with the inner product given by
\begin{equation}
  \left\langle f,g\right\rangle=\int\limits_{-\infty}^{+\infty}dz\,u(z)^3\,f(z)\,g(z).
\end{equation}
The basis is given by the set of functions
\begin{equation}
  \phi_0(z)\equiv\sqrt{\frac{\pi}{\kappa}}\pz w(z),\qquad
  \phi_n(z)\equiv\frac{\pz\psi_n(z)}{\sqrt{-\lambda_n}},\quad n\ge1.
\end{equation}
The completeness relation for this basis:
\begin{equation}
  \pz w(z)-\kappa\sum_{n=1}^\infty u(z')^3\,\frac{\pz\psi_n(z)\,\partial_{z'}\psi_n(z')}{\lambda_n}=\delta(z-z').
\end{equation}

\section{Numerical Results Related to the Eigensystem~\texorpdfstring{\eqref{eq:spec_prob}}{(4.14)}}\label{app:num}

In this Appendix some numerical results which involve the eigenvalues and eigenfunctions
of the spectral problem~\eqref{eq:spec_prob} are presented.

Several lowest eigenvalues have been known since the original works on the SS model~\cite{Sakai:2004cn,
Sakai:2005yt}. We have extended these results significantly: in Table~\ref{tab:1} the
first 39 eigenvalues \(\lambda_n\) are listed. The definition of the spectral problem in
this work has an opposite sign for the eigenvalues compared to~\cite{Sakai:2004cn}. The
eigensystem was computed using the built-in functionality of Wolfram Mathmeatica which
relies on the Arnoldi method. To ensure the accuracy of the results, the discretization
parameter of the method was varied and fine-tuned such that the results are stable with
three digit precision.

Several values of the couplings used in the discussion of the vector meson interactions
in Section~\ref{sect:vmes_int} have been computed in Table~\ref{tab:2}. The couplings are
given in the units of \(\kappa^{-1}\) due to the normalization condition~\eqref{eq:eigf_ortnorm}
of the eigenfunctions.
\begin{table}[tbp]
  \centering
  \(\begin{array}{c||c|c|c|c|c|c|c|c|c|c}
    n & 1 & 2 & 3 & 4 & 5 & 6 & 7 & 8 & 9 & 10 \\
    \hline
    \lambda_n & -0.669 & -1.57 & -2.87 & -4.55 & -6.59 & -9.01 & -11.8 & -15 & -18.5 & -22.4 \\
    \hhline{=#=|=|=|=|=|=|=|=|=|=}
    n & 11 & 12 & 13 & 14 & 15 & 16 & 17 & 18 & 19 & 20 \\
    \hline
    \lambda_n & -26.7 & -31.3 & -36.3 & -41.7 & -47.5 & -53.6 & -60.1 & -67 & -74.3 & -81.9 \\
    \hhline{=#=|=|=|=|=|=|=|=|=|=}
    n & 21 & 22 & 23 & 24 & 25 & 26 & 27 & 28 & 29 & 30 \\
    \hline
    \lambda_n & -89.9 & -98.2 & -107 & -116 & -126 & -135 & -146 & -156 & -167 & -179 \\
    \hhline{=#=|=|=|=|=|=|=|=|=|=}
    n & 31 & 32 & 33 & 34 & 35 & 36 & 37 & 38 & 39 & \\
    \cline{1-10}
    \lambda_n & -190 & -202 & -215 & -228 & -241 & -254 & -268 & -283 & -297
  \end{array}\)
  \caption{Eigenvalues of the spectral problem.}
  \label{tab:1}
\end{table}
\begin{table}[tbp]
  \centering
  \(\begin{array}{c|c||r|r|r|r|r|r|r}
   \kappa a^{(i)}_{j,\dots} & \text{\diagbox{\(n\)}{\(m\)}} & \multicolumn{1}{c|}{1} & \multicolumn{1}{c|}{2} & \multicolumn{1}{c|}{3} & \multicolumn{1}{c|}{4} & \multicolumn{1}{c|}{5} & \multicolumn{1}{c|}{6} & \multicolumn{1}{c}{7} \\
   \hhline{=|=#=|=|=|=|=|=|=}
   \multirow{7}{*}{\(\kappa a^{(2)}_{2,nm}\)} & 1 & -0.101 & -0.186 & -0.158 & -0.0609 & 0.0666 & 0.00644 & -0.0119 \\
   \cline{2-9}
   & 2 & 0.186 & -0.0503 & 0.272 & -0.177 & -0.0908 & 0.085 & 0.00864 \\
   \cline{2-9}
   & 3 & 0.0656 & -0.272 & -0.0548 & -0.356 & 0.216 & 0.119 & -0.106 \\
   \cline{2-9}
   & 4 & 0.0609 & 0.105 & 0.356 & -0.0514 & -0.439 & 0.251 & 0.145 \\
   \cline{2-9}
   & 5 & -0.0382 & 0.0908 & -0.141 & 0.439 & -0.0509 & -0.521 & 0.287 \\
   \cline{2-9}
   & 6 & -0.00644 & -0.0586 & -0.119 & -0.178 & 0.521 & -0.0503 & -0.602 \\
   \cline{2-9}
   & 7 & 0.00568 & -0.00864 & 0.078 & -0.145 & -0.214 & 0.602 & -0.05 \\
   \hhline{=|=#=|=|=|=|=|=|=}
   \multirow{7}{*}{\(\kappa a^{(2)}_{7,nm}\)} & 1 & 0.5 & -0.264 & 0 & -0.02 & 0 & 0.00098 & 0 \\
   \cline{2-9}
   & 2 & -0.264 & 0.5 & 0.265 & 0 & -0.023 & 0 & 0.0011 \\
   \cline{2-9}
   & 3 & 0 & 0.265 & 0.5 & -0.271 & 0 & 0.0246 & 0 \\
   \cline{2-9}
   & 4 & -0.02 & 0 & -0.271 & 0.5 & -0.273 & 0 & 0.0255 \\
   \cline{2-9}
   & 5 & 0 & -0.023 & 0 & -0.273 & 0.5 & -0.274 & 0 \\
   \cline{2-9}
   & 6 & 0.00098 & 0 & 0.0246 & 0 & -0.274 & 0.5 & -0.275 \\
   \cline{2-9}
   & 7 & 0 & 0.0011 & 0 & 0.0255 & 0 & -0.275 & 0.5
  \end{array}\)
  \caption{Couplings \(a^{(2)}_{2,nm}\) and \(a^{(2)}_{7,nm}\).}
  \label{tab:2}
\end{table}

\section{Spatial Parity Transformations}\label{app:parity}

In QCD, parity transformation implies an exchange between left and right components of
the fermions. In the massless case, when helicity matches chirality, this can be easily
understood by noting that the parity flips the direction of the momentum, but not the
spin. In the holographic setup the left and right symmetry groups are associated with
\(z\to+\infty\) and \(z\to-\infty\) boundaries, so the proper parity transformation must
also involve the sign flip of the \(z\) coordinate,
\begin{equation}
  P\text{:}\quad\lb x^1,x^2,x^3,z\rb\to\lb-x^1,-x^2,-x^3,-z\rb.
\end{equation}
The bulk gauge field (a covariant vector) under \(P\) transforms as
\begin{equation}
  A^P_0=A_0,\qquad A^P_1=-A_1,\qquad
  A^P_2=-A_2,\qquad A^P_3=-A_3,\qquad
  A^P_z=-A_z.
\end{equation}
Then from the definition~\eqref{eq:eta_def} of \(\eta'\)-meson as the integral of the
fifth component of the abelian part of the bulk field it follows
\begin{equation}
  \eta'^P=P\lb\int\limits_{-\infty}^{+\infty}dz\,a^\beta_z(x^\mu,z)\rb=
  -\int\limits_{-\infty}^{+\infty}dz\,a^\beta_z(x^\mu,z)=-\eta',
\end{equation}
which confirms that the \(\eta'\)-meson is a pseudoscalar. Similarly, applying parity
tranformation to the definition~\eqref{eq:Sigma_split} of \(\Sigma\) as the integral of
the fifth component of the non-abelian part of the bulk field, demonstrates that the
parity transformation is equivalent to the complex conjugation, \(\Sigma^P=\Sigma^\ast\).
Together with the exponential parametrization~\eqref{eq:sigma_exp} of \(\Sigma\) it leads
to
\begin{equation}
  \Sigma^P=\exp\lb\frac{2i}{f_\pi}\lb\pi^a\rb^PT^a\rb,\quad
  \Sigma^\ast=\exp\lb-\frac{2i}{f_\pi}\pi^aT^a\rb\qrq\lb\pi^a\rb^P=-\pi^a,
\end{equation}
which confirms that the \(\pi\)-mesons are pseudoscalar.

The parity of the vector mesons~\eqref{eq:vtilde_decomp}, introduced in~\eqref{eq:lo_sols_mom},
is connected to the properties of the eigenfunctions (see also the original work~\cite{Sakai:2004cn}).
Since~\eqref{eq:spec_prob} is invariant under the change \(z\to-z\), the eigenfunctions
\(\psi_n\) must be either even or odd functions of \(z\). As such, in order to maintain
the parity transformation property of the bulk gauge vector field, the vector mesons on
the boundary must have the following parity,
\begin{equation}
  \begin{aligned}
    &\text{if }\psi_n\text{ even}&\qrq&\eta_P=-1\qrq V_{\mu,n}\text{ is a vector meson,}\\
    &\text{if }\psi_n\text{ odd}&\qrq&\eta_P=+1\qrq V_{\mu,n}\text{ is an axial-vector meson.}
  \end{aligned}
\end{equation}
where \(\eta_P\) is the parity sign defined as
\begin{equation}
  V^P_{\mu,n}=\eta_PV_{\mu,n},\qquad\mu\in\left\lbrace1,2,3\right\rbrace.
\end{equation}
Specific values for each meson multiplet can be found if parity of the eigenfunctions is
known. The available numerical evidence shows that \(A_1\) is an even function and that
the parity of eigenfunctions is alternating at least for the first 39 eigenfunctions.
Thus, the lowest meson multiplet has odd parity~\cite{Sakai:2004cn}.

\section{Longitudinal Green Function}\label{app:long_gf}

In this Appendix the longitudinal Green function \(G_l\) introduced in~\eqref{eq:gf_intro}
is derived. The Green function is defined by the equation
\begin{equation}\label{eq:gf_t_eq}
  K_l\lsb G_l(z,z')\rsb=\frac{1}{L^4}\pz\lb u(z)^3\pz G_l(z,z')\rb=\delta(z-z'),
\end{equation}
with the homogeneous boundary conditions. Because of the factor \(u(z)^3=1+z^2\), the
differential operator is not translationally invariant. Thus the Green function cannot be
expressed as a function of \(z-z'\). However, the operator is symmetric,
\begin{equation}
  \left\langle f,K_l\lsb g\rsb\right\rangle_{\text{Dom}\lb K_l\rb}=
  \left\langle K_l\lsb f\rsb,g\right\rangle_{\text{Dom}\lb K_l\rb},\qquad
  \forall f,g\in\text{Dom}\lb K_l\rb,
\end{equation}
so the Green function is expected to be symmetric,
\begin{equation}
  G_l(z,z')=G_l(z',z).
\end{equation}
For \(z\ne z'\) the Green function is a solution of the corresponding homogeneous
equation. This solution was found in~\eqref{eq:long_sol}, so the following can be taken
as an ansatz for the Green function:
\begin{equation}
  G_l(z,z')=\begin{dcases}
    C_1(z')\,w(z)+C_2(z'),&z<z',\\
    C_3(z')\,w(z)+C_4(z'),&z>z',
  \end{dcases}
\end{equation}
where \(w(z)\) is defined in~\eqref{eq:w_def} and the integration ``constants'' \(C_i(z')\)
will be found from the boundary and (dis)continuity conditions. From the boundary
conditions~\eqref{eq:gf_bcond} it follows
\begin{equation}\label{eq:gl_temp}
  \begin{dcases}
    C_2(z')=0,\\[0.2cm]
    C_3(z')+C_4(z')=0,
  \end{dcases}\qrq
  G_l(z,z')=\begin{dcases}
    C_1(z')\,w(z),&z<z',\\[0.2cm]
    C_3(z')\lb w(z)-1\rb,&z>z'.
  \end{dcases}
\end{equation}
At \(z=z'\), the Green function must be continuous which relates the coefficients \(C_1\)
and \(C_3\),
\begin{equation}\label{eq:c1_c3_eq}
  C_1(z')\,w(z')=C_3(z')\lb w(z')-1\rb\qrq
  C_3(z')=C_1(z')\,\frac{w(z')}{w(z')-1}.
\end{equation}
Integrating the equation~\eqref{eq:gf_t_eq} once over \(z\) from \(-\infty\) to
\(+\infty\) gives
\begin{equation}
  \left.\vphfrac u(z)^3\pz G_l(z,z')\right|^{+\infty}_{-\infty}=L^4.
\end{equation}
Substituting the derivative of~\eqref{eq:gl_temp},
\begin{equation}
  \pz G_l(z,z')=\begin{dcases}
    \frac{C_1(z')}{\pi u(z)^3},&z<z',\\
    \frac{C_3(z')}{\pi u(z)^3},&z>z',
  \end{dcases}
\end{equation}
where we used~\eqref{eq:w_prop}, provides another relation for \(C_1(z')\) and \(C_3(z')\),
\begin{equation}
  C_3(z')-C_1(z')=\pi L^4.
\end{equation}
Combined with~\eqref{eq:c1_c3_eq}:
\begin{equation}
  \begin{dcases}
    C_3(z')=C_1(z')\,\frac{w(z')}{w(z')-1},\\[0.2cm]
    C_3(z')=C_1(z')+\pi L^4,
  \end{dcases}\qrq
  \begin{dcases}
    C_1(z')=\pi L^4\lb w(z')-1\rb,\\[0.2cm]
    C_3(z')=\pi L^4w(z').
  \end{dcases}
\end{equation}
Finally, the result quoted in~\eqref{eq:long_gf},
\begin{equation}
  G_l(z,z')=\pi L^4\lsb
  \theta(z'-z)\lb w(z')-1\rb w(z)+\theta(z-z')\,w(z')\lb w(z)-1\rb\rsb.
\end{equation}

\section{Higher Order Solutions}\label{app:ho_sols}

In this Appendix we present the explicit expressions for the solutions of the abelian
and non-abelian second and third order equations~\eqref{eq:pert_eq}. As mentioned in
Section~\ref{sect:ho}, to construct higher order solutions the sources \(j^{\nu,(j)}\)
and \(J^{\nu,a,(j)}\) are split into \(z\)- and \(q\)-dependent factors. Furthermore, the
sources contain multiple terms at each order \(j\in\lbrace2,3\rbrace\). More explicitly,
this factorization takes the following form,
\begin{equation}\label{eq:src_fact}
  j^{\nu,(j)}(q,z)=\sum_{i,\dots}j^{\nu,(j)}_{i,\dots}(q)\,g^{(j)}_{i,\dots}(z),\qquad
  J^{\nu,a,(j)}(q,z)=\sum_{i,\dots}J^{\nu,a,(j)}_{i,\dots}(q)\,g^{(j)}_{i,\dots}(z).
\end{equation}
For any given \(i\)-th term, there frequently appear summations over additional indices,
which for generality are denoted with ellipsis. This Appendix presents expressions for
the sources' factors, after the factorization. If a factor \(j^{\nu,(j)}_{i,\dots}\) is
not explicitly defined below, then it is simply zero. In what follows we utilize the
following notations:

\noindent the \emph{higher order longitudinal function},
\begin{equation}
  w^{(j)}_{i,\dots}(z)\equiv\int\limits_{-\infty}^{+\infty}dz'\,G_l(z,z')\,g^{(j)}_{i,\dots}(z'),
\end{equation}
the \emph{higher order transverse functions},
\begin{equation}\label{eq:ho_s}
  s^{(j)}_{i,\dots,k}(z)\equiv\int\limits_{-\infty}^{+\infty}dz'\,\psi_k(z)\,\psi_k(z')\,g^{(j)}_{i,\dots}(z')=
 \psi_k(z)\,c^{(j)}_{i,\dots,k},
\end{equation}
the corresponding \emph{\(c\)-coefficients},
\begin{equation}\label{eq:cij}
  c^{(j)}_{i,\dots,k}\equiv\int\limits_{-\infty}^{+\infty}dz\,\psi_k(z)\,g^{(j)}_{i,\dots}(z),
\end{equation}
and the \emph{full higher order transverse functions},
\begin{equation}\label{eq:full_ho_s}
  \hat{s}^{(j)}_{i,\dots}(q,z)\equiv-\frac{1}{\kappa L^4}\int\limits_{-\infty}^{+\infty}dz'\,G_t(z,z',q)\,
  g^{(j)}_{i,\dots}(z')=\int\limits_{-\infty}^{+\infty}dz'\sum_{k=1}^\infty\frac{\psi_k(z)\,\psi_k(z')}{q^2-\lambda_k}g^{(j)}_{i,\dots}(z').
\end{equation}
Here, \(G_l\) and \(G_t\) are the longitudinal~\eqref{eq:long_gf} and transverse~\eqref{eq:trans_gf}
Green functions, \(\psi_k\) are the eigenfunctions of the transverse spectral problem~\eqref{eq:spec_prob}.
The higher order transverse functions~\eqref{eq:ho_s} obey several properties similar to
those of the eigenfunctions,
\begin{equation}
  \kappa\int\frac{dz}{u(z)}s^{(j_1)}_{i_1,\dots,k}(z)\,s^{(j_2)}_{i_2,\dots,l}(z)=
  \delta_{kl}c^{(j_1)}_{i_1,\dots,k}c^{(j_2)}_{i_2,\dots,l},\qquad
  \kappa\int\frac{dz}{u(z)}\psi_k(z)\,s^{(j)}_{i,\dots,l}(z)=
  \delta_{kl}c^{(j)}_{i,\dots,l},
\end{equation}
\begin{equation}
  \begin{gathered}
    \kappa\int\limits_{-\infty}^{+\infty}dz\,u(z)^3\pz s^{(j_1)}_{i_1,\dots,k}(z)\,\pz s^{(j_2)}_{i_2,\dots,l}(z)=
    -\lambda_k\delta_{kl}c^{(j_1)}_{i_1,\dots,k}c^{(j_2)}_{i_2,\dots,l},\\[0.2cm]
    \kappa\int\limits_{-\infty}^{+\infty}dz\,u(z)^3\pz\psi_k(z)\,\pz s^{(j)}_{i,\dots,l}(z)=
    -\lambda_k\delta_{kl}c^{(j)}_{i,\dots,l}.
  \end{gathered}
\end{equation}

\noindent \(\bullet\) \textbf{Second order solutions }\(\bm{\lb j=2\rb}\)\textbf{.}

\begin{subequations}
  \begin{equation}
    a^{(2)}_\mu(q,z)=\sum_{i,\dots}D^\lambda_\mu(q)\,j^{(2)}_{\lambda,i,\dots}(q)\,w^{(2)}_{i,\dots}(z)+
    \sum_{k,i,\dots}\frac{-\kappa L^4\lb\delta^\lambda_\mu-D^\lambda_\mu(q)\rb}{q^2-\lambda_k}
    j^{(2)}_{\lambda,i,\dots}(q)\,s^{(2)}_{i,\dots,k}(z),
  \end{equation}
  \begin{multline}\label{eq:nlo_sol_n_ab}
    A^{a,(2)}_\mu=\sum_{i,\dots}D^\lambda_\mu(q)\,J^{a,(2)}_{\lambda,i,\dots}(q)\,w^{(2)}_{i,\dots}(z)+
    \sum_{\substack{k,i,\dots\\i\neq5}}\frac{-\kappa L^4\lb\delta^\lambda_\mu-D^\lambda_\mu(q)\rb}{q^2-\lambda_k}
    J^{a,(2)}_{\lambda,i,\dots}(q)\,s^{(2)}_{i,\dots,k}(z)\\-
    \kappa L^4\lb\delta^\lambda_\mu-D^\lambda_\mu(q)\rb J^{a,(2)}_{\lambda,5}(q)\,\hat{s}^{(2)}_5(q,z).
  \end{multline}
\end{subequations}
The second order \(z\)-dependent factors of the sources are defined as\footnote{A certain
mathematical issue related to the function \(g^{(2)}_5(z)\) is discussed in Appendix~\ref{app:op_reord}.}
\begin{equation}\label{eq:nlo_g}
  \begin{gathered}
    g^{(2)}_{1,n}\equiv\pz w\psi_n,\qquad
    g^{(2)}_{2,nm}\equiv\psi_n\pz\psi_m,\qquad
    g^{(2)}_3\equiv w\pz w,\qquad
    g^{(2)}_{4,n}\equiv w\pz\psi_n,\\[0.2cm]
    g^{(2)}_5\equiv\frac{w^2}{u},\qquad
    g^{(2)}_{6,n}\equiv\frac{w\psi_n}{u},\qquad
    g^{(2)}_{7,nm}\equiv\frac{\psi_n\psi_m}{u}.
  \end{gathered}
\end{equation}
The second order \(q\)-dependent factors:
\begin{equation}
  \begin{split}
    j^{\nu,(2)}_{1,n}&\equiv-\frac{iN_c\veps^{z\mu\nu\rho\sigma}}{8N_f(2\pi)^2\pi^2\kappa L^4}\int d^4p\,
    p_\mu\,\text{STr}\lsb V_{\rho,n}(p)\,B_\sigma(q-p)\rsb,\\[0.2cm]
    j^{\nu,(2)}_{2,nm}&\equiv-\frac{iN_c\veps^{z\mu\nu\rho\sigma}}{8N_f(2\pi)^2\pi^2\kappa L^4}\int d^4p\,
    p_\mu\,\text{STr}\lsb V_{\rho,n}(p)\,V_{\sigma,m}(q-p)\rsb,\\[0.2cm]
    j^{\nu,(2)}_3&\equiv-\frac{iN_c\veps^{z\mu\nu\rho\sigma}}{8N_f(2\pi)^2\pi^2\kappa L^4}\int d^4p\,
    p_\mu\,\text{STr}\lsb B_\rho(p)\,B_\sigma(q-p)\rsb,\\[0.2cm]
    j^{\nu,(2)}_{4,n}&\equiv-\frac{iN_c\veps^{z\mu\nu\rho\sigma}}{8N_f(2\pi)^2\pi^2\kappa L^4}\int d^4p\,
    p_\mu\,\text{STr}\lsb B_\rho(p)\,V_{\sigma,n}(q-p)\rsb,
  \end{split}
\end{equation}
\begin{equation}
  \begin{split}
    J^{\nu,a,(2)}_{1,n}&\equiv-\frac{iN_c\veps^{z\mu\nu\rho\sigma}}{4(2\pi)^2\pi^2\kappa L^4}\int d^4p\,
    p_\mu\,\text{STr}\lsb V_{\rho,n}(p)\,B_\sigma(q-p)\,T^a\rsb,\\[0.2cm]
    J^{\nu,a,(2)}_{2,nm}&\equiv-\frac{iN_c\veps^{z\mu\nu\rho\sigma}}{4(2\pi)^2\pi^2\kappa L^4}\int d^4p\,
    p_\mu\,\text{STr}\lsb V_{\rho,n}(p)\,V_{\sigma,m}(q-p)\,T^a\rsb,\\[0.2cm]
    J^{\nu,a,(2)}_3&\equiv-\frac{iN_c\veps^{z\mu\nu\rho\sigma}}{4(2\pi)^2\pi^2\kappa L^4}\int d^4p\,
    p_\mu\,\text{STr}\lsb B_\rho(p)\,B_\sigma(q-p)\,T^a\rsb,\\[0.2cm]
    J^{\nu,a,(2)}_{4,n}&\equiv-\frac{iN_c\veps^{z\mu\nu\rho\sigma}}{4(2\pi)^2\pi^2\kappa L^4}\int d^4p\,
    p_\mu\,\text{STr}\lsb B_\rho(p)\,V_{\sigma,n}(q-p)\,T^a\rsb,
  \end{split}
\end{equation}
\begin{equation}
  J^{\nu,a,(2)}_5\equiv\frac{if_{abc}}{(2\pi)^2L^4}\int d^4p
  \lb2\eta^{\mu\rho}\eta^{\nu\sigma}-\eta^{\mu\sigma}\eta^{\nu\rho}-
  \eta^{\mu\nu}\eta^{\rho\sigma}\rb
  p_\rho B^b_\mu(q-p)\,B^c_\sigma(p),
\end{equation}
\begin{multline}
  J^{\nu,a,(2)}_{6,n}\equiv\frac{if_{abc}}{(2\pi)^2L^4}\int d^4p
  \lb2\eta^{\mu\rho}\eta^{\nu\sigma}-\eta^{\mu\sigma}\eta^{\nu\rho}-
  \eta^{\mu\nu}\eta^{\rho\sigma}\rb\\\times
  p_\rho\lb V^b_{\mu,n}(q-p)\,B^c_\sigma(p)+B^b_\mu(q-p)\,V^c_{\sigma,n}(p)\rb,
\end{multline}
\begin{equation}
  J^{\nu,a,(2)}_{7,nm}\equiv\frac{if_{abc}}{(2\pi)^2L^4}\int d^4p
  \lb2\eta^{\mu\rho}\eta^{\nu\sigma}-\eta^{\mu\sigma}\eta^{\nu\rho}\rb
  p_\rho V^b_{\mu,n}(q-p)\,V^c_{\sigma,m}(p).
\end{equation}

\noindent \(\bullet\) \textbf{Third order solutions }\(\bm{\lb j=3\rb}\)\textbf{.}

Since the sources for the third order equations,~\eqref{eq:J3_ab} and~\eqref{eq:J3_n_ab},
involve the second order solutions, the factorization of the sources (in the meaning
described at the beginning of this Appendix) requires that those solutions factorize too.
For the most part, the second order solutions are indeed constructed in the factorized
form, except the last term in the non-abelian solution~\eqref{eq:nlo_sol_n_ab}. To
accommodate for the fact that this term is not factorizable, several additional notations
are introduced:
\begin{equation}\label{eq:KL_not}
  \begin{gathered}
    l_{\lambda,i,\dots}(q,z)\equiv\int\limits_{-\infty}^{+\infty}dz'\,G_l(z,z')\,j^{(3)}_{\lambda,i,\dots}(q,z'),
    \ k_{\lambda,i,\dots}(q,z)\equiv\int\limits_{-\infty}^{+\infty}dz'\,G_t(z,z',q)\,j^{(3)}_{\lambda,i,\dots}(q,z'),\\[0.2cm]
    L^a_{\lambda,i,\dots}(q,z)\equiv\int\limits_{-\infty}^{+\infty}dz'\,G_l(z,z')\,J^{a,(3)}_{\lambda,i,\dots}(q,z'),
    \ K^a_{\lambda,i,\dots}(q,z)\equiv\int\limits_{-\infty}^{+\infty}dz'\,G_t(z,z',q)\,J^{a,(3)}_{\lambda,i,\dots}(q,z'),
  \end{gathered}
\end{equation}
where \(i\in\lsb35,42\rsb\). Then the third order solutions can be written as
\begin{subequations}
  \begin{multline}
    a^{(3)}_\mu(q,z)=\sum_{i,\dots}D^\lambda_\mu(q)\,j^{(3)}_{\lambda,i,\dots}(q)\,w^{(3)}_{i,\dots}(z)+
    \sum_{k,i,\dots}\frac{-\kappa L^4\lb\delta^\lambda_\mu-D^\lambda_\mu(q)\rb}{q^2-\lambda_k}
    j^{(3)}_{\lambda,i,\dots}(q)\,s^{(3)}_{i,\dots,k}(z)\\+
    \sum_{\substack{i,\dots\\i\in\lsb35,38\rsb}}\lb D^\lambda_\mu(q)\,l_{\mu,i,\dots}(q,z)+\lb\delta^\lambda_\mu-D^\lambda_\mu(q)\rb k_{\mu,i,\dots}(q,z)\rb,
  \end{multline}
  \begin{multline}
    A^{a,(3)}_\mu(q,z)=\sum_{i,\dots}D^\lambda_\mu(q)\,J^{a,(3)}_{\lambda,i,\dots}(q)\,w^{(3)}_{i,\dots}(z)+
    \sum_{\substack{k,i,\dots\\i\neq31}}\frac{-\kappa L^4\lb\delta^\lambda_\mu-D^\lambda_\mu(q)\rb}{q^2-\lambda_k}
    J^{a,(3)}_{\lambda,i,\dots}(q)\,s^{(3)}_{i,\dots,k}(z)\\-
    \kappa L^4\lb\delta^\lambda_\mu-D^\lambda_\mu(q)\rb J^{a,(3)}_{\lambda,31}(q)\,\hat{s}^{(3)}_{31}(q,z)\\+
    \sum_{\substack{i,\dots\\i\in\lsb35,42\rsb}}\lb D^\lambda_\mu(q)\,L^a_{\lambda,i,\dots}(q,z)+\lb\delta^\lambda_\mu-D^\lambda_\mu(q)\rb K^a_{\lambda,i,\dots}(q,z)\rb.
  \end{multline}
\end{subequations}
The third order \(z\)-dependent factors of the sources are defined as\footnote{A certain
mathematical issue related to the function \(g^{(3)}_{31}(z)\) is discussed in Appendix~\ref{app:op_reord}.}
\begin{equation}\label{eq:nnlo_g}
  \begin{gathered}
    g^{(3)}_{1,l}\equiv(\pz w)^2\psi_l,\quad
    g^{(3)}_{2,lm}\equiv\pz w\psi_l\pz\psi_m,\quad
    g^{(3)}_{3,lmn}\equiv\frac{\psi_l\psi_m\psi_n}{u^4},\\
    g^{(3)}_{4,lmn}\equiv\psi_l\pz\psi_m\pz\psi_n,\quad
    g^{(3)}_{5,lm}\equiv\frac{w\psi_l\psi_m}{u^4},\quad
    g^{(3)}_{6,l}\equiv\frac{w^2\psi_l}{u^4},\\
    g^{(3)}_7\equiv\frac{w^3}{u^4},\quad
    g^{(3)}_8\equiv w\lb\pz w\rb^2,\quad
    g^{(3)}_{9,l}\equiv w\pz w\pz\psi_l,\\
    g^{(3)}_{10,lm}\equiv w\pz\psi_l\pz\psi_m,\quad
    g^{(3)}_{11}\equiv\pz\lb u^4(\pz w)^3\rb,\quad
    g^{(3)}_{12,l}\equiv\pz\lb u^4(\pz w)^2\pz\psi_l\rb,\\
    g^{(3)}_{13,lm}\equiv\pz\lb\pz w\psi_l\psi_m\rb,\quad
    g^{(3)}_{14,lm}\equiv\pz\lb u^4\pz w\pz\psi_l\pz\psi_m\rb,\quad
    g^{(3)}_{15,lmn}\equiv\pz\lb\psi_l\psi_m\pz\psi_n\rb,\\
    g^{(3)}_{16,lmn}\equiv\pz\lb u^4\pz\psi_l\pz\psi_m\pz\psi_n\rb,\quad
    g^{(3)}_{17,l}\equiv\pz\lb w\pz w\psi_l\rb,\quad
    g^{(3)}_{18}\equiv\pz\lb w^2\pz w\rb,\\
    g^{(3)}_{19,lm}\equiv\pz\lb w\psi_l\pz\psi_m\rb,\quad
    g^{(3)}_{20,l}\equiv\pz\lb w^2\pz\psi_l\rb,\quad
    g^{(3)}_{21,n,i,\dots}\equiv\psi_n\pz w^{(2)}_{i,\dots},\\
    g^{(3)}_{22,n,k,j,\dots}\equiv\psi_n\pz s^{(2)}_{j,\dots,k},\quad
    g^{(3)}_{23,k,j,\dots}\equiv\pz ws^{(2)}_{j,\dots,k},\quad
    g^{(3)}_{24,n,k,j,\dots}\equiv\pz\psi_ns^{(2)}_{j,\dots,k},\\
    g^{(3)}_{25,i,\dots}\equiv w\pz w^{(2)}_{i,\dots},\quad
    g^{(3)}_{26,k,i,\dots}\equiv w\pz s^{(2)}_{i,\dots,k},\quad
    g^{(3)}_{27,i,\dots}\equiv\frac{ww^{(2)}_{i,\dots}}{u},\\
    g^{(3)}_{28,j,k,\dots}\equiv\frac{ws^{(2)}_{j,\dots,k}}{u},\quad
    g^{(3)}_{29,n,i,\dots}\equiv\frac{\psi_nw^{(2)}_{i,\dots}}{u},\quad
    g^{(3)}_{30,n,j,k,\dots}\equiv\frac{\psi_ns^{(2)}_{j,\dots,k}}{u},\\
    g^{(3)}_{31}\equiv\frac{w^3}{u},\qquad
    g^{(3)}_{32,l}\equiv\frac{w^2\psi_l}{u},\quad
    g^{(3)}_{33,lm}\equiv\frac{w\psi_l\psi_m}{u},\quad
    g^{(3)}_{34,lmn}\equiv\frac{\psi_l\psi_m\psi_n}{u}.
  \end{gathered}
\end{equation}
The third order \(q\)-dependent factors:
\begin{equation}\label{eq:J3_T_ab}
  \begin{split}
    j^{\nu,(3)}_{i,\dots}&\equiv-\frac{i(\pi\alpha')^2}{4N_fL^4}\eta^{\mu\rho}\eta^{\nu\sigma}q_\mu\lb
    4N_fX^{[i]}_{\rho\sigma,1,\dots}+2X^{[i],aa}_{\rho\sigma,3,\dots}+d_{abc}X^{[i],abc}_{\rho\sigma,4,\dots}\rb,\quad
    \text{for }i\in\lsb1,10\rsb,\\
    j^{\nu,(3)}_{i,\dots}&\equiv-\frac{(\pi\alpha')^2}{4N_fL^4}\eta^{\nu\sigma}\lb
    4N_f\mathcal{X}^{[i-10]}_{\sigma,1,\dots}+2\mathcal{X}^{[i-10],aa}_{\sigma,3,\dots}+d_{abc}\mathcal{X}^{[i-10],abc}_{\sigma,4,\dots}\rb,\quad
    \text{for }i\in\lsb11,20\rsb,
  \end{split}
\end{equation}
\begin{equation}
  \begin{split}
    j^{\nu,(3)}_{21,n,i,\dots}&\equiv-\frac{iN_c\veps^{z\mu\nu\rho\sigma}}{8N_f(2\pi)^2\pi^2\kappa L^4}\int d^4p\,p_\mu
    D^\lambda_\sigma(q-p)\,\text{STr}\lsb V_{\rho,n}(p)\,J^{(2)}_{\lambda,i,\dots}(q-p)\rsb,\\[0.2cm]
    j^{\nu,(3)}_{22,n,k,j,\dots}&\equiv\frac{iN_c\veps^{z\mu\nu\rho\sigma}}{8N_f(2\pi)^2\pi^2}\int d^4p\,p_\mu
    \frac{\delta^\lambda_\sigma-D^\lambda_\sigma(q-p)}{(q-p)^2-\lambda_k}
    \text{STr}\lsb V_{\rho,n}(p)\,J^{(2)}_{\lambda,j,\dots}(q-p)\rsb,
  \end{split}
\end{equation}
\begin{equation}
  \begin{split}
    j^{\nu,(3)}_{23,k,j,\dots}&\equiv\frac{iN_c\veps^{z\mu\nu\rho\sigma}}{8N_f(2\pi)^2\pi^2}\int d^4p\,
    \frac{p_\mu}{p^2-\lambda_k}
    \text{STr}\lsb J^{(2)}_{\rho,j,\dots}(p)\,B_\sigma(q-p)\rsb,\\[0.2cm]
    j^{\nu,(3)}_{24,n,k,j,\dots}&\equiv\frac{iN_c\veps^{z\mu\nu\rho\sigma}}{8N_f(2\pi)^2\pi^2}\int d^4p\,
    \frac{p_\mu}{p^2-\lambda_k}
    \text{STr}\lsb J^{(2)}_{\rho,j,\dots}(p)\,V_{\sigma,n}(q-p)\rsb,
  \end{split}
\end{equation}
\begin{equation}
  \begin{split}
    j^{\nu,(3)}_{25,i,\dots}&\equiv-\frac{iN_c\veps^{z\mu\nu\rho\sigma}}{8N_f(2\pi)^2\pi^2\kappa L^4}\int d^4p\,p_\mu
    D^\lambda_\sigma(q-p)\,\text{STr}\lsb B_\rho(p)\,J^{(2)}_{\lambda,i,\dots}(q-p)\rsb,\\[0.2cm]
    j^{\nu,(3)}_{26,k,i,\dots}&\equiv\frac{iN_c\veps^{z\mu\nu\rho\sigma}}{8N_f(2\pi)^2\pi^2}\int d^4p\,p_\mu
    \frac{\delta^\lambda_\sigma-D^\lambda_\sigma(q-p)}{(q-p)^2-\lambda_k}
    \text{STr}\lsb B_\rho(p)\,J^{(2)}_{\lambda,i,\dots}(q-p)\rsb,
  \end{split}
\end{equation}
\begin{equation}
  \begin{gathered}
    \begin{multlined}[t][0.8\linewidth]
      j^{\nu,(3)}_{35}\equiv\frac{iN_c\veps^{z\mu\nu\rho\sigma}}{8N_f(2\pi)^2\pi^2}\int d^4p\,p_\mu
      \lb\delta^\lambda_\sigma-D^\lambda_\sigma(q-p)\rb\\\times
      \text{STr}\lsb B_\rho(p)\,J^{(2)}_{\lambda,5}(q-p)\rsb
      w\pz\hat{s}^{(2)}_5(q-p,z),
    \end{multlined}\\[0.2cm]
    j^{\nu,(3)}_{36}\equiv-\frac{iN_c\veps^{z\mu\nu\rho\sigma}}{8N_f(2\pi)^2\pi^2}\int d^4p\,p_\mu\,
    \text{STr}\lsb B_\rho(q-p)\,J^{(2)}_{\sigma,5}(p)\rsb
    \pz w\hat{s}^{(2)}_5(p,z),
  \end{gathered}
\end{equation}
\begin{equation}
  \begin{gathered}
    \begin{multlined}[t][0.8\linewidth]
      j^{\nu,(3)}_{37,n}\equiv\frac{iN_c\veps^{z\mu\nu\rho\sigma}}{8N_f(2\pi)^2\pi^2}\int d^4p\,p_\mu
      \lb\delta^\lambda_\sigma-D^\lambda_\sigma(q-p)\rb\\\times
      \text{STr}\lsb V_{\rho,n}(p)\,J^{(2)}_{\lambda,5}(q-p)\rsb
     \psi_n\pz\hat{s}^{(2)}_5(q-p,z),
    \end{multlined}\\[0.2cm]
    j^{\nu,(3)}_{38,n}\equiv-\frac{iN_c\veps^{z\mu\nu\rho\sigma}}{8N_f(2\pi)^2\pi^2}\int d^4p\,p_\mu\,
    \text{STr}\lsb V_{\rho,n}(q-p)\,J^{(2)}_{\sigma,5}(p)\rsb
    \pz\psi_n\hat{s}^{(2)}_5(p,z),
  \end{gathered}
\end{equation}
\begin{multline}\label{eq:J3_T_1_n_ab}
  J^{\nu,a,(3)}_{i,\dots}\equiv-\frac{i(\pi\alpha')^2}{2L^4}\eta^{\mu\rho}\eta^{\nu\sigma}q_\mu\lsb
  2X^{[i],a}_{\rho\sigma,2,\dots}+d_{abc}X^{[i],bc}_{\rho\sigma,3,\dots}\right.\\\left.+
  4\,\text{STr}\lb T^aT^bT^cT^d\rb X^{[i],bcd}_{\rho\sigma,4,\dots}\rsb,\quad
  \text{for }i\in\lsb1,10\rsb,
\end{multline}
\begin{multline}\label{eq:J3_T_2_n_ab}
  J^{\nu,a,(3)}_{i,\dots}\equiv-\frac{(\pi\alpha')^2}{2L^4}\eta^{\nu\sigma}\lsb
  2\mathcal{X}^{[i-10],a}_{\sigma,2,\dots}+d_{abc}\mathcal{X}^{[i-10],bc}_{\sigma,3,\dots}\right.\\\left.+
  4\,\text{STr}\lb T^aT^bT^cT^d\rb\mathcal{X}^{[i-10],bcd}_{\sigma,4,\dots}\rsb,\quad
  \text{for }i\in\lsb11,20\rsb,
\end{multline}
\begin{equation}
  \begin{split}
    J^{\nu,a,(3)}_{21,n,i,\dots}&\equiv-\frac{iN_c\veps^{z\mu\nu\rho\sigma}}{4(2\pi)^2\pi^2\kappa L^4}\int d^4p\,p_\mu
    D^\lambda_\sigma(q-p)\,\text{STr}\lsb V_{\rho,n}(p)\,J^{(2)}_{\lambda,i,\dots}(q-p)\,T^a\rsb,\\[0.2cm]
    J^{\nu,a,(3)}_{22,n,k,j,\dots}&\equiv\frac{iN_c\veps^{z\mu\nu\rho\sigma}}{4(2\pi)^2\pi^2}\int d^4p\,p_\mu
    \frac{\delta^\lambda_\sigma-D^\lambda_\sigma(q-p)}{(q-p)^2-\lambda_k}
    \text{STr}\lsb V_{\rho,n}(p)\,J^{(2)}_{\lambda,j,\dots}(q-p)\,T^a\rsb,
  \end{split}
\end{equation}
\begin{equation}
  \begin{split}
    J^{\nu,a,(3)}_{23,k,j,\dots}&\equiv\frac{iN_c\veps^{z\mu\nu\rho\sigma}}{4(2\pi)^2\pi^2}\int d^4p\,
    \frac{p_\mu}{p^2-\lambda_k}
    \text{STr}\lsb J^{(2)}_{\rho,j,\dots}(p)\,B_\sigma(q-p)\,T^a\rsb,\\[0.2cm]
    J^{\nu,a,(3)}_{24,n,k,j,\dots}&\equiv\frac{iN_c\veps^{z\mu\nu\rho\sigma}}{4(2\pi)^2\pi^2}\int d^4p\,
    \frac{p_\mu}{p^2-\lambda_k}
    \text{STr}\lsb J^{(2)}_{\rho,j,\dots}(p)\,V_{\sigma,n}(q-p)\,T^a\rsb,
  \end{split}
\end{equation}
\begin{equation}
  \begin{split}
    J^{\nu,a,(3)}_{25,i,\dots}&\equiv-\frac{iN_c\veps^{z\mu\nu\rho\sigma}}{4(2\pi)^2\pi^2\kappa L^4}\int d^4p\,p_\mu
    D^\lambda_\sigma(q-p)\,\text{STr}\lsb B_\rho(p)\,J^{(2)}_{\lambda,i,\dots}(q-p)\,T^a\rsb,\\[0.2cm]
    J^{\nu,a,(3)}_{26,k,i,\dots}&\equiv\frac{iN_c\veps^{z\mu\nu\rho\sigma}}{4(2\pi)^2\pi^2}\int d^4p\,
    \frac{p_\mu\lb\delta^\lambda_\sigma-D^\lambda_\sigma(q-p)\rb}{(q-p)^2-\lambda_k}
    \text{STr}\lsb B_\rho(p)\,J^{(2)}_{\lambda,i,\dots}(q-p)\,T^a\rsb,
  \end{split}
\end{equation}
\begin{multline}
  J^{\nu,a,(3)}_{27,i,\dots}\equiv\frac{if_{abc}}{(2\pi)^2L^4}\lb2\eta^{\mu\rho}\eta^{\nu\sigma}-\eta^{\mu\sigma}\eta^{\nu\rho}-
  \eta^{\mu\nu}\eta^{\rho\sigma}\rb\int d^4p\lb
  p_\rho B^b_\mu(q-p)\,D^\vk_\sigma(p)\,J^{c,(2)}_{\vk,i,\dots}(p)\right.\\\left.+
  p_\rho B^c_\sigma(p)\,D^\vk_\mu(q-p)\,J^{b,(2)}_{\vk,i,\dots}(q-p)\rb,
\end{multline}
\begin{multline}
  J^{\nu,a,(3)}_{28,j,k\dots}\equiv-\frac{if_{abc}\kappa}{(2\pi)^2}\lb2\eta^{\mu\rho}\eta^{\nu\sigma}-\eta^{\mu\sigma}\eta^{\nu\rho}-
  \eta^{\mu\nu}\eta^{\rho\sigma}\rb\int d^4p\lb
  \frac{p_\rho\lb\delta^\vk_\sigma-D^\vk_\sigma(p)\rb}{p^2-\lambda_k}
  B^b_\mu(q-p)\,J^{c,(2)}_{\vk,j,\dots}(p)\right.\\\left.+
  \frac{p_\rho\lb\delta^\vk_\mu-D^\vk_\mu(q-p)\rb}{(q-p)^2-\lambda_k}
  B^c_\sigma(p)\,J^{b,(2)}_{\vk,j,\dots}(q-p)\rb,
\end{multline}
\begin{multline}
  J^{\nu,a,(3)}_{29,n,i\dots}\equiv\frac{if_{abc}}{(2\pi)^2L^4}\lb2\eta^{\mu\rho}\eta^{\nu\sigma}-\eta^{\mu\sigma}\eta^{\nu\rho}-
  \eta^{\mu\nu}\eta^{\rho\sigma}\rb\int d^4p\lb
  p_\rho V^b_{\mu,n}(q-p)\,D^\vk_\sigma(p)\,J^{c,(2)}_{\vk,i,\dots}(p)\right.\\\left.+
  p_\rho V^c_{\sigma,n}(p)\,D^\vk_\mu(q-p)\,J^{b,(2)}_{\vk,i,\dots}(q-p)\rb,
\end{multline}
\begin{multline}
  J^{\nu,a,(3)}_{30,n,j,k,\dots}\equiv-\frac{if_{abc}\kappa}{(2\pi)^2}\lb2\eta^{\mu\rho}\eta^{\nu\sigma}-\eta^{\mu\sigma}\eta^{\nu\rho}-
  \eta^{\mu\nu}\eta^{\rho\sigma}\rb\\\times\int d^4p\lb
  \frac{p_\rho\lb\delta^\vk_\sigma-D^\vk_\sigma(p)\rb}{p^2-\lambda_k}
  V^b_{\mu,n}(q-p)\,J^{c,(2)}_{\vk,j,\dots}(p)\right.\\\left.+
  \frac{p_\rho\lb\delta^\vk_\mu-D^\vk_\mu(q-p)\rb}{(q-p)^2-\lambda_k}
  V^c_{\sigma,n}(p)\,J^{b,(2)}_{\vk,j,\dots}(q-p)\rb,
\end{multline}
\begin{equation}
  J^{\nu,a,(3)}_{31}\equiv-\frac{f_{abc}f_{cde}}{(2\pi)^4L^4}\int d^4p_1\,d^4p_2\,
  \eta^{\mu\rho}\eta^{\nu\sigma}
  B^b_\mu(p_2)\,B^d_\rho(p_1-p_2)\,B^e_\sigma(q-p_1),
\end{equation}
\begin{multline}
  J^{\nu,a,(3)}_{32,l}\equiv-\frac{f_{abc}f_{cde}}{(2\pi)^4L^4}\int d^4p_1\,d^4p_2\,\eta^{\mu\rho}\eta^{\nu\sigma}
  \lb B^b_\mu(p_2)\,B^d_\rho(p_1-p_2)\,V^e_{\sigma,l}(q-p_1)\right.\\\left.+
  B^b_\mu(p_2)\,V^d_{\rho,l}(p_1-p_2)\,B^e_\sigma(q-p_1)+
  V^b_{\mu,l}(p_2)\,B^d_\rho(p_1-p_2)\,B^e_\sigma(q-p_1)\rb,
\end{multline}
\begin{multline}
  J^{\nu,a,(3)}_{33,lm}\equiv-\frac{f_{abc}f_{cde}}{(2\pi)^4L^4}\int d^4p_1\,d^4p_2\,\eta^{\mu\rho}\eta^{\nu\sigma}
  \lb B^b_\mu(p_2)\,V^d_{\rho,l}(p_1-p_2)\,V^e_{\sigma,m}(q-p_1)\right.\\\left.+
  V^b_{\mu,l}(p_2)\,B^d_\rho(p_1-p_2)\,V^e_{\sigma,m}(q-p_1)+
  V^b_{\mu,l}(p_2)\,V^d_{\rho,m}(p_1-p_2)\,B^e_\sigma(q-p_1)\rb,
\end{multline}
\begin{equation}
  J^{\nu,a,(3)}_{34,lmn}\equiv-\frac{f_{abc}f_{cde}}{(2\pi)^4L^4}\int d^4p_1\,d^4p_2\,\eta^{\mu\rho}\eta^{\nu\sigma}
  V^b_{\mu,l}(p_2)\,V^d_{\rho,m}(p_1-p_2)\,V^e_{\sigma,n}(q-p_1),
\end{equation}
\begin{equation}
  \begin{gathered}
    \begin{multlined}[t][0.8\linewidth]
      J^{\nu,a,(3)}_{35}\equiv\frac{iN_c\veps^{z\mu\nu\rho\sigma}}{4(2\pi)^2\pi^2}\int d^4p\,p_\mu
      \lb\delta^\lambda_\sigma-D^\lambda_\sigma(q-p)\rb\\\times
      \text{STr}\lsb B_\rho(p)\,J^{(2)}_{\lambda,5}(q-p)\,T^a\rsb
      w\pz\hat{s}^{(2)}_5(q-p,z),
    \end{multlined}\\[0.2cm]
    J^{\nu,a,(3)}_{36}\equiv-\frac{iN_c\veps^{z\mu\nu\rho\sigma}}{4(2\pi)^2\pi^2}\int d^4p\,p_\mu\,
    \text{STr}\lsb B_\rho(q-p)\,J^{(2)}_{\sigma,5}(p)\,T^a\rsb
    \pz w\hat{s}^{(2)}_5(p,z),
  \end{gathered}
\end{equation}
\begin{equation}
  \begin{gathered}
    \begin{multlined}[t][0.8\linewidth]
      J^{\nu,a,(3)}_{37,n}\equiv\frac{iN_c\veps^{z\mu\nu\rho\sigma}}{4(2\pi)^2\pi^2}\int d^4p\,p_\mu
      \lb\delta^\lambda_\sigma-D^\lambda_\sigma(q-p)\rb\\\times
      \text{STr}\lsb V_{\rho,n}(p)\,J^{(2)}_{\lambda,5}(q-p)\,T^a\rsb
     \psi_n\pz\hat{s}^{(2)}_5(q-p,z),
    \end{multlined}\\[0.2cm]
    J^{\nu,a,(3)}_{38,n}\equiv-\frac{iN_c\veps^{z\mu\nu\rho\sigma}}{4(2\pi)^2\pi^2}\int d^4p\,p_\mu\,
    \text{STr}\lsb V_{\rho,n}(q-p)\,J^{(2)}_{\sigma,5}(p)\,T^a\rsb
    \pz\psi_n\hat{s}^{(2)}_5(p,z),
  \end{gathered}
\end{equation}
\begin{multline}
  J^{\nu,a,(3)}_{39}\equiv-\frac{if_{abc}\kappa}{(2\pi)^2}\lb2\eta^{\mu\rho}\eta^{\nu\sigma}-\eta^{\mu\sigma}\eta^{\nu\rho}-
  \eta^{\mu\nu}\eta^{\rho\sigma}\rb\\\times\int d^4p\,p_\rho
  \lb\delta^\lambda_\sigma-D^\lambda_\sigma(p)\rb
  B^b_\mu(q-p)\,J^{c,(2)}_{\lambda,5}(p)\,\frac{w}{u}\hat{s}^{(2)}_5(p,z),
\end{multline}
\begin{multline}
  J^{\nu,a,(3)}_{40}\equiv-\frac{if_{abc}\kappa}{(2\pi)^2}\lb2\eta^{\mu\rho}\eta^{\nu\sigma}-\eta^{\mu\sigma}\eta^{\nu\rho}-
  \eta^{\mu\nu}\eta^{\rho\sigma}\rb\\\times\int d^4p\,p_\rho
  \lb\delta^\tau_\mu-D^\tau_\mu(q-p)\rb
  B^c_\sigma(p)\,J^{b,(2)}_{\tau,5}(q-p)\,\frac{w}{u}\hat{s}^{(2)}_5(q-p,z),
\end{multline}
\begin{multline}
  J^{\nu,a,(3)}_{41,n}\equiv-\frac{if_{abc}\kappa}{(2\pi)^2}\lb2\eta^{\mu\rho}\eta^{\nu\sigma}-\eta^{\mu\sigma}\eta^{\nu\rho}-
  \eta^{\mu\nu}\eta^{\rho\sigma}\rb\\\times\int d^4p\,p_\rho
  \lb\delta^\lambda_\sigma-D^\lambda_\sigma(p)\rb
  V^b_{\mu,n}(q-p)\,J^{c,(2)}_{\lambda,5}(p)\,\frac{\psi_n}{u}\hat{s}^{(2)}_5(p,z),
\end{multline}
\begin{multline}
  J^{\nu,a,(3)}_{42,n}\equiv-\frac{if_{abc}\kappa}{(2\pi)^2}\lb2\eta^{\mu\rho}\eta^{\nu\sigma}-\eta^{\mu\sigma}\eta^{\nu\rho}-
  \eta^{\mu\nu}\eta^{\rho\sigma}\rb\\\times\int d^4p\,p_\rho
  \lb\delta^\tau_\mu-D^\tau_\mu(q-p)\rb
  V^c_{\sigma,n}(p)\,J^{b,(2)}_{\tau,5}(q-p)\,\frac{\psi_n}{u}\hat{s}^{(2)}_5(q-p,z).
\end{multline}
Some of the functions used in the sources~\eqref{eq:J3_T_ab},~\eqref{eq:J3_T_1_n_ab},
and~\eqref{eq:J3_T_2_n_ab} are zero,
\begin{multline}
  X^{[5]}_{\rho\sigma,1,lm}=X^{[6]}_{\rho\sigma,1,l}=X^{[7]}_{\rho\sigma,1}=
  X^{[8]}_{\rho\sigma,1}=X^{[9]}_{\rho\sigma,1,l}=X^{[10]}_{\rho\sigma,1,lm}=
  \mathcal{X}^{[7]}_{\sigma,1,l}=\mathcal{X}^{[8]}_{\sigma,1}=
  \mathcal{X}^{[9]}_{\sigma,1,lm}\\=\mathcal{X}^{[10]}_{\sigma,1,l}=
  X^{[6],a}_{\rho\sigma,2,l}=X^{[7],a}_{\rho\sigma,2}=
  \mathcal{X}^{[8],a}_{\sigma,2}=\mathcal{X}^{[10],a}_{\sigma,2,l}=
  X^{[7],ab}_{\rho\sigma,3}\equiv0,
\end{multline}
while the rest are defined as
\begin{equation}
  X^{[1]}_{\rho\sigma,1,l}\equiv\frac{2\eta^{\vk\lambda}}{L^4}\lb
  v_{\rho\sigma,l}b_\vk b_\lambda+2v_{\vk\rho,l}b_\sigma b_\lambda-
  2v_{\lambda\sigma,l}b_\rho b_\vk\rb,
\end{equation}
\begin{equation}
  X^{[2]}_{\rho\sigma,1,lm}\equiv\frac{4\eta^{\vk\lambda}}{L^4}\lb
  v_{\rho\sigma,l}b_\vk v_{\lambda,m}+v_{\vk\rho,l}b_\sigma v_{\lambda,m}-
  v_{\lambda\sigma,l}b_\rho v_{\vk,m}+
  v_{\vk\rho,l}v_{\sigma,m}b_\lambda-v_{\lambda\sigma,l}v_{\rho,m}b_\vk\rb,
\end{equation}
\begin{equation}
  X^{[3]}_{\rho\sigma,1,lmn}\equiv\frac{\eta^{\vk\lambda}\eta^{\iota\tau}}{L^4}\lb
  v_{\rho\sigma,l}v_{\vk\iota,m}v_{\lambda\tau,n}-
  4v_{\vk\rho,l}v_{\iota\sigma,m}v_{\lambda\tau,n}\rb,
\end{equation}
\begin{equation}
  X^{[4]}_{\rho\sigma,1,lmn}\equiv\frac{2\eta^{\vk\lambda}}{L^4}\lb
  v_{\rho\sigma,l}v_{\vk,m}v_{\lambda,n}+2v_{\vk\rho,l}v_{\sigma,m}v_{\lambda,n}-
  2v_{\lambda\sigma,l}v_{\rho,m}v_{\vk,n}\rb,
\end{equation}
\begin{equation}
  \mathcal{X}^{[1]}_{\sigma,1}\equiv-\frac{2\eta^{\vk\lambda}}{L^4}
  b_\sigma b_\vk b_\lambda,
\end{equation}
\begin{equation}
  \mathcal{X}^{[2]}_{\sigma,1,l}\equiv-\frac{\eta^{\vk\lambda}}{L^4}\lb
  2b_\vk b_\lambda v_{\sigma,l}+4b_\sigma b_\vk v_{\lambda,l}\rb,
\end{equation}
\begin{equation}
  \mathcal{X}^{[3]}_{\sigma,1,lm}\equiv\frac{\eta^{\vk\lambda}\eta^{\iota\tau}}{L^4}\lb
  v_{\vk\iota,l}v_{\lambda\tau,m}b_\sigma+4v_{\iota\sigma,l}v_{\lambda\tau,m}b_\vk\rb,
\end{equation}
\begin{equation}
  \mathcal{X}^{[4]}_{\sigma,1,lm}\equiv-\frac{\eta^{\vk\lambda}}{L^4}\lb
  2b_\sigma v_{\vk,l}v_{\lambda,m}+4v_{\sigma,l}b_\vk v_{\lambda,m}\rb,
\end{equation}
\begin{equation}
  \mathcal{X}^{[5]}_{\sigma,1,lmn}\equiv\frac{\eta^{\vk\lambda}\eta^{\iota\tau}}{L^4}\lb
  v_{\vk\iota,l}v_{\lambda\tau,m}v_{\sigma,n}+4v_{\iota\sigma,l}v_{\lambda\tau,m}v_{\vk,n}\rb,
\end{equation}
\begin{equation}
  \mathcal{X}^{[6]}_{\sigma,1,lmn}\equiv-\frac{2\eta^{\vk\lambda}}{L^4}
  v_{\sigma,l}v_{\vk,n}v_{\lambda,m},
\end{equation}
\begin{multline}
  X^{[1],a}_{\rho\sigma,2,l}\equiv\frac{2\eta^{\vk\lambda}}{L^4}\lb
  V^a_{\rho\sigma,l}b_\lambda b_\vk+2v_{\rho\sigma,l}b_\lambda B^a_\vk+
  2V^a_{\vk\rho,l}b_\sigma b_\lambda+2v_{\vk\rho,l}b_\lambda B^a_\sigma\right.\\\left.+
  2v_{\vk\rho,l}b_\sigma B^a_\lambda-2v_{\lambda\sigma,l}b_\vk B^a_\rho-
  2V^a_{\lambda\sigma,l}b_\rho b_\vk-2v_{\lambda\sigma,l}b_\rho B^a_\vk\rb,
\end{multline}
\begin{multline}
  X^{[2],a}_{\rho\sigma,2,lm}\equiv\frac{2\eta^{\vk\lambda}}{L^4}\lb
  V^a_{\rho\sigma,l}b_\lambda v_{\vk,m}+2v_{\rho\sigma,l}b_\lambda V^a_{\vk,m}+
  2V^a_{\vk\rho,l}b_\sigma v_{\lambda,m}\right.\\\left.+
  2v_{\vk\rho,l}b_\lambda V^a_{\sigma,m}+2v_{\vk\rho,l}b_\sigma V^a_{\lambda,m}-
  2v_{\lambda\sigma,l}b_\vk V^a_{\rho,m}-2V^a_{\lambda\sigma,l}b_\rho v_{\vk,m}\right.\\\left.-
  2v_{\lambda\sigma,l}b_\rho V^a_{\vk,m}+V^a_{\rho\sigma,l}v_{\lambda,m}b_\vk+
  2v_{\rho\sigma,l}v_{\lambda,m}B^a_\vk+2V^a_{\vk\rho,l}v_{\sigma,m}b_\lambda\right.\\\left.+
  2v_{\vk\rho,l}v_{\lambda,m}B^a_\sigma+2v_{\vk\rho,l}v_{\sigma,m}B^a_\lambda-
  2v_{\lambda\sigma,l}v_{\vk,m}B^a_\rho-2V^a_{\lambda\sigma,l}v_{\rho,m}b_\vk-
  2v_{\lambda\sigma,l}v_{\rho,m}B^a_\vk\rb,
\end{multline}
\begin{multline}
  X^{[3],a}_{\rho\sigma,2,lmn}\equiv\frac{\eta^{\vk\lambda}\eta^{\iota\tau}}{L^4}\lb
  v_{\vk\iota,l}v_{\lambda\tau,m}V^a_{\rho\sigma,n}+2v_{\vk\iota,l}v_{\rho\sigma,m}V^a_{\lambda\tau,n}-
  4v_{\iota\sigma,l}v_{\lambda\tau,m}V^a_{\vk\rho,n}\right.\\\left.-
  4v_{\vk\rho,l}v_{\lambda\tau,m}V^a_{\iota\sigma,n}-4v_{\vk\rho,l}v_{\iota\sigma,m}V^a_{\lambda\tau,n}\rb,
\end{multline}
\begin{multline}
  X^{[4],a}_{\rho\sigma,2,lmn}\equiv\frac{2\eta^{\vk\lambda}}{L^4}\lb
  V^a_{\rho\sigma,l}v_{\lambda,m}v_{\vk,n}+2v_{\rho\sigma,l}v_{\lambda,m}V^a_{\vk,n}+
  2V^a_{\vk\rho,l}v_{\sigma,m}v_{\lambda,n}+2v_{\vk\rho,l}v_{\lambda,m}V^a_{\sigma,n}\right.\\\left.+
  2v_{\vk\rho,l}v_{\sigma,m}V^a_{\lambda,n}-2v_{\lambda\sigma,l}v_{\vk,m}V^a_{\rho,n}-
  2V^a_{\lambda\sigma,l}v_{\rho,m}v_{\vk,n}-2v_{\lambda\sigma,l}v_{\rho,m}V^a_{\vk,n}\rb,
\end{multline}
\begin{multline}
  X^{[5],a}_{\rho\sigma,2,lm}\equiv\frac{\eta^{\vk\lambda}\eta^{\iota\tau}}{L^4}\lb
  v_{\vk\iota,l}v_{\lambda\tau,m}B^a_{\rho\sigma}+2v_{\vk\iota,l}v_{\rho\sigma,m}B^a_{\lambda\tau}-
  4v_{\iota\sigma,l}v_{\lambda\tau,m}B^a_{\vk\rho}\right.\\\left.-
  4v_{\vk\rho,l}v_{\lambda\tau,m}B^a_{\iota\sigma}-4v_{\vk\rho,l}v_{\iota\sigma,m}B^a_{\lambda\tau}\rb,
\end{multline}
\begin{equation}
  X^{[8],a}_{\rho\sigma,2}\equiv\frac{2\eta^{\vk\lambda}}{L^4}\lb
  B^a_{\rho\sigma}b_\lambda b_\vk+2B^a_{\vk\rho}b_\sigma b_\lambda-
  2B^a_{\lambda\sigma}b_\rho b_\vk\rb,
\end{equation}
\begin{multline}
  X^{[9],a}_{\rho\sigma,2,l}\equiv\frac{2\eta^{\vk\lambda}}{L^4}\lb
  B^a_{\rho\sigma}b_\lambda v_{\vk,l}+2B^a_{\vk\rho}b_\sigma v_{\lambda,l}-
  2B^a_{\lambda\sigma}b_\rho v_{\vk,l}\right.\\\left.+
  B^a_{\rho\sigma}v_{\lambda,l}b_\vk+2B^a_{\vk\rho}v_{\sigma,l}b_\lambda-
  2B^a_{\lambda\sigma}v_{\rho,l}b_\vk\rb,
\end{multline}
\begin{equation}
  X^{[10],a}_{\rho\sigma,2,lm}\equiv\frac{2\eta^{\vk\lambda}}{L^4}\lb
  B^a_{\rho\sigma}v_{\lambda,l}v_{\vk,m}+2B^a_{\vk\rho}v_{\sigma,l}v_{\lambda,m}-
  2B^a_{\lambda\sigma}v_{\rho,l}v_{\vk,m}\rb,
\end{equation}
\begin{equation}
  \mathcal{X}^{[1],a}_{\sigma,2}\equiv-\frac{2\eta^{\vk\lambda}}{L^4}\lb
  b_\lambda b_\vk B^a_\sigma+2b_\vk b_\sigma B^a_\lambda\rb,
\end{equation}
\begin{equation}
  \mathcal{X}^{[2],a}_{\sigma,2,l}\equiv-\frac{2\eta^{\vk\lambda}}{L^4}\lb
  b_\lambda b_\vk V^a_{\sigma,l}+2b_\vk b_\sigma V^a_{\lambda,l}+
  b_\lambda v_{\vk,l}B^a_\sigma+
  2b_\vk v_{\sigma,l}B^a_\lambda+v_{\lambda,l}b_\vk B^a_\sigma+
  2v_{\vk,l}b_\sigma B^a_\lambda\rb,
\end{equation}
\begin{multline}
  \mathcal{X}^{[3],a}_{\sigma,2,lm}\equiv\frac{\eta^{\vk\lambda}\eta^{\iota\tau}}{L^4}\lb
  v_{\vk\iota,l}v_{\lambda\tau,m}B^a_\sigma+2v_{\vk\iota,l}V^a_{\lambda\tau,m}b_\sigma+
  4v_{\iota\sigma,l}v_{\lambda\tau,m}B^a_\vk\right.\\\left.+
  4v_{\lambda\tau,l}V^a_{\iota\sigma,m}b_\vk+4v_{\iota\sigma,l}V^a_{\lambda\tau,m}b_\vk\rb,
\end{multline}
\begin{multline}
  \mathcal{X}^{[4],a}_{\sigma,2,lm}\equiv-\frac{2\eta^{\vk\lambda}}{L^4}\lb
  b_\lambda v_{\vk,l}V^a_{\sigma,m}+2b_\vk v_{\sigma,l}V^a_{\lambda,m}+
  v_{\lambda,l}b_\vk V^a_{\sigma,m}\right.\\\left.+
  2v_{\vk,l}b_\sigma V^a_{\lambda,m}+v_{\lambda,l}v_{\vk,m}B^a_\sigma+
  2v_{\vk,l}v_{\sigma,m}B^a_\lambda\rb,
\end{multline}
\begin{multline}
  \mathcal{X}^{[5],a}_{\sigma,2,lmn}\equiv\frac{\eta^{\vk\lambda}\eta^{\iota\tau}}{L^4}\lb
  v_{\vk\iota,l}v_{\lambda\tau,m}V^a_{\sigma,n}+2v_{\vk\iota,l}V^a_{\lambda\tau,m}v_{\sigma,n}+
  4v_{\iota\sigma,l}v_{\lambda\tau,m}V^a_{\vk,n}\right.\\\left.+
  4v_{\lambda\tau,l}V^a_{\iota\sigma,m}v_{\vk,n}+4v_{\iota\sigma,l}V^a_{\lambda\tau,m}v_{\vk,n}\rb,
\end{multline}
\begin{equation}
  \mathcal{X}^{[6],a}_{\sigma,2,lmn}\equiv-\frac{2\eta^{\vk\lambda}}{L^4}\lb
  v_{\lambda,l}v_{\vk,m}V^a_{\sigma,n}+2v_{\vk,l}v_{\sigma,m}V^a_{\lambda,n}\rb,
\end{equation}
\begin{equation}
  \mathcal{X}^{[7],a}_{\sigma,2,l}\equiv\frac{\eta^{\vk\lambda}\eta^{\iota\tau}}{L^4}\lb
  2v_{\vk\iota,l}B^a_{\lambda\tau}b_\sigma+4v_{\lambda\tau,l}B^a_{\iota\sigma}b_\vk+
  4v_{\iota\sigma,l}B^a_{\lambda\tau}b_\vk\rb,
\end{equation}
\begin{equation}
  \mathcal{X}^{[9],a}_{\sigma,2,lm}\equiv\frac{\eta^{\vk\lambda}\eta^{\iota\tau}}{L^4}\lb
  2v_{\vk\iota,l}B^a_{\lambda\tau}v_{\sigma,m}+4v_{\lambda\tau,l}B^a_{\iota\sigma}v_{\vk,m}+
  4v_{\iota\sigma,l}B^a_{\lambda\tau}v_{\vk,m}\rb,
\end{equation}
\begin{multline}
  X^{[1],ab}_{\rho\sigma,3,l}\equiv\frac{2\eta^{\vk\lambda}}{L^4}\lb
  v_{\rho\sigma,l}B^a_\vk B^b_\lambda+2b_\lambda B^a_\vk V^b_{\rho\sigma,l}+
  2v_{\vk\rho,l}B^a_\lambda B^b_\sigma+2b_\sigma B^a_\lambda V^b_{\vk\rho,l}\right.\\\left.+
  2b_\lambda V^a_{\vk\rho,l}B^b_\sigma-2b_\rho B^a_\vk V^b_{\lambda\sigma,l}-
  2v_{\lambda\sigma,l}B^a_\vk B^b_\rho-2b_\vk B^a_\rho V^b_{\lambda\sigma,l}\rb,
\end{multline}
\begin{multline}
  X^{[2],ab}_{\rho\sigma,3,lm}\equiv\frac{2\eta^{\vk\lambda}}{L^4}\lb
  v_{\rho\sigma,l}B^a_\vk V^b_{\lambda,m}+2b_\lambda V^a_{\vk,m}V^b_{\rho\sigma,l}+
  2v_{\vk\rho,l}B^a_\lambda V^b_{\sigma,m}\right.\\\left.+
  2b_\sigma V^a_{\lambda,m}V^b_{\vk\rho,l}+2b_\lambda V^a_{\vk\rho,l}V^b_{\sigma,m}-
  2b_\rho V^a_{\vk,m}V^b_{\lambda\sigma,l}-2v_{\lambda\sigma,l}B^a_\vk V^b_{\rho,m}\right.\\\left.-
  2b_\vk V^a_{\rho,m}V^b_{\lambda\sigma,l}+v_{\rho\sigma,l}V^a_{\vk,m}B^b_\lambda+
  2v_{\lambda,m}B^a_\vk V^b_{\rho\sigma,l}+2v_{\vk\rho,l}V^a_{\lambda,m}B^b_\sigma\right.\\\left.+
  2v_{\sigma,m}B^a_\lambda V^b_{\vk\rho,l}+2v_{\lambda,m}V^a_{\vk\rho,l}B^b_\sigma-
  2v_{\rho,m}B^a_\vk V^b_{\lambda\sigma,l}-2v_{\lambda\sigma,l}V^a_{\vk,m}B^b_\rho-
  2v_{\vk,m}B^a_\rho V^b_{\lambda\sigma,l}\rb,
\end{multline}
\begin{multline}
  X^{[3],ab}_{\rho\sigma,3,lmn}\equiv\frac{\eta^{\vk\lambda}\eta^{\iota\tau}}{L^4}\lb
  v_{\rho\sigma,l}V^a_{\lambda\tau,m}V^b_{\vk\iota,n}+2v_{\vk\iota,l}V^a_{\lambda\tau,m}V^b_{\rho\sigma,n}-
  4v_{\vk\rho,l}V^a_{\lambda\tau,m}V^b_{\iota\sigma,n}\right.\\\left.-
  4v_{\iota\sigma,l}V^a_{\lambda\tau,m}V^b_{\vk\rho,n}-4v_{\lambda\tau,l}V^a_{\vk\rho,m}V^b_{\iota\sigma,n}\rb,
\end{multline}
\begin{multline}
  X^{[4],ab}_{\rho\sigma,3,lmn}\equiv\frac{2\eta^{\vk\lambda}}{L^4}\lb
  v_{\rho\sigma,l}V^a_{\vk,m}V^b_{\lambda,n}+2v_{\lambda,m}V^a_{\vk,n}V^b_{\rho\sigma,l}+
  2v_{\vk\rho,l}V^a_{\lambda,m}V^b_{\sigma,n}\right.\\\left.+
  2v_{\sigma,m}V^a_{\lambda,n}V^b_{\vk\rho,l}+2v_{\lambda,m}V^a_{\vk\rho,l}V^b_{\sigma,n}-
  2v_{\rho,m}V^a_{\vk,n}V^b_{\lambda\sigma,l}\right.\\\left.-
  2v_{\lambda\sigma,l}V^a_{\vk,m}V^b_{\rho,n}-2v_{\vk,m}V^a_{\rho,n}V^b_{\lambda\sigma,l}\rb,
\end{multline}
\begin{multline}
  X^{[5],ab}_{\rho\sigma,3,lm}\equiv\frac{\eta^{\vk\lambda}\eta^{\iota\tau}}{L^4}\lb
  v_{\rho\sigma,l}B^a_{\lambda\tau}V^b_{\vk\iota,m}+2v_{\vk\iota,l}B^a_{\lambda\tau}V^b_{\rho\sigma,m}-
  4v_{\vk\rho,l}B^a_{\lambda\tau}V^b_{\iota\sigma,m}\right.\\\left.-
  4v_{\iota\sigma,l}B^a_{\lambda\tau}V^b_{\vk\rho,m}-4v_{\lambda\tau,l}B^a_{\vk\rho}V^b_{\iota\sigma,m}+
  v_{\rho\sigma,l}V^a_{\lambda\tau,m}B^b_{\vk\iota}\right.\\\left.+
  2v_{\vk\iota,l}V^a_{\lambda\tau,m}B^b_{\rho\sigma}-4v_{\vk\rho,l}V^a_{\lambda\tau,m}B^b_{\iota\sigma}-
  4v_{\iota\sigma,l}V^a_{\lambda\tau,m}B^b_{\vk\rho}-
  4v_{\lambda\tau,l}V^a_{\vk\rho,m}B^b_{\iota\sigma}\rb,
\end{multline}
\begin{multline}
  X^{[6],ab}_{\rho\sigma,3,l}\equiv\frac{\eta^{\vk\lambda}\eta^{\iota\tau}}{L^4}\lb
  v_{\rho\sigma,l}B^a_{\lambda\tau}B^b_{\vk\iota}+2v_{\vk\iota,l}B^a_{\lambda\tau}B^b_{\rho\sigma}-
  4v_{\vk\rho,l}B^a_{\lambda\tau}B^b_{\iota\sigma}\right.\\\left.-
  4v_{\iota\sigma,l}B^a_{\lambda\tau}B^b_{\vk\rho}-4v_{\lambda\tau,l}B^a_{\vk\rho}B^b_{\iota\sigma}\rb,
\end{multline}
\begin{equation}
  X^{[8],ab}_{\rho\sigma,3}\equiv\frac{2\eta^{\vk\lambda}}{L^4}\lb
  2b_\lambda B^a_\vk B^b_{\rho\sigma}+2b_\sigma B^a_\lambda B^b_{\vk\rho}+
  2b_\lambda B^a_{\vk\rho}B^b_\sigma-2b_\rho B^a_\vk B^b_{\lambda\sigma}-
  2b_\vk B^a_\rho B^b_{\lambda\sigma}\rb,
\end{equation}
\begin{multline}
  X^{[9],ab}_{\rho\sigma,3,l}\equiv\frac{2\eta^{\vk\lambda}}{L^4}\lb
  2b_\lambda V^a_{\vk,l}B^b_{\rho\sigma}+2b_\sigma V^a_{\lambda,l}B^b_{\vk\rho}+
  2b_\lambda B^a_{\vk\rho}V^b_{\sigma,l}-2b_\rho V^a_{\vk,l}B^b_{\lambda\sigma}\right.\\\left.-
  2b_\vk V^a_{\rho,l}B^b_{\lambda\sigma}+2v_{\lambda,l}B^a_\vk B^b_{\rho\sigma}+
  2v_{\sigma,l}B^a_\lambda B^b_{\vk\rho}+2v_{\lambda,l}B^a_{\vk\rho}B^b_\sigma\right.\\\left.-
  2v_{\rho,l}B^a_\vk B^b_{\lambda\sigma}-2v_{\vk,l}B^a_\rho B^b_{\lambda\sigma}\rb,
\end{multline}
\begin{multline}
  X^{[10],ab}_{\rho\sigma,3,lm}\equiv\frac{2\eta^{\vk\lambda}}{L^4}\lb
  2v_{\lambda,l}V^a_{\vk,m}B^b_{\rho\sigma}+2v_{\sigma,l}V^a_{\lambda,m}B^b_{\vk\rho}+
  2v_{\lambda,l}B^a_{\vk\rho}V^b_{\sigma,m}\right.\\\left.-
  2v_{\rho,l}V^a_{\vk,m}B^b_{\lambda\sigma}-2v_{\vk,l}V^a_{\rho,m}B^b_{\lambda\sigma}\rb,
\end{multline}
\begin{equation}
  \mathcal{X}^{[1],ab}_{\sigma,3}\equiv-\frac{2\eta^{\vk\lambda}}{L^4}\lb
  b_\sigma B^a_\vk B^b_\lambda+2b_\lambda B^a_\vk B^b_\sigma\rb,
\end{equation}
\begin{multline}
  \mathcal{X}^{[2],ab}_{\sigma,3,l}\equiv-\frac{2\eta^{\vk\lambda}}{L^4}\lb
  b_\sigma B^a_\vk V^b_{\lambda,l}+2b_\lambda B^a_\vk V^b_{\sigma,l}+
  b_\sigma V^a_{\vk,l}B^b_\lambda+2b_\lambda V^a_{\vk,l}B^b_\sigma\right.\\\left.+
  v_{\sigma,l}B^a_\vk B^b_\lambda+2v_{\lambda,l}B^a_\vk B^b_\sigma\rb,
\end{multline}
\begin{multline}
  \mathcal{X}^{[3],ab}_{\sigma,3,lm}\equiv\frac{\eta^{\vk\lambda}\eta^{\iota\tau}}{L^4}\lb
  b_\sigma V^a_{\lambda\tau,l}V^b_{\vk\iota,m}+2v_{\vk\iota,l}V^a_{\lambda\tau,m}B^b_\sigma\right.\\\left.+
  4b_\vk V^a_{\lambda\tau,l}V^b_{\iota\sigma,m}+4v_{\iota\sigma,l}V^a_{\lambda\tau,m}B^b_\vk+
  4v_{\lambda\tau,l}B^a_\vk V^b_{\iota\sigma,m}\rb,
\end{multline}
\begin{multline}
  \mathcal{X}^{[4],ab}_{\sigma,3,lm}\equiv-\frac{2\eta^{\vk\lambda}}{L^4}\lb
  b_\sigma V^a_{\vk,l}V^b_{\lambda,m}+2b_\lambda V^a_{\vk,l}V^b_{\sigma,m}+
  v_{\sigma,l}B^a_\vk V^b_{\lambda,m}+2v_{\lambda,l}B^a_\vk V^b_{\sigma,m}\right.\\\left.+
  v_{\sigma,l}V^a_{\vk,m}B^b_\lambda+2v_{\lambda,l}V^a_{\vk,m}B^b_\sigma\rb,
\end{multline}
\begin{multline}
  \mathcal{X}^{[5],ab}_{\sigma,3,lmn}\equiv\frac{\eta^{\vk\lambda}\eta^{\iota\tau}}{L^4}\lb
  v_{\sigma,n}V^a_{\lambda\tau,l}V^b_{\vk\iota,m}+2v_{\vk\iota,l}V^a_{\lambda\tau,m}V^b_{\sigma,n}\right.\\\left.+
  4v_{\vk,n}V^a_{\lambda\tau,l}V^b_{\iota\sigma,m}+4v_{\iota\sigma,l}V^a_{\lambda\tau,m}V^b_{\vk,n}+
  4v_{\lambda\tau,l}V^a_{\vk,n}V^b_{\iota\sigma,m}\rb,
\end{multline}
\begin{equation}
  \mathcal{X}^{[6],ab}_{\sigma,3,lmn}\equiv-\frac{2\eta^{\vk\lambda}}{L^4}\lb
  v_{\sigma,l}V^a_{\vk,m}V^b_{\lambda,n}+2v_{\lambda,l}V^a_{\vk,m}V^b_{\sigma,n}\rb,
\end{equation}
\begin{multline}
  \mathcal{X}^{[7],ab}_{\sigma,3,l}\equiv\frac{\eta^{\vk\lambda}\eta^{\iota\tau}}{L^4}\lb
  2v_{\vk\iota,l}B^a_{\lambda\tau}B^b_\sigma+4v_{\iota\sigma,l}B^a_{\lambda\tau}B^b_\vk+
  4v_{\lambda\tau,l}B^a_\vk B^b_{\iota\sigma}\right.\\\left.+
  b_\sigma B^a_{\lambda\tau}V^b_{\vk\iota,l}+4b_\vk B^a_{\lambda\tau}V^b_{\iota\sigma,l}+
  b_\sigma V^a_{\lambda\tau,l}B^b_{\vk\iota}+4b_\vk V^a_{\lambda\tau,l}B^b_{\iota\sigma}\rb,
\end{multline}
\begin{equation}
  \mathcal{X}^{[8],ab}_{\sigma,3}\equiv\frac{\eta^{\vk\lambda}\eta^{\iota\tau}}{L^4}\lb
  b_\sigma B^a_{\lambda\tau}B^b_{\vk\iota}+4b_\vk B^a_{\lambda\tau}B^b_{\iota\sigma}\rb,
\end{equation}
\begin{multline}
  \mathcal{X}^{[9],ab}_{\sigma,3,lm}\equiv\frac{\eta^{\vk\lambda}\eta^{\iota\tau}}{L^4}\lb
  2v_{\vk\iota,l}B^a_{\lambda\tau}V^b_{\sigma,m}+4v_{\iota\sigma,l}B^a_{\lambda\tau}V^b_{\vk,m}+
  4v_{\lambda\tau,l}V^a_{\vk,m}B^b_{\iota\sigma}+v_{\sigma,m}B^a_{\lambda\tau}V^b_{\vk\iota,l}\right.\\\left.+
  4v_{\vk,m}B^a_{\lambda\tau}V^b_{\iota\sigma,l}+v_{\sigma,m}V^a_{\lambda\tau,l}B^b_{\vk\iota}+
  4v_{\vk,m}V^a_{\lambda\tau,l}B^b_{\iota\sigma}\rb,
\end{multline}
\begin{equation}
  \mathcal{X}^{[10],ab}_{\sigma,3,l}\equiv\frac{\eta^{\vk\lambda}\eta^{\iota\tau}}{L^4}\lb
  v_{\sigma,l}B^a_{\lambda\tau}B^b_{\vk\iota}+4v_{\vk,l}B^a_{\lambda\tau}B^b_{\iota\sigma}\rb,
\end{equation}
\begin{equation}
  X^{[1],abc}_{\rho\sigma,4,l}\equiv\frac{2\eta^{\vk\lambda}}{L^4}\lb
  B^a_\vk B^b_\lambda V^c_{\rho\sigma,l}+2 B^a_\lambda V^b_{\vk\rho,l}B^c_\sigma-
  2 B^a_\vk B^b_\rho V^c_{\lambda\sigma,l}\rb,
\end{equation}
\begin{multline}
  X^{[2],abc}_{\rho\sigma,4,lm}\equiv\frac{2\eta^{\vk\lambda}}{L^4}\lb
  B^a_{z\vk}V^b_{\lambda,m}V^c_{\rho\sigma,l}+
  2 B^a_{z\lambda}V^b_{\vk\rho,l}V^c_{\sigma,m}-
  2 B^a_{z\vk}V^b_{\rho,m}V^c_{\lambda\sigma,l}\right.\\\left.+
  V^a_{\vk,m}B^b_\lambda V^c_{\rho\sigma,l}+
  2V^a_{\lambda,m}V^b_{\vk\rho,l}B^c_\sigma-
  2V^a_{\vk,m}B^b_\rho V^c_{\lambda\sigma,l}\rb,
\end{multline}
\begin{equation}
  X^{[3],abc}_{\rho\sigma,4,lmn}\equiv\frac{\eta^{\vk\lambda}\eta^{\iota\tau}}{L^4}\lb
  V^a_{\lambda\tau,l}V^b_{\vk\iota,m}V^c_{\rho\sigma,n}-4V^a_{\lambda\tau,l}V^b_{\vk\rho,m}V^c_{\iota\sigma,n}\rb,
\end{equation}
\begin{equation}
  X^{[4],abc}_{\rho\sigma,4,lmn}\equiv\frac{2\eta^{\vk\lambda}}{L^4}\lb
  V^a_{\vk,m}V^b_{\lambda,n}V^c_{\rho\sigma,l}+2V^a_{\lambda,m}V^b_{\vk\rho,l}V^c_{\sigma,n}-
  2V^a_{\vk,m}V^b_{\rho,n}V^c_{\lambda\sigma,l}\rb,
\end{equation}
\begin{multline}
  X^{[5],abc}_{\rho\sigma,4,lm}\equiv\frac{\eta^{\vk\lambda}\eta^{\iota\tau}}{L^4}\lb
  B^a_{\lambda\tau}V^b_{\vk\iota,l}V^c_{\rho\sigma,m}-4B^a_{\lambda\tau}V^b_{\vk\rho,l}V^c_{\iota\sigma,m}+
  V^a_{\lambda\tau,l}B^b_{\vk\iota}V^c_{\rho\sigma,m}\right.\\\left.-
  4V^a_{\lambda\tau,l}B^b_{\vk\rho}V^c_{\iota\sigma,m}+V^a_{\lambda\tau,l}V^b_{\vk\iota,m}B^c_{\rho\sigma}-
  4V^a_{\lambda\tau,l}V^b_{\vk\rho,m}B^c_{\iota\sigma}\rb,
\end{multline}
\begin{multline}
  X^{[6],abc}_{\rho\sigma,4,l}\equiv\frac{\eta^{\vk\lambda}\eta^{\iota\tau}}{L^4}\lb
  B^a_{\lambda\tau}B^b_{\vk\iota}V^c_{\rho\sigma,l}-
  4B^a_{\lambda\tau}B^b_{\vk\rho}V^c_{\iota\sigma,l}+
  B^a_{\lambda\tau}V^b_{\vk\iota,l}B^c_{\rho\sigma}\right.\\\left.-
  4B^a_{\lambda\tau}V^b_{\vk\rho,l}B^c_{\iota\sigma}+
  V^a_{\lambda\tau,l}B^b_{\vk\iota}B^c_{\rho\sigma}-
  4V^a_{\lambda\tau,l}B^b_{\vk\rho}B^c_{\iota\sigma}\rb,
\end{multline}
\begin{equation}
  X^{[7],abc}_{\rho\sigma,4}\equiv\frac{\eta^{\vk\lambda}\eta^{\iota\tau}}{L^4}\lb
  B^a_{\lambda\tau}B^b_{\vk\iota}B^c_{\rho\sigma}-
  4B^a_{\lambda\tau}B^b_{\vk\rho}B^c_{\iota\sigma}\rb,
\end{equation}
\begin{equation}
  X^{[8],abc}_{\rho\sigma,4}\equiv\frac{2\eta^{\vk\lambda}}{L^4}\lb
  B^a_\vk B^b_\lambda B^c_{\rho\sigma}+2B^a_\lambda B^b_{\vk\rho}B^c_\sigma-
  2 B^a_\vk B^b_\rho B^c_{\lambda\sigma}\rb,
\end{equation}
\begin{multline}
  X^{[9],abc}_{\rho\sigma,4,l}\equiv\frac{2\eta^{\vk\lambda}}{L^4}\lb
  B^a_{z\vk}V^b_{\lambda,l}B^c_{\rho\sigma}+2B^a_{z\lambda}B^b_{\vk\rho}V^c_{\sigma,l}-
  2 B^a_{z\vk}V^b_{\rho,l}B^c_{\lambda\sigma}\right.\\\left.+
  V^a_{\vk,l}B^b_\lambda B^c_{\rho\sigma}+2V^a_{\lambda,l}B^b_{\vk\rho}B^c_\sigma-
  2V^a_{\vk,l}B^b_\rho B^c_{\lambda\sigma}\rb,
\end{multline}
\begin{equation}
  X^{[10],abc}_{\rho\sigma,4,lm}\equiv\frac{2\eta^{\vk\lambda}}{L^4}\lb
  V^a_{\vk,l}V^b_{\lambda,m}B^c_{\rho\sigma}+2V^a_{\lambda,l}B^b_{\vk\rho}V^c_{\sigma,m}-
  2V^a_{\vk,l}V^b_{\rho,m}B^c_{\lambda\sigma}\rb,
\end{equation}
\begin{equation}
  \mathcal{X}^{[1],abc}_{\sigma,4}\equiv-\frac{2\eta^{\vk\lambda}}{L^4}B^a_\vk B^b_\lambda B^c_\sigma,
\end{equation}
\begin{equation}
  \mathcal{X}^{[2],abc}_{\sigma,4,l}\equiv-\frac{2\eta^{\vk\lambda}}{L^4}\lb
  B^a_\vk B^b_\lambda V^c_{\sigma,l}+B^a_\vk V^b_{\lambda,l}B^c_\sigma+
  V^a_{\vk,l}B^b_\lambda B^c_\sigma\rb,
\end{equation}
\begin{equation}
  \mathcal{X}^{[3],abc}_{\sigma,4,lm}\equiv\frac{\eta^{\vk\lambda}\eta^{\iota\tau}}{L^4}\lb
  V^a_{\lambda\tau,l}V^b_{\vk\iota,m}B^c_\sigma+4V^a_{\lambda\tau,l}B^b_\vk V^c_{\iota\sigma,m}\rb,
\end{equation}
\begin{equation}
  \mathcal{X}^{[4],abc}_{\sigma,4,lm}\equiv-\frac{2\eta^{\vk\lambda}}{L^4}\lb
  B^a_\vk V^b_{\lambda,l}V^c_{\sigma,m}+V^a_{\vk,l}B^b_\lambda V^c_{\sigma,m}+
  V^a_{\vk,l}V^b_{\lambda,m}B^c_\sigma\rb,
\end{equation}
\begin{equation}
  \mathcal{X}^{[5],abc}_{\sigma,4,lmn}\equiv\frac{\eta^{\vk\lambda}\eta^{\iota\tau}}{L^4}\lb
  V^a_{\lambda\tau,l}V^b_{\vk\iota,m}V^c_{\sigma,n}+4V^a_{\lambda\tau,l}V^b_{\vk,n}V^c_{\iota\sigma,m}\rb,
\end{equation}
\begin{equation}
  \mathcal{X}^{[6],abc}_{\sigma,4,lmn}\equiv-\frac{2\eta^{\vk\lambda}}{L^4}
  V^a_{\vk,l}V^b_{\lambda,m}V^c_{\sigma,n},
\end{equation}
\begin{equation}
  \mathcal{X}^{[7],abc}_{\sigma,4,l}\equiv\frac{\eta^{\vk\lambda}\eta^{\iota\tau}}{L^4}\lb
  B^a_{\lambda\tau}V^b_{\vk\iota,l}B^c_\sigma+4B^a_{\lambda\tau}B^b_\vk V^c_{\iota\sigma,l}+
  V^a_{\lambda\tau,l}B^b_{\vk\iota}B^c_\sigma+4V^a_{\lambda\tau,l}B^b_\vk B^c_{\iota\sigma}\rb,
\end{equation}
\begin{equation}
  \mathcal{X}^{[8],abc}_{\sigma,4}\equiv\frac{\eta^{\vk\lambda}\eta^{\iota\tau}}{L^4}\lb
  B^a_{\lambda\tau}B^b_{\vk\iota}B^c_\sigma+4B^a_{\lambda\tau}B^b_\vk B^c_{\iota\sigma}\rb,
\end{equation}
\begin{equation}
  \mathcal{X}^{[9],abc}_{\sigma,4,lm}\equiv\frac{\eta^{\vk\lambda}\eta^{\iota\tau}}{L^4}\lb
  B^a_{\lambda\tau}V^b_{\vk\iota,l}V^c_{\sigma,m}+4B^a_{\lambda\tau}V^b_{\vk,m}V^c_{\iota\sigma,l}+
  V^a_{\lambda\tau,l}B^b_{\vk\iota}V^c_{\sigma,m}+4V^a_{\lambda\tau,l}V^b_{\vk,m}B^c_{\iota\sigma}\rb,
\end{equation}
\begin{equation}
  \mathcal{X}^{[10],abc}_{\sigma,4,l}\equiv\frac{\eta^{\vk\lambda}\eta^{\iota\tau}}{L^4}\lb
  B^a_{\lambda\tau}B^b_{\vk\iota}V^c_{\sigma,l}+4B^a_{\lambda\tau}V^b_{\vk,l}B^c_{\iota\sigma}\rb.
\end{equation}
In the above the following notations have been used:
\begin{equation}
  \begin{aligned}
    B^a_{\mu\nu}&\equiv-2i\,\text{Tr}\lb\lsb B_\mu,B_\nu\rsb T^a\rb,\qquad
    &&V_{\mu\nu,n}\equiv v_{\mu\nu,n}\mathbb{1}_{N_f}+V^a_{\mu\nu,n}T^a,\\[0.2cm]
    v_{\mu\nu,n}&\equiv\partial_\mu v_{\nu,n}-\partial_\nu v_{\mu,n},\qquad
    &&V^a_{\mu\nu,n}\equiv\partial_\mu V^a_{\nu,n}-\partial_\nu V^a_{\mu,n}.
  \end{aligned}
\end{equation}

\section{On Order of Operations in Higher Order Transverse Solutions}\label{app:op_reord}

In this Appendix we discuss a complication that occurs in writing the higher order
transverse solutions using the Green function method. Consider a source term from the
factorization~\eqref{eq:src_fact} of an abelian source of the order \((j)\),
\begin{equation}
  j^{(j)}_\lambda(q,z')=\mathcal{J}^{(j)}_\lambda(q)\,g^{(j)}(z').
\end{equation}
Then the transverse part of the abelian solution corresponding to this source can be
written as
\begin{multline}\label{eq:ho_sol_trans}
  a^{(j)}_{\mu,t}(q,z)=
  \lb\delta^\lambda_\mu-D^\lambda_\mu(q)\rb\mathcal{J}^{(j)}_\lambda(q)\int\limits_{-\infty}^{+\infty}dz'\,G_t(z,z',q)\,g^{(j)}(z')\\=
  -\kappa L^4\lb\delta^\lambda_\mu-D^\lambda_\mu(q)\rb\mathcal{J}^{(j)}_\lambda(q)\int\limits_{-\infty}^{+\infty}dz'\,\sum_{n=1}^\infty\frac{\psi_n(z)\,\psi_n(z')}{q^2-\lambda_n}\,g^{(j)}(z').
\end{multline}
In principle we are not allowed to swap the order of summation and integration in~\eqref{eq:ho_sol_trans},
unless convergence of the series can be proven. This requires knowing more about the
analytical properties of the eigensystem and it also depends on the properties of the
function \(g^{(j)}\). We have devised a method which detects when changing the order of
operations is problematic.

Consider the \(z\)-dependent part of the longitudinal solution corresponding to the
transverse solution~\eqref{eq:ho_sol_trans},
\begin{equation}\label{eq:ho_w_int}
  w^{(j)}(z)\equiv\int\limits_{-\infty}^{+\infty}dz'\,G_l(z,z')\,g^{(j)}(z')=
  \kappa L^4\int\limits_{-\infty}^{+\infty}dz'\,\sum_n\frac{\psi_n(z)\,\psi_n(z')}{\lambda_n}\,g^{(j)}(z'),
\end{equation}
where it was used that \(G_t=G_l\) for \(q=0\), see~\eqref{eq:tl_gf}. If the order of
summation and integration can be exchanged, then~\eqref{eq:ho_w_int} could be rewritten
as
\begin{equation}\label{eq:ho_w_sum}
  w^{(j)}(z)=\kappa L^4\sum_n\frac{\psi_n(z)\,c^{(j)}_n}{\lambda_n},\qquad
  c^{(j)}_n\equiv\int\limits_{-\infty}^{+\infty}dz\,\psi_n(z)\,g^{(j)}(z),
\end{equation}
where \(c^{(j)}_n\) is a constant.\footnote{These constants are indeed finite for all
\(g^{(2)}_{i,\dots}\) and \(g^{(3)}_{i,\dots}\) listed in~\eqref{eq:nlo_g} and~\eqref{eq:nnlo_g}.}
In the limit \(z\to+\infty\), according to the asymptotics~\eqref{eq:trans_sus_as} of the
eigenfunctions, the right-hand side of~\eqref{eq:ho_w_sum} behaves as \(1/z\). Using the
expression~\eqref{eq:long_gf} for the longitudinal Green function, the left-hand side
of~\eqref{eq:ho_w_sum} can be shown to behave as
\begin{equation}\label{eq:as_int}
  w^{(j)}(z)\underset{z\to+\infty}{\approx}
  -\frac{L^4}{z}\int\limits_{-\infty}^{z}dz'\,w(z')\,g^{(j)}(z').
\end{equation}
If the integral in~\eqref{eq:as_int} converges in the \(z\to+\infty\) limit, then the
asymptotics on both sides of~\eqref{eq:ho_w_sum} match. This implies that changing the
order of operations was allowed.

In small number of cases (when \(g^{(j)}\) is either \(g^{(2)}_5\) or \(g^{(3)}_{31}\)),
the integral in~\eqref{eq:as_int} is divergent, which affects the asymptotics of \(w^{(j)}(z)\).
This means that the two sides of~\eqref{eq:ho_w_sum} have incompatible asymptotics, so it
is prohibited to exchange the order of summation and integration not only in~\eqref{eq:ho_w_int},
but also in the transverse component of the solution~\eqref{eq:ho_sol_trans}. For
majority of the second~\eqref{eq:nlo_g} and third~\eqref{eq:nnlo_g} order functions
\(g^{(j)}_{i,\dots}\), the above analysis shows that exchanging the order of summation and
integration is allowed.

\section{Single-Trace Operators from Effective Action Terms}\label{app:lec_terms}

In this Appendix we provide the details of the derivation of the single-trace operators~\eqref{eq:L3}
from two effective action terms.

\noindent \(\bullet\) \(\bm{S_{\text{eff},2}}\)

The first term that generates single-trace operators is \(S_{\text{eff},2}\), which is
copied below from~\eqref{eq:Seff2},
\begin{multline}\label{eq:Seff2_init}
  S_{\text{eff},2}=\frac{if_{abc}\kappa}{2(2\pi)^2L^4}\lim_{z\to+\infty}u(z)^3w(z)\int d^4q\,d^4p\,dz'
  \lb2\eta^{\mu\rho}\eta^{\nu\sigma}-\eta^{\mu\sigma}\eta^{\nu\rho}-
  \eta^{\mu\nu}\eta^{\rho\sigma}\rb\\\times
  p_\rho B^a_\sigma(p)\,B^b_\mu(q-p)\,B^c_\vk(-q)\,
  \pz\lb D^\vk_\nu(q)\,G_l(z,z')+\lb\delta^\vk_\nu-D^\vk_\nu(q)\rb G_t(z,z',q)\rb
  g^{(2)}_5(z')\\+
  \frac{f_{abc}f_{cde}\kappa}{2(2\pi)^4L^4}\lim_{z\to+\infty}u(z)^3w(z)\int d^4q\,d^4p_1\,d^4p_2\,dz'\,
  \eta^{\mu\rho}\eta^{\nu\sigma}
  B^a_\vk(-q)\,B^b_\mu(p_2)\,B^d_\rho(p_1-p_2)\,B^e_\sigma(q-p_1)\\\times
  \pz\lb D^\vk_\nu(q)\,G_l(z,z')+\lb\delta^\vk_\nu-D^\vk_\nu(q)\rb G_t(z,z',q)\rb
  g^{(3)}_{31}(z')\\+
  \frac{if_{abc}\kappa\mathcal{I}_1}{2(2\pi)^2}\int d^4q\,d^4p\,
  \eta^{\mu\sigma}\eta^{\nu\rho}
  p_\rho B^a_\sigma(p)\,B^b_\mu(q-p)\,B^c_\nu(-q)\\+
  \frac{if_{abc}\kappa\mathcal{I}_2}{2(2\pi)^2}\int d^4q\,d^4p
  \eta^{\mu\sigma}\eta^{\nu\rho}
  p_\rho B^a_\sigma(p)\,B^b_\mu(q-p)\,B^c_\nu(-q)\\-
  \frac{f_{abc}f_{cde}\kappa\mathcal{I}_3}{4(2\pi)^4}\int d^4q\,d^4p_1\,d^4p_2\,
  \eta^{\mu\sigma}\eta^{\nu\rho}
  B^a_\mu(-q)\,B^b_\nu(p_2)\,B^d_\rho(p_1-p_2)\,B^e_\sigma(q-p_1),
\end{multline}
where \(\mathcal{I}_k\) are given by~\eqref{eq:Ik} and the functions \(g^{(2)}_5\) and
\(g^{(3)}_{31}\) are defined as (see~\eqref{eq:nlo_g} and~\eqref{eq:nnlo_g}),
\begin{equation}
  g^{(2)}_5(z)=\frac{w(z)^2}{u(z)},\qquad
  g^{(3)}_{31}(z)=\frac{w(z)^3}{u(z)}.
\end{equation}
The integrals with three \(B\)-fields in~\eqref{eq:Seff2_init} will be rewritten below in
terms of four \(B\)-fields via~\eqref{eq:der_to_B}. After performing the local expansion
and truncating it to the leading term and using the relation~\eqref{eq:tl_gf} between the
longitudinal and transverse Green functions, the expression is converted into coordinate
space via~\eqref{eq:int_x_to_q_3_4},
\begin{multline}\label{eq:Seff2_1}
  \left.S_{\text{eff},2}\vphfrac\right|_\text{loc}=\frac{f_{abc}\kappa}{2L^4}\int d^4x\,
  \lb2\eta^{\mu\rho}\eta^{\nu\sigma}-\eta^{\mu\sigma}\eta^{\nu\rho}\rb
  \partial_\rho B^a_\sigma B^b_\mu B^c_\nu\\\times
  \lim_{z\to+\infty}u(z)^3w(z)\int\limits_{-\infty}^{+\infty}dz'\,\pz G_l(z,z')\,g^{(2)}_5(z')\\+
  \frac{f_{abc}f_{cde}\kappa}{2L^4}\int d^4x\,
  \eta^{\mu\rho}\eta^{\nu\sigma}
  B^a_\nu B^b_\mu B^d_\rho B^e_\sigma
  \lim_{z\to+\infty}u(z)^3w(z)\int\limits_{-\infty}^{+\infty}dz'\,\pz G_l(z,z')\,g^{(3)}_{31}(z')\\+
  \frac{f_{abc}\kappa\mathcal{I}_1}{2}\int d^4x\,
  \eta^{\mu\sigma}\eta^{\nu\rho}
  \partial_\rho B^a_\sigma B^b_\mu B^c_\nu+
  \frac{f_{abc}\kappa\mathcal{I}_2}{2}\int d^4x\,
  \eta^{\mu\sigma}\eta^{\nu\rho}
  \partial_\rho B^a_\sigma B^b_\mu B^c_\nu\\-
  \frac{f_{abc}f_{cde}\kappa\mathcal{I}_3}{4}\int d^4x\,
  \eta^{\mu\sigma}\eta^{\nu\rho}
  B^a_\mu B^b_\nu B^d_\rho B^e_\sigma,
\end{multline}
where the integral containing \(\eta^{\mu\nu}\) in the first term vanished as a result of
contraction of the symmetric metric with the antisymmetric tensors. Inserting the
derivative of the longitudinal Green function~\eqref{eq:long_gf},
\begin{multline}
  \pz G_l(z,z')=\pi L^4\lsb
  -\delta(z'-z)\lb w(z')-1\rb w(z)+
  \theta(z'-z)\frac{w(z')-1}{\pi u(z)^3}\right.\\\left.+
  \delta(z-z')\lb w(z)-1\rb w(z')+
  \theta(z-z')\frac{w(z')}{\pi u(z)^3}\rsb,
\end{multline}
the \(z\to+\infty\) limits in the first two integrals can be written as
\begin{equation}
  \lim_{z\to+\infty}u(z)^3\,w(z)\int\limits_{-\infty}^{+\infty}dz'\,\pz G_l(z,z')\,g^{(j)}_i(z')=
  L^4\lim_{z\to+\infty}w(z)\int\limits_{-\infty}^zdz'\,w(z')\,g^{(j)}_i(z').
\end{equation}
Taking these limits with the help of~\eqref{eq:w_prop} is then equivalent to computing
the integrals \(\mathcal{I}_k\), defined in~\eqref{eq:Ik}:
\begin{equation}
  \lim_{z\to+\infty}w(z)\int\limits_{-\infty}^zdz'\,w(z')\,g^{(2)}_5(z')=\mathcal{I}_2,\qquad
  \lim_{z\to+\infty}w(z)\int\limits_{-\infty}^zdz'\,w(z')\,g^{(3)}_{31}(z')=\mathcal{I}_3.
\end{equation}
This result makes it possible to write~\eqref{eq:Seff2_1} more compactly:
\begin{multline}
  \left.S_{\text{eff},2}\vphfrac\right|_\text{loc}=\frac{f_{abc}\kappa}{2}\lb\mathcal{I}_1-2\mathcal{I}_2\rb\int d^4x\,
  \eta^{\mu\sigma}\eta^{\nu\rho}
  \partial_\rho B^a_\sigma B^b_\mu B^c_\nu\\+
  \frac{f_{abc}f_{cde}\kappa\mathcal{I}_3}{4}\int d^4x\,
  \eta^{\mu\sigma}\eta^{\nu\rho}
  B^a_\mu B^b_\nu B^d_\rho B^e_\sigma.
\end{multline}
After performing the antisymmetrization of the derivative of \(B^a\) in the first
integral and inserting the definition of the structure constants~\eqref{eq:d_tens} and
the formula~\eqref{eq:ff}, it is possible to rewrite the expression in terms of the
commutators,
\begin{multline}
  \left.S_{\text{eff},2}\vphfrac\right|_\text{loc}=-\frac{i\kappa}{2}\lb\mathcal{I}_1-2\mathcal{I}_2\rb\int d^4x\,
  \eta^{\mu\sigma}\eta^{\nu\rho}
  \lb\partial_\rho B^a_\sigma-\partial_\sigma B^a_\rho\rb
  \text{Tr}\lb\lsb B_\mu,B_\nu\rsb T^a\rb\\-
  \frac{\kappa}{2}\mathcal{I}_3\int d^4x\,
  \eta^{\mu\sigma}\eta^{\nu\rho}
  \text{Tr}\lb\lsb B_\mu,B_\nu\rsb\lsb B_\rho,B_\sigma\rsb\rb.
\end{multline}
Converting the antisymmetrized derivatives of \(B^a_\mu\) into a commutator via~\eqref{eq:der_to_B}
and applying~\eqref{eq:gen_comp} in the first integral, this effective action term can be
written as
\begin{equation}\label{eq:Seff2_2}
  \left.S_{\text{eff},2}\vphfrac\right|_\text{loc}=-\frac{\kappa}{2}C\int d^4x\,
  \eta^{\mu\sigma}\eta^{\nu\rho}\,
  \text{Tr}\lb\lsb B_\mu,B_\nu\rsb\lsb B_\rho,B_\sigma\rsb\rb,\qquad
  C\equiv\mathcal{I}_1-2\mathcal{I}_2+\mathcal{I}_3.
\end{equation}
All divergences from the divergent integrals~\eqref{eq:Ik} have cancelled out in the
numerical coefficient,
\begin{equation}\label{eq:C_const}
  C=\int\limits_{-\infty}^{+\infty}dz\,\frac{w(z)^2-2w(z)^3+w(z)^4}{u(z)}\approx0.157.
\end{equation}
The effective action term~\eqref{eq:Seff2_2} can be further written as
\begin{equation}
  \left.S_{\text{eff},2}\vphfrac\right|_\text{loc}=-\frac{\kappa}{2}C\int d^4x\,
  B^a_\mu B^{\mu,b}B^c_\nu B^{\nu,d}\,
  \text{Tr}\lb\lsb T^a,T^d\rsb\lsb T^c,T^b\rsb\rb.
\end{equation}
Expanding the commutators, employing the cyclic property of the trace and renaming the
generator indices \(a\leftrightarrow b\), \(c\leftrightarrow d\) in some places gives
\begin{equation}
  \left.S_{\text{eff},2}\vphfrac\right|_\text{loc}=-\kappa C\int d^4x\,
  B^a_\mu B^{\mu,b}B^c_\nu B^{\nu,d}\,
  \text{Tr}\lb T^aT^bT^cT^d-T^aT^cT^bT^d\rb.
\end{equation}
After applying the relations~\eqref{eq:L3_to_B},
\begin{equation}\label{eq:Seff2_4}
  \begin{gathered}
    \left.S_{\text{eff},2}\vphfrac\right|_\text{loc}=L^{(l)}\int d^4x\lsb
    \text{Tr}\lb\partial_\mu\Sigma^{-1}\partial^\mu\Sigma\partial_\nu\Sigma^{-1}\partial^\nu\Sigma\rb-
    \text{Tr}\lb\partial_\mu\Sigma^{-1}\partial_\nu\Sigma\partial^\mu\Sigma^{-1}\partial^\nu\Sigma\rb\rsb,\\[0.2cm]
    L^{(l)}\equiv-\kappa C.
  \end{gathered}
\end{equation}

\noindent \(\bullet\) \(\bm{S_{\text{eff},3}}\)

Another pair of the single-trace operators~\eqref{eq:L3} arises from the first integral
in \(S_{\text{eff},3}\) with \(J^{\nu,(3)}_{11}\) (for the complete expressions see~\eqref{eq:Seff3}
and~\eqref{eq:J3_T_2_n_ab}, respectively),
\begin{multline}
  \left.S_{\text{eff},3}\vphfrac\right|_{BJ}=\sum_k\kappa^2L^4\mathcal{C}_kc^{(3)}_{11,k}\int d^4q\,
  \lb\frac{\delta^\vk_\nu}{2\lambda_k}-\frac{D^\vk_\nu(q)}{\lambda_k}+
  \frac{\delta^\vk_\nu-D^\vk_\nu(q)}{q^2-\lambda_k}\rb\\\times
  \text{STr}\lsb B_\vk(-q)\,J^{\nu,(3)}_{11}(q)\rsb,
\end{multline}
The constants \(\mathcal{C}_k\) are given by~\eqref{eq:Cn} and \(c^{(3)}_{11,k}\)
by~\eqref{eq:cij} together with~\eqref{eq:nnlo_g},
\begin{equation}
  c^{(3)}_{11,k}=\int\limits_{-\infty}^{+\infty}dz\,\psi_k(z)\,g^{(3)}_{11}(z)=
  \int\limits_{-\infty}^{+\infty}dz\,\psi_k(z)\,\pz\lb u(z)^4(\pz w(z))^3\rb,
\end{equation}
where \(\psi_k\) are the eigenfunctions of the spectral problem~\eqref{eq:spec_prob}.
After performing the local approximation and using the relation~\eqref{eq:ac} (which
removes the sum over \(k\)),
\begin{equation}\label{eq:Seff3_1}
  \left.S_{\text{eff},3}\vphfrac\right|_{BJ,\text{loc}}=-\frac{\kappa L^4}{4}a^{(3)}_{11}\int d^4q\,
  B^a_\nu(-q)\,J^{\nu,a,(3)}_{11}(q),
\end{equation}
where the symmetrized trace was expanded and the purely abelian term was omitted, since
it cannot contribute to LECs. The constant \(a^{(3)}_{11}\) is given by~\eqref{eq:aij}
together with~\eqref{eq:nnlo_g}
\begin{equation}\label{eq:a3_11}
  a^{(3)}_{11}=\int\limits_{-\infty}^{+\infty}dz\,w(z)\,g^{(3)}_{11}(z)=
  \int\limits_{-\infty}^{+\infty}dz\,w(z)\,\pz\lb u(z)^4(\pz w(z))^3\rb=
  -\frac{7\Gamma\lb\frac{1}{6}\rb}{40\pi^{7/2}\Gamma\lb\frac{2}{3}\rb}\approx-0.013.
\end{equation}
The source \(J^{a,(3)}_{\sigma,11}\) is given by~\eqref{eq:J3_T_2_n_ab}; its non-abelian
contribution
\begin{equation}\label{eq:J11}
  \left.J^{\nu,a,(3)}_{11}\vphfrac\right|_\text{non-ab}=\frac{4\eta^{\vk\lambda}(\pi\alpha')^2}{L^8}\eta^{\nu\sigma}\,
  \text{STr}\lb T^aT^bT^cT^d\rb B^b_\vk B^c_\lambda B^d_\sigma.
\end{equation}
Substituting~\eqref{eq:J11} into~\eqref{eq:Seff3_1} and converting into coordinate space
(by means of~\eqref{eq:int_x_to_q_2}),
\begin{equation}
  \left.S_{\text{eff},3}\vphfrac\right|_{BJ,\text{loc}}=-\frac{\kappa^2}{2\tilde{T}_8}a^{(3)}_{11}\int d^4x\,
  \eta^{\nu\sigma}\eta^{\vk\lambda}
  B^a_\nu B^b_\vk B^c_\lambda B^d_\sigma\,\text{STr}\lb T^aT^bT^cT^d\rb.
\end{equation}
Substituting~\eqref{eq:4gen_str} for the symmetrized trace and utilizing the symmetries
of the expression and the cyclic property of the trace gives
\begin{equation}
  \left.S_{\text{eff},3}\vphfrac\right|_{BJ,\text{loc}}=-\frac{\kappa^2}{6\tilde{T}_8}a^{(3)}_{11}\int d^4x\,
  B^a_\mu B^{\mu,b}B^c_\nu B^{\nu,d}\,
  \text{Tr}\lb2T^aT^bT^cT^d+T^aT^cT^bT^d\rb.
\end{equation}
After applying the relations~\eqref{eq:L3_to_B},
\begin{equation}\label{eq:Seff3_3}
  \begin{gathered}
    \left.S_{\text{eff},3}\vphfrac\right|_{BJ,\text{loc}}=L^{(nl)}\int d^4x\lsb
    2\text{Tr}\lb\partial_\mu\Sigma^{-1}\partial^\mu\Sigma\partial_\nu\Sigma^{-1}\partial^\nu\Sigma\rb+
    \text{Tr}\lb\partial_\mu\Sigma^{-1}\partial_\nu\Sigma\partial^\mu\Sigma^{-1}\partial^\nu\Sigma\rb\rsb,\\[0.2cm]
    L^{(nl)}\equiv-\frac{\kappa^2}{6\tilde{T}_8}a^{(3)}_{11}.
  \end{gathered}
\end{equation}

\section{Vector Meson Redefinition and Scattering Amplitude}\label{app:ss_redef}

To emphasize the point about the bare vs.~effective LECs made in Section~\ref{sect:lecs},
in this Appendix we review the original identification of LECs in~\cite{Sakai:2004cn,Sakai:2005yt}
from the point of view of their effective values (a similar analysis was performed in~\cite{Hoyos:2022ptd}).
The effective Lagrangian was obtained in~\cite{Sakai:2004cn,Sakai:2005yt} from the
leading order term of the DBI action expansion and takes the form,
\begin{multline}\label{eq:ss_chi_lagr}
  \mathcal{L}=-\text{tr}\lb\partial_\mu\Pi\partial^\mu\Pi\rb-
  \frac{1}{3f^2_\pi}\,\text{tr}\lsb\Pi,\partial_\mu\Pi\rsb^2+
  \frac{1}{2e^2_Sf^4_\pi}\,\text{tr}\lsb\partial_\mu\Pi,\partial_\nu\Pi\rsb^2\\+
  \sum_n2\,\text{tr}\lb\partial_{[\mu}v^n_{\nu]}\rb^2+
  \sum_nm^2_n\,\text{tr}\lb v^n_\mu\rb^2+
  \sum_n\frac{2b_n}{f^2_\pi}\,\text{tr}\lb\partial_{[\mu}v^n_{\nu]}\lsb\partial^\mu\Pi,\partial^\nu\Pi\rsb\rb,
\end{multline}
where \(\Pi\equiv\pi^aT^a\) is the \(\pi\)-meson matrix field and \(v^n_\mu\) are the
vector mesons. The numeric coefficients in the couplings of the quartic and trilinear
terms in the notation of the present work can be written as,
\begin{equation}
  b_n\equiv4\kappa\int\limits_{-\infty}^{+\infty}dz\,\frac{\psi_{2n-1}(z)\,w(z)\lb1-w(z)\rb}{u(z)},\qquad
  e^{-2}_S\equiv16\kappa\int\limits_{-\infty}^{+\infty}dz\,\frac{w(z)^2\lb1-w(z)\rb^2}{u(z)}.
\end{equation}
The \(\pi\pi\) elastic scattering tree level amplitude is generated by the last two terms
on the first line of~\eqref{eq:ss_chi_lagr}. The trilinear interaction term on the second
line does not contribute because of the derivative count. The result takes the form~\eqref{eq:scat_amp}
with
\begin{equation}\label{eq:eff_LECs_ss}
  \mathbb{L}_1=-\mathbb{L}_2=-\frac{1}{16e^2_S}.
\end{equation}
Consider a redefinition of the vector mesons,\footnote{For the reasoning behind this
redefinition we refer the reader to~\cite{Sakai:2004cn,Sakai:2005yt}.}
\begin{equation}\label{eq:redef_ss}
  v^n_\mu=\hat{v}^n_\mu-\frac{b_n}{2f^2_\pi}\lsb\Pi,\partial_\mu\Pi\rsb.
\end{equation}
After substituting~\eqref{eq:redef_ss} back into~\eqref{eq:ss_chi_lagr},
\begin{multline}
  \mathcal{L}=-\text{tr}\lb\partial_\mu\Pi\partial^\mu\Pi\rb+
  \lb\sum_n\frac{b^2_nm^2_n}{4f^4_\pi}-\frac{1}{3f^2_\pi}\rb\text{tr}\lsb\Pi,\partial_\mu\Pi\rsb^2+
  \lb\frac{1}{2e^2_Sf^4_\pi}-\sum_n\frac{b^2_n}{2f^4_\pi}\rb\text{tr}\lsb\partial_\mu\Pi,\partial_\nu\Pi\rsb^2\\+
  \sum_n2\,\text{tr}\lb\partial_{[\mu}\hat{v}^n_{\nu]}\rb^2+
  \sum_nm^2_n\,\text{tr}\lb\hat{v}^n_\mu\rb^2-
  \sum_n\frac{b_nm^2_n}{f^2_\pi}\,\text{tr}\lb\hat{v}^n_\mu\lsb\Pi,\partial_\mu\Pi\rsb\rb.
\end{multline}
Apparently,
\begin{equation}\label{eq:ss_coeffs_rels}
  \sum_nm^2_nb^2_n=\frac{4}{3}f^2_\pi,\qquad
  \sum_nb^2_n=e^{-2}_S,
\end{equation}
so the final form of the effective Lagrangian after the redefinition~\eqref{eq:redef_ss}:
\begin{equation}\label{eq:chi_lagr_redef_ss}
  \mathcal{L}=-\text{tr}\lb\partial_\mu\Pi\partial^\mu\Pi\rb+
  \sum_n2\,\text{tr}\lb\partial_{[\mu}\hat{v}^n_{\nu]}\rb^2+
  \sum_nm^2_n\,\text{tr}\lb\hat{v}^n_\mu\rb^2-
  \sum_n\frac{b_nm^2_n}{f^2_\pi}\,\text{tr}\lb\hat{v}^n_\mu\lsb\Pi,\partial_\mu\Pi\rsb\rb.
\end{equation}
From this it was concluded that the values of LECs are zero, since there are no
corresponding terms in the Lagrangian~\eqref{eq:chi_lagr_redef_ss}. However, a
straightforward calculation of the amplitude shows that it is given by~\eqref{eq:scat_amp_M}
with\footnote{A somewhat similar result can be found in~\cite{Hoyos:2022ptd}.
The difference between~\eqref{Astu} and the result quoted in~\cite{Hoyos:2022ptd} is due
to an additional four pion term present in the action of~\cite{Hoyos:2022ptd}.}
\begin{equation}\label{Astu}
  A(s,t,u)=-\sum_{n=1}^\infty\frac{b^2_nm^4_n}{4f^4_\pi}\lb\frac{s-u}{t-m^2_n}+\frac{s-t}{u-m^2_n}\rb.
\end{equation}
Expanding this expression for low energies, \(s,t,u\ll m^2_n\), and applying the
relations~\eqref{eq:ss_coeffs_rels} gives
\begin{equation}
  A(s,t,u)\approx\frac{s}{f^2_\pi}-\frac{s^2}{2e^2_Sf^4_\pi}+\frac{t^2+u^2}{4e^2_Sf^4_\pi}.
\end{equation}
Comparing with~\eqref{eq:scat_amp_A} reveals that the effective LECs for the Lagrangian~\eqref{eq:chi_lagr_redef_ss}
are still given by~\eqref{eq:eff_LECs_ss}. This amplitude was produced entirely by the
last term in~\eqref{eq:chi_lagr_redef_ss}. What we have demonstrated is that any
scattering experiment involving pions, as expected, is blind to the definition of the
vector mesons. In order to identify the vector mesons as physical particles one requires
an experiment with the vector mesons being external particles.

\section{Traces of Generator Products}\label{app:traces}

In this Appendix we summarize results for (symmetrized) traces of products of different
number of \(\mathfrak{su}(N_f)\) generators. The formulas below are derived from the
results collected in the review~\cite{Haber:2019sgz}.

\noindent \(\bullet\) \textbf{Symmetrized traces of one and two generators.}

\begin{equation}\label{eq:1_2gen_tr}
  \text{STr}\,T^a=\text{Tr}\,T^a=0,\qquad
  \text{STr}\lb T^aT^b\rb=\text{Tr}\lb T^aT^b\rb=\frac{1}{2}\delta_{ab},
\end{equation}
where the constant \(1/2\) is a conventional choice for the defining (fundamental)
representation. For any \(\mathfrak{su}(N_f)\) matrices \(A\) and \(B\),
\begin{equation}\label{eq:gen_comp}
  A^a=2\text{Tr}\lb AT^a\rb,\qquad
  \text{Tr}\lb AT^a\rb T^a=\frac{1}{2}A,\qquad
  \text{Tr}\lb AT^a\rb\text{Tr}\lb BT^a\rb=\frac{1}{2}\text{Tr}\lb AB\rb.
\end{equation}

\clearpage

\noindent \(\bullet\) \textbf{Symmetrized trace of three generators.}

Trace of a product of three generators is
\begin{equation}\label{eq:d_tens}
  \begin{gathered}
    \text{Tr}\lb T^aT^bT^c\rb=\frac{1}{4}\lb d_{abc}+if_{abc}\rb,\\[0.2cm]
    d_{abc}\equiv 2\text{Tr}\lb\lbrace T^a,T^b\rbrace T^c\rb,\qquad
    f_{abc}=-2i\,\text{Tr}\lb\lsb T^a,T^b\rsb T^c\rb,
  \end{gathered}
\end{equation}
where \(d_{abc}\) is a totally symmetric tensor. From this and the cycling property of
the trace,
\begin{equation}\label{eq:3gen_symtr}
  \text{STr}\lb T^aT^bT^c\rb=
  \frac{1}{2}\lsb\text{Tr}\lb T^aT^bT^c\rb+\text{Tr}\lb T^bT^aT^c\rb\rsb=
  \frac{1}{4}d_{abc}.
\end{equation}
For \(N_f=2\) the \(d\)-tensor is a zero tensor, so the symmetrized trace is equal to
zero too.

\noindent \(\bullet\) \textbf{Symmetrized trace of four generators.}

Due to the cycling property of the trace,
\begin{multline}\label{eq:4gen_str}
  \text{STr}\lb T^aT^bT^cT^d\rb=\frac{1}{6}\,\text{Tr}\lb
  T^aT^bT^cT^d+T^aT^bT^dT^c+T^aT^cT^bT^d\right.\\\left.+
  T^aT^dT^bT^c+T^aT^dT^cT^b+T^bT^aT^cT^d\rb.
\end{multline}
In terms of the \(f\)- and \(d\)-tensors, the trace of four generators is
\begin{multline}\label{eq:4gen_tr}
  \text{Tr}\lb T^aT^bT^cT^d\rb=\frac{1}{4N_f}\lb\delta_{ab}\delta_{cd}-\delta_{ac}\delta_{bd}+\delta_{ad}\delta_{bc}\rb+
  \frac{1}{8}\lb d_{abe}d_{cde}-d_{ace}d_{bde}+d_{ade}d_{bce}\rb\\+
  \frac{i}{8}\lb d_{abe}f_{cde}+d_{ace}f_{bde}+d_{ade}f_{bce}\rb.
\end{multline}
From the following two identities,
\begin{equation}
  f_{abe}d_{cde}+f_{ace}d_{bde}+f_{ade}d_{bce}=0,\qquad
  f_{abe}d_{cde}+f_{cbe}d_{ade}+f_{dbe}d_{ace}=0,
\end{equation}
it follows that (by subtracting them),
\begin{equation}
  f_{ace}d_{bde}+f_{ade}d_{bce}-f_{cbe}d_{ade}-f_{dbe}d_{ace}=0,
\end{equation}
which together with the antisymmetry of \(f_{abc}\) leads to
\begin{equation}
  \text{STr}\lb T^aT^bT^cT^d\rb=
  \frac{1}{12N_f}\lb\delta_{ab}\delta_{cd}+\delta_{ac}\delta_{bd}+\delta_{ad}\delta_{bc}\rb+
  \frac{1}{24}\lb d_{abe}d_{cde}+d_{ace}d_{bde}+d_{ade}d_{bce}\rb.
\end{equation}
The expression simplifies for two particular values of \(N_f\), \(N_f=2\) and \(N_f=3\),
which, incidentally, are the two physically interesting choices. While for \(N_f=2\) the
\(d\)-tensors vanish, for \(N_f=3\), the following identity holds:
\begin{equation}\label{eq:d_tens_rel}
  d_{abe}d_{cde}+d_{ace}d_{bde}+d_{ade}d_{bce}=
  \frac{1}{3}\lb\delta_{ab}\delta_{cd}+\delta_{ac}\delta_{bd}+\delta_{ad}\delta_{bc}\rb.
\end{equation}
In both of these cases the result for the symmetrized trace
\begin{equation}
  N_f=2,3\text{:}\qquad
  \text{STr}\lb T^aT^bT^cT^d\rb=
  \frac{1}{24}\lb\delta_{ab}\delta_{cd}+\delta_{ac}\delta_{bd}+\delta_{ad}\delta_{bc}\rb.
\end{equation}
The product of two structure constants with one index contracted can be related to the
trace of the product of two commutators~\cite{Haber:2019sgz},
\begin{equation}\label{eq:ff}
  \begin{gathered}
    f_{abc}f_{cde}=-2\,\text{Tr}\lb\lsb T^a,T^b\rsb\lsb T^d,T^e\rsb\rb=
    \frac{2}{N_f}\lb\delta_{ad}\delta_{be}-\delta_{ae}\delta_{bd}\rb+
    d_{adf}d_{bef}-d_{aef}d_{bdf},\\[0.2cm]
    N_f=2\text{:}\qquad
    f_{abc}f_{cde}=\delta_{ad}\delta_{be}-\delta_{ae}\delta_{bd}.
  \end{gathered}
\end{equation}

\noindent \(\bullet\) \textbf{Usefull expressions for the symmetrized traces.}

For two \(U(N_f)\) matrices, \(A\) and \(B\), decomposed into abelian and non-abelian
parts,
\begin{equation}
  A=a\mathbb{1}_{N_f}+A^aT^a,\qquad
  B=b\mathbb{1}_{N_f}+B^aT^a,
\end{equation}
using~\eqref{eq:1_2gen_tr} and~\eqref{eq:3gen_symtr},
\begin{equation}
  2\,\text{STr}\lsb AB\rsb=2N_fab+A^aB^a,\qquad
  2\,\text{STr}\lsb ABT^a\rsb=aB^a+A^ab+\frac{1}{2}d_{abc}A^bB^c.
\end{equation}

\section{Fourier Transform and Longitudinal Projector}\label{app:fourier_proj}

In this Appendix we summarize notations related to the Fourier transform with respect to
4D coordinates and briefly discuss the longitudinal projector.

\noindent \(\bullet\) \textbf{Notations and definitions.}

The Fourier transform \(\mathcal{F}\) and its inverse \(\mathcal{F}^{-1}\) are defined as
\begin{equation}
  \begin{gathered}
    F(q,z)\equiv\mathcal{F}\lsb F(x,z)\rsb=
    \frac{1}{(2\pi)^2}\int\limits_{-\infty}^{+\infty}d^4x\,F(x,z)\,e^{-iq\cdot x},\\[0.2cm]
    F(x,z)\equiv\mathcal{F}^{-1}\lsb F(q,z)\rsb=
    \frac{1}{(2\pi)^2}\int\limits_{-\infty}^{+\infty}d^4q\,F(q,z)\,e^{iq\cdot x}.
  \end{gathered}
\end{equation}
where \(q\cdot x\equiv q_\mu x^\mu\). The Fourier transform of the derivative (and its
inverse transform):
\begin{equation}
  \mathcal{F}\lsb\partial_\mu F(x,z)\rsb=iq_\mu F(q,z),\qquad
  \mathcal{F}^{-1}\lsb q_\mu F(q,z)\rsb=-i\partial_\mu F(x,z).
\end{equation}
The notation for the momentum squared:
\begin{equation}\label{eq:momsq}
  q^2\equiv\eta^{\mu\nu}q_\mu q_\nu.
\end{equation}
The integral representation for the delta-function,
\begin{equation}
  \int d^4x\,e^{iqx}=(2\pi)^4\delta(q).
\end{equation}

\noindent \(\bullet\) \textbf{Convolution.}

\begin{equation}\label{eq:int_x_to_q_2}
  \mathcal{F}\lsb FG\rsb(q)=\frac{1}{(2\pi)^2}\int d^4p\,F(p)\,G(q-p),\qquad
  \int d^4x\,A(x)\,B(x)=\int d^4q\,A(q)\,B(-q),
\end{equation}
\begin{equation}\label{eq:int_x_to_q_3_4}
  \begin{gathered}
    \int d^4x\,A(x)\,B(x)\,C(x)=
    \frac{1}{(2\pi)^2}\int d^4q\,d^4p\,A(-q)\,B(p)\,C(q-p),\\[0.2cm]
    \int d^4x\,A(x)\,B(x)\,C(x)\,D(x)=
    \frac{1}{(2\pi)^4}\int d^4q\,d^4p\,d^4p'\,A(-q)\,B(p-p')\,C(p')\,D(q-p).
  \end{gathered}
\end{equation}

\noindent \(\bullet\) \textbf{Longitudinal projector }\(\bm{D^\nu_\mu}\)\textbf{.}

\begin{equation}\label{eq:proj_def}
  D^\nu_\mu(q)\equiv\frac{\eta^{\sigma\nu}q_\mu q_\sigma}{q^2},\qquad
  \eta^{\mu\nu}D^\rho_\mu D^\sigma_\nu=\eta^{\mu\sigma}D^\rho_\mu,\qquad
  \veps^{z\mu\nu\rho\sigma}q_\mu D^\lambda_\sigma=0,
\end{equation}
\begin{equation}
  \begin{split}
    \eta^{\mu\nu}D^\rho_\mu\lb\delta^\sigma_\nu-D^\sigma_\nu\rb&=
    \eta^{\mu\nu}q_\mu\lb\delta^\sigma_\nu-D^\sigma_\nu\rb=0,\\[0.2cm]
    \eta^{\mu\nu}\lb\delta^\rho_\mu-D^\rho_\mu\rb\lb\delta^\sigma_\nu-D^\sigma_\nu\rb&=
    \eta^{\nu\rho}\lb\delta^\sigma_\nu-D^\sigma_\nu\rb.
  \end{split}
\end{equation}
For the abelian part of the \(B\)-field~\eqref{eq:lo_sols_amp},
\begin{equation}\label{eq:d_ab_prop}
  b_\lambda D^\lambda_\mu=b_\mu,\qquad
  \eta^{\mu\nu}b_\mu D^\sigma_\nu=\eta^{\mu\sigma}b_\mu,\qquad
  \veps^{z\mu\nu\rho\sigma}q_\mu b_\nu=0.
\end{equation}

\end{document}